\documentclass[twocolumn]{aastex631}
\usepackage{comment}
\usepackage{enumitem}
\newcommand{\rvir}{R_{\rm vir}}

\newcommand{\ztwo}{$z=2$}
\newcommand{\zzero}{$z=0$}

\newcommand{\thel}{$\theta_\mathrm{L}$}
\newcommand{\phil}{$\phi_\mathrm{L}$}
\newcommand{\philthel}{$\phi_\mathrm{L}\,$-$\,\theta_\mathrm{L}$}

\usepackage{color}
\usepackage{graphics}
\usepackage{amsmath}
\usepackage{multirow}
\usepackage{lipsum}
\graphicspath{{./}{./}}
\newcommand{\mgii}{\ion{Mg}{2}}

\defcitealias{peeples19}{FOGGIE~I}
\defcitealias{corlies20}{FOGGIE~II}
\defcitealias{zheng20}{FOGGIE~III}
\defcitealias{simons20}{FOGGIE~IV}
\defcitealias{lochhaas21}{FOGGIE~V}
\defcitealias{lochhaas23}{FOGGIE~VI}
\defcitealias{wright24}{FOGGIE~VII}
\defcitealias{Acharyya24}{FOGGIE~VIII}

\usepackage{ragged2e}
\shorttitle{FOGGIE: The Angular Momentum Evolution of Milky Way-like Galaxies and their CGM}
\shortauthors{Simons et al.}

\begin{document}

\title{\large \textbf{Figuring Out Gas \& Galaxies in Enzo (FOGGIE). IX: The Angular Momentum Evolution of Milky Way-like Galaxies and their Circumgalactic Gas}}

\correspondingauthor{Raymond C.\ Simons}
\email{rcsimons@providence.edu}

\author[0000-0002-6386-7299]{Raymond C.\ Simons}
\affiliation{Department of Engineering and Physics, Providence College, 1 Cunningham Sq, Providence, RI 02918 USA}

\author[0000-0003-1455-8788]{Molly S.\ Peeples}
\affiliation{Space Telescope Science Institute, 3700 San Martin Dr., Baltimore, MD 21218}
\affiliation{Center for Astrophysical Sciences, William H.\ Miller III Department of Physics \& Astronomy, Johns Hopkins University, 3400 N.\ Charles Street, Baltimore, MD 21218}

\author[0000-0002-7982-412X]{Jason Tumlinson}
\affiliation{Space Telescope Science Institute, 3700 San Martin Dr., Baltimore, MD 21218}
\affiliation{Center for Astrophysical Sciences, William H.\ Miller III Department of Physics \& Astronomy, Johns Hopkins University, 3400 N.\ Charles Street, Baltimore, MD 21218}

\author[0000-0002-2786-0348]{Brian W.\ O'Shea}
\affiliation{Department of Computational Mathematics, Science, and Engineering, 
Department of Physics and Astronomy,
and Facility for Rare Isotope Beams,  
Michigan State University}

\author[0000-0003-1785-8022]{Cassandra Lochhaas}
\affiliation{Space Telescope Science Institute, 3700 San Martin Dr., Baltimore, MD 21218}
\affiliation{Center for Astrophysics, Harvard \& Smithsonian, 60 Garden St., Cambridge, MA 02138}
\affiliation{NASA Hubble Fellow}

\author[0000-0002-1685-5818]{Anna C.\ Wright}
\affiliation{Center for Astrophysical Sciences, William H.\ Miller III Department of Physics \& Astronomy, Johns Hopkins University, 3400 N.\ Charles Street, Baltimore, MD 21218}
\affiliation{Center for Computational Astrophysics, Flatiron Institute, 162 Fifth Avenue, New York, NY 10010, USA}

\author[0000-0003-4804-7142]{Ayan Acharyya}
\affiliation{Center for Astrophysical Sciences, William H.\ Miller III Department of Physics \& Astronomy, Johns Hopkins University, 3400 N.\ Charles Street, Baltimore, MD 21218}
\affiliation{INAF - Astronomical Observatory of Padova, vicolo dell’Osservatorio 5, IT-35122 Padova, Italy}

\author[0000-0001-7472-3824]{Ramona Augustin}
\affiliation{Space Telescope Science Institute, 3700 San Martin Dr., Baltimore, MD 21218}
\affiliation{Leibniz-Institut f{\"u}r Astrophysik Potsdam (AIP), An der Sternwarte 16, 14482 Potsdam, Germany}

\author[0000-0003-1296-8775]{Kathleen A. Hamilton-Campos}
\affiliation{Center for Astrophysical Sciences, William H.\ Miller III Department of Physics \& Astronomy, Johns Hopkins University, 3400 N.\ Charles Street, Baltimore, MD 21218}
\affiliation{Applied Physics Laboratory,
11100 Johns Hopkins Rd,
Laurel, MD 20723, USA}

\author[0000-0002-6804-630X]{Britton D.\ Smith}
\affiliation{Institute for Astronomy, University of Edinburgh, Royal Observatory, EH9 3HJ, UK}

\author[0000-0001-9158-0829]{Nicolas Lehner}
\affiliation{Department of Physics and Astronomy, University of Notre Dame, Notre Dame, IN 46556, USA}

\author[0000-0002-0355-0134]{Jessica K.\ Werk}
\affiliation{Department of Astronomy, University of Washington, Seattle, WA 98195}

\author[0000-0003-4158-5116]{Yong Zheng}
\affiliation{Department of Physics, Applied Physics and Astronomy, Rensselaer Polytechnic Institute, Troy, NY 12180}

\begin{abstract} 
We investigate the co-evolution of the angular momentum of Milky Way-like galaxies, their circumgalactic gas, and their dark matter halos using zoom-in simulations from the Figuring Out Gas \& Galaxies in Enzo (FOGGIE) suite. We examine how the magnitude and orientation of the angular momentum varies over time within the halo and between the components of mass. From $z\sim2$ to today, and in general across the simulated halos, the specific angular momenta of the central galaxies and the cool gas in their circumgalactic media ($T<10^5\,\mathrm{K}$) increase together. Over that same period, the specific angular momenta of the hot ($>\,10^6\,\mathrm{K}$) and dark components of the halo change minimally. By $z\sim 1$, the central galaxies have generally lost association with the angular momentum of their full dark matter halo---both in magnitude and orientation. We find a wide distribution of angular momentum orientations in the halo, varying by up to $180^{\circ}$ over small ($\sim$tens of kpc) scales and between the different components of mass. The {\emph{net}} angular momenta of the galaxies, their circumgalactic gas, and their dark matter halos are generally misaligned with one another at all cosmic times. The present-day orientation of the central galaxies are established at late times (after $z=1$), after the rates of cosmic accretion and mergers decline and the disks are able to settle and stabilize their orientation.
\end{abstract}

\section{Introduction}

Angular momentum is one of the key physical properties of astrophysical systems, along with mass and energy. In a closed and non-expanding system, the total angular momentum is conserved. In an open system like a galaxy and its halo, angular momentum can be gained from or lost to the local environment similar to mass and energy. Like mass and energy, angular momentum is distributed among the constituents of the system---every element of mass carries its own unique angular momentum which contributes to the total. Angular momentum can be exchanged and re-distributed around the system through long-range action (e.g., gravitational torques), direct collisions, or through bulk flows. However, unlike the scalar quantities of mass and energy, angular momentum is a vector described by both a magnitude and direction. To {\emph{fully}} describe the angular momentum of a system, it is useful to consider it as a distribution of vector elements which each carry a magnitude and direction.

In a galaxy, the spatial distribution of the angular momentum vector determines basic properties of its identity---including its kinematics, size, and morphology. Understanding how angular momentum is gained, lost, and transported around galaxies and their halos is critical for explaining how galaxies take shape as they grow.

%Milky Way focused
%potential high time variability
%full distribution of angular momentum

Modern numerical simulations are now able to produce galaxies with kinematic and structural properties that generally reflect those of the observed galaxy population over $0\,\lesssim\,z\,\lesssim\,2$ (e.g., \citealt{2009ApJ...703..785D, 2014MNRAS.443.3675M, welker14, kassin14, genel15, 2015MNRAS.446..521S, Lagos17, 2017MNRAS.467.2430M, 2017MNRAS.467..179G, 2017MNRAS.464..635M, GK18, 2018MNRAS.477.1536E, 2019MNRAS.490.3196P, Bird21, RG22, 2023arXiv230314210M, Semenov23a, Semenov23b, Grand24, McCluskey24}). Such simulations broadly indicate that there are four primary channels for angular momentum transfer that affect the baryons and dark matter (to varying degrees): inflow, outflow, mergers, and gravitational torques. The cold filamentary accretion of baryons from the cosmic web through the circumgalactic medium and onto the galactic disk \citep{keres05,Brooks09, keres09,dekel09} preferentially adds angular momentum to the central galaxy.  It is now well-established that the baryons inside (simulated) halos can carry a higher specific angular momentum ($j$) than that of the dark matter \citep{Kimm11,2013ApJ...769...74S, 2015MNRAS.449.2087D}, mainly as a result of cold filamentary accretion of baryons. Moreover, outflows can also increase the galactic $j$ by preferentially carrying away low-angular momentum gas which may later re-accrete with higher $j$ (e.g., \citealt{Brook12, ubler14}). It is clear that the circumgalactic medium (CGM) surrounding galaxies is an important arena for angular momentum exchange.

In the last few years, numerical simulations have begun to resolve the CGM with enough detail to capture small-scale fracturing and mixing that are important for modeling its dynamical and kinematic evolution \citep{Peeples19, vandevoort19, hummels19, Kopenhafer23, lochhaas23}. The CGM, qualitatively defined as the volume between edge of the interstellar medium and the virial radius, comprises the majority of the total mass of baryons in the halo \citep{werk14, tpw17}. The CGM is also a significant reservoir of angular momentum. For instance, the Milky Way's hot circumgalactic halo carries a level of angular momentum that is comparable to that of the stellar disk  \citep{2016ApJ...822...21H}, and the Milky Way is still actively accreting angular momentum from the CGM through cold cloud accretion (e.g., \citealt{Lockman23}).

Outside of the Milky Way, kinematic measurements of CGM gas have now been carried out from the present-day to $z\sim2$ (e.g., \citealt{Ho17, augustin18, Lopez18, lochhaas19, Zabl19, Martin19, Lopez20, lehner21}) --- allowing us to probe the character of the baryonic angular momentum in the halos of galaxy populations over a wide span in cosmic time. Despite empirical and numerical progress, there are major open theoretical questions regarding the role that the CGM plays in regulating the angular momentum evolution of galaxies, and how the angular momenta of the CGM gas, central galaxy, and dark matter halo are related.

In this paper, we study the spatially-resolved time evolution of the angular momentum content of Milky Way-mass galaxies and their halos. To that end, we use the FOGGIE simulations of six Milky Way-like halos to probe: (1) how angular momentum is distributed in physical space and among the components of mass in the halo, (2) how the angular momentum of the central galaxy is related to the distribution of angular momenta in the CGM and DM halo, and (3) how angular momentum is re-distributed around the FOGGIE halos over time. We examine the simulation outputs at a hyper-fine cadence of $\Delta t\sim 5$\, Myr. By design, the FOGGIE simulations enhance resolution in the diffuse CGM, with smaller physical resolution and much finer mass resolution than most zoom simulations. We consider the evolution of both the magnitude and orientation of angular momentum, and we account for the fact that the gaseous and DM components fill out a distribution of angular momentum vectors. Finally, we also study how the angular momentum is distributed as a function of gas temperature and amongst the phases of mass---gas, stars, and dark matter.

The outline of this paper is as follows. In \S\ref{sec:simulations}, we review the details of the simulations. In \S\ref{sec:measurements}, we discuss the measurements of angular momentum in the simulations. In \S\ref{sec:AM_magnitude} and \S\ref{sec:AM_orientation}, respectively, we examine the evolution of the magnitude and orientation of angular momentum in the FOGGIE halos as a function of time, position in the halo, and component of mass. Then in \S\ref{sec:inflowing_outflowing}, we measure the characteristics of the angular momentum of inflowing and outflowing gas and relate that back to the time evolution of the central disks. Finally, in \S\ref{sec:discussion} we discuss our results and in \S\ref{sec:conclusions} we briefly conclude.

\section{FOGGIE Simulations}\label{sec:simulations}

\subsection{Description of the Simulations}\label{subsec:desc_sims}

We base our analysis on six zoom-in galaxy formation simulations of Milky Way-like halos from the Figuring Out Gas \& Galaxies in Enzo (FOGGIE) suite. The FOGGIE simulations were run using the open-source adaptive mesh refinement code Enzo \citep{bryan14, brummel-smith19}. These simulations are distinguished from most other zoom-ins in that they employ a refinement scheme that allocates high resolution over a large volume of the circumgalactic medium (CGM). 

The FOGGIE simulations have been used to study, interpret, and produce observable predictions for a range of CGM and galaxy science: absorption \citep{Peeples19} and emission-line \citep{corlies20} observations of the CGM of external galaxies, absorption-line observations of the CGM of the Milky Way \citep{zheng20}, ram pressure stripping in the group environment \citep{simons20}, the kinematic support and dynamical equilibrium of circumgalactic gas \citep{lochhaas21, lochhaas23}, the origin and evolution of stellar halos in galaxies \citep{wright24}, the spatially-resolved evolution of metallicity in galaxies  \citep{Acharyya24}, the physical and chemical properties of gas clumps in the CGM (R.\ Augustin et al., in prep), and accretion onto galaxies from the CGM (C.\ Lochhaas et al., in prep).

The FOGGIE simulations were introduced in \citet{Peeples19}. The first generation of simulations (discussed further in \citealt{corlies20, zheng20}) implement a novel ``forced refinement'' resolution scheme. In this scheme, every gas cell in a fixed box centered on and moving with the galaxy is forced to a minimum comoving spatial resolution. The FOGGIE simulations use a force-refined box with length $200\,\mathrm{kpc}\,h^{-1}$ comoving on a side. At each timestep of the simulations, cells within this box are allowed to further refine based on density---the scheme followed by most adaptive mesh simulations. By forcing a minimum resolution for every cell in the track box, the FOGGIE simulations capture small-scale gas physics (cooling, mixing, clumping) in the warm and diffuse gas in the CGM that would otherwise be poorly resolved in a refinement scheme based on density alone. 

The second-generation of the simulations were introduced in \citet{simons20} and implement a similar but more computationally-efficient ``forced + {\emph{cooling}}'' refinement scheme. As before, cells inside the tracking box are force-refined to a minimum resolution. For this generation of simulations, we use a resolution of 1100 pc comoving ($n_{\mathrm{ref}}=9$ in our $100 h^{-1}$\,Mpc comoving box with a $256^3$ root grid). This is one level more coarse than the level of force-refinement implemented in the first generation of simulations---and thus less computationally-expensive. To compensate for the lower force-refinement level, an additional ``cooling refinement'' criterion is added to better resolve thermally unstable gas. With cooling refinement, cells are refined so that their size is smaller than the cooling length (sound speed $\times$ cooling time)---up to a maximum refinement of $n_{\mathrm{ref}}=11$ or 274 comoving parsecs. As in the first generation, cells are also allowed to refine up to $n_{\mathrm{ref}}=11$ based on density. Nearly all of the gas in the interstellar medium is resolved to this maximum level. With the forced + cooling refinement scheme, we are able to safely reduce the forced resolution of the full refinement region without under-resolving gas in the portions of the CGM with short(-er) cooling times. Taken together, the second-generation of simulations are more computationally-efficient than the first-generation. They allocate resolution to resolve cooling where it is needed (to a maximum refinement level), while not over-resolving the diffuse hot gas (with long cooling times) where it is not needed. At \ztwo{}, and in all six halos, the cooling length is resolved in more than 90\% of the volume and 99\% of the mass in the track box \citep{simons20}.

In this paper, we use the second-generation FOGGIE simulations.  We refer the reader to previous work (FOGGIE IV--VII) for more details. In addition to the refinement scheme described above and the evolution to $z=2$ presented in \citet{simons20}, \citet{lochhaas21,lochhaas23} present the thermodynamic properties of the CGM to $z=0$. We refer the reader to Figure 17 of \citet{lochhaas23} for the relevance of high spatial resolution in capturing the effects of turbulence and the kinematics of the hot gas. Likewise, we refer the reader to \citet{wright24} for a thorough discussion of the merger histories of these halos. As of the analysis presented in this paper, four of the six halos (``Tempest'', ``Squall'', ``Blizzard'', and ``Maelstrom'') had been evolved down to $z=0$. The ``Hurricane'' halo had reached $z=0.2$ and the ``Cyclone'' halo had reached $z=0.95$.  All snapshot outputs of the simulations are written out and stored with a high cadence---every 5.4 Myr of cosmic time.

The initial conditions of the FOGGIE halos were selected from a 100 Mpc $h^{-1}$ dark matter-only pathfinder simulation. The six halos were selected to be similar to that of the Milky Way in terms of \zzero{} halo mass ($\sim\,1\,\times\,10^{12}$ M$_{\odot}$; \citealt{2016ARA&A..54..529B}). The virial masses of the FOGGIE halos span $0.1-0.8\,\times\,10^{12}$ M$_{\odot}$ at \ztwo{} and $0.6-1.5\,\times\,10^{12}$ M$_{\odot}$ at \zzero{}. The halos were also selected because they undergo their last major merger (10:1 mass ratio or lower) at or before $z=2$, which is when the Milky Way is thought to have had its last major merger \citep{2018Natur.563...85H}. Despite this selection, the production run of the Squall halo does experience a 2:1 major merger at $z\sim0.7$. The $z\,<\,2$ accretion and merger histories of the FOGGIE central galaxies are discussed further in a later section of this paper (\S7.1).

The satellite galaxies (subhalos) of the FOGGIE halos were identified in \citep{wright24} using the {\tt{ROCKSTAR}} halo finder. \citet{wright24} produced satellite catalogs for of the simulation snapshots studied in this paper. We adopt these catalogs to identify gas in the virial volume that is spatially-associated with a nearby satellite for each snapshot (defined here as mass within 1 pkpc of the center of any satellite). We refer to \citep{wright24} for full details on the satellite/subhalo identification procedure.

\subsection{$R_{200}$, $M_{200}$, and Gas Temperature Classifications}\label{subsec:defs_sims}

In this paper, we will sometimes normalize the radial distance of a location in the halo by the virial radius ($\rvir$) of the halo. We define the virial radius as that of a sphere that is centered on the halo center and that, on average, contains 200 times the critical density of the Universe at that redshift. In the literature, this definition is also commonly referred to as $R_{\mathrm{200}}$. In the four halos that had been run to \zzero{} by the time of this analysis, $\rvir$ varies from 50--80 proper kpc at \ztwo{} and 170--225 at \zzero. 

\citet{lochhaas21} compute a more strict definition of $R_{\mathrm{vir}}$ that allows for the overdensity factor (e.g., 200 in $R_{200}$) to evolve with redshift \citep{bryan98}. Under this definition, they show that the virial radius of the Tempest halo is larger than $R_{200}$ by $\sim 50$\,kpc at \zzero{} and by smaller amounts at higher redshifts. Since the precise definition of R$_{\mathrm{vir}}$ is not important for the conclusions of this paper, and to facilitate comparisons with previous works, we opt to use the more common (albeit less physically-motivated) definition of $R_{\mathrm{vir}} \equiv R_{200}$. 

At both $z=2$ and $z=0$, the half-length of the high-resolution track box is slightly smaller than the $R_{\mathrm{vir}}$ of the halos. However, the distance from the center of the trackbox to its corner (i.e., $\sqrt{3}\,\times\,100\,\mathrm{kpc}\,h^{-1}$ comoving) is larger than $R_{\mathrm{vir}}$. As such, the virial volume falls partially outside the tracking box.  We note that, empirically, the CGM does not end at the virial radius. Strong HI absorption ($N_{\mathrm{HI}}>10^{14}$ cm$^{-2}$) associated with the bound CGM around galaxies is regularly detected out to $\sim2$--$3\times R_{\mathrm{vir}}$ \citep{wilde21, wilde23}. 

Throughout this paper, we assign baryons to one of five categories: stars, cold gas ($<15{,}000\,\mathrm{K}$), warm gas ($15{,}000\,-\,10^5\,\mathrm{K}$), warm-hot gas ($10^5\,-\,10^6\,\mathrm{K}$), and hot gas ($>10^6\,\mathrm{K}$). These distinctions in temperature correspond to gas with generally distinct kinematics and which occupy different regions of the halo \citep{simons20, lochhaas21, lochhaas23}. For instance, in general across all six halos and at all cosmic times studied here, cold gas is found in the interstellar medium of the central and satellite galaxies and in clumps in the inner CGM. Warm gas is found in the wide filaments feeding the halo from the large-scale cosmic web. Warm and warm-hot gas fill the majority of the volume of the halo. This warmer volume-filling gas is generally a factor of $\sim2\times$ lower in temperature than that expected from classic virial equilibrium arguments \citep{lochhaas23} due to its high dynamical support from bulk non-thermal motions (rotation and turbulence). Hot gas in the FOGGIE halos is generally found only in the star-formation driven outflows. These outflows will either cool and mix into the volume-filling warm and warm-hot circumgalactic medium or escape the halo entirely.

\begin{deluxetable*}{rccccccccccccccccccc}
\tablecolumns{11}
\tablewidth{0pt}
\tablecaption{Properties of FOGGIE Halos inside $R_{\mathrm{vir}}$ \label{table:angmom_rvir}}
\tablehead{
\colhead{Halo} & 
\multicolumn{4}{c}{$M_{\mathrm{vir}}$} &
\multicolumn{4}{c}{$R_{\mathrm{vir}}$} &
\multicolumn{4}{c}{$L_{\mathrm{vir, bary}}$} &
\multicolumn{3}{|c}{\multirow{2}{*}{\fontsize{13}{15}\selectfont$\frac{L_{\mathrm{vir,bary, z=0}}}{L_{\mathrm{vir, bary, z=2}}}$}} &
\multicolumn{3}{c}{\multirow{2}{*}{\fontsize{13}{15}\selectfont$\frac{(R_{\mathrm{vir}}\, M_{\mathrm{vir}}\, v_{\mathrm{vir}})_{z=0}}{(R_{\mathrm{vir}}\, M_{\mathrm{vir}}\, v_{\mathrm{vir}})_{z=2}}$}}
\\
\colhead{} & 
\multicolumn{4}{c}{($10^{12}\,\mathrm{M}_{\odot}$)} &
\multicolumn{4}{c}{(pkpc)} &
\multicolumn{4}{c}{($10^{12}\,\mathrm{kpc}\,\mathrm{km}\,\mathrm{s}^{-1}$)} &
\multicolumn{3}{|c}{} &
\multicolumn{3}{c}{} \\
\colhead{} & 
\multicolumn{4}{c}{\fontsize{9}{11}\selectfont$z = 0$\,\,\,\,\,$z = 2$} & 
\multicolumn{4}{c}{\fontsize{9}{11}\selectfont$z = 0$\,\,\,\,\,$z = 2$} &
\multicolumn{4}{c}{\fontsize{9}{11}\selectfont$z = 0$\,\,\,\,\,$z = 2$} &
\multicolumn{3}{|c}{} &
\multicolumn{3}{c}{}}
\startdata
Blizzard  & & 1.14 & 0.48 & & & 220.3 & 77.2 & & \,\,& 298.2 & 26.2& & & 11.4 & & &\,\,6.2 \\
Squall    & & 0.80 & 0.16 & & & 195.9 & 55.3 & & \,\,& 45.1 & 7.7 & & & 5.9   & & &\,\,21.0\\
Maelstrom & & 1.01 & 0.33 & & & 211.9 & 69.9 & & \,\,& 131.1 & 19.0 & & & 7.0 & & &\,\,9.3\\
Tempest   & & 0.51 & 0.14 & & & 169.3 & 51.8 & & \,\,& 65.4& 2.7& & & 24.2    & & &\,\,12.6\\
\enddata
\tablecomments{The virial radius ($R_{\mathrm{vir}}$) is defined as that of a sphere enclosing a total mass ($M_{\mathrm{vir}}$) whose average density is $200\times$ the critical density of the universe at that redshift (i.e., R$_{200}$ and M$_{200}$). $L_{\mathrm{vir, bary}}$ is the total angular momentum of the baryonic mass inside the virial volume.}

\end{deluxetable*}

\section{Measuring Angular Momentum in the Simulations}\label{sec:measurements}

In this section, we briefly discuss the simulation measurements that are used in this paper. In \S\ref{subsec:AM_description}, we define and discuss the measurements of total and specific angular momentum. In \S\ref{subsec:philthel}, we introduce and motivate an analytical parameter space (``\philthel") that is used in later sections of this paper. As discussed below, this space allows us to track how the orientations of the {\emph{full distribution of angular momentum vectors}} in the halo evolves in 3D space. 

\subsection{Angular Momentum}\label{subsec:AM_description}

In each simulation output generated by {\tt{Enzo}}, every cell (representing gas) and particle (representing stars or dark matter) is defined with a proper position and velocity with respect to the fixed Cartesian axes of the simulation box: $\vec{r'}=(x', y', z')$ and $\vec{v'} = (\dot{x'}, \dot{y'}, \dot{z'})$, respectively. For convenience, we establish a new reference frame that is centered on, and at rest with, the central galaxy in the halo of interest. We define the center as the location of the peak in dark matter density in the inner halo. During the periods of the simulation which are relatively calm (e.g., low accretion and merger activity), this definition is generally consistent with the center-of-mass of the baryonic component of the inner halo ($\lesssim$ 20 proper kpc)---differing by less than the size of a fully-refined simulation cell (274 comoving pc). However, during mergers, the peak in the dark matter density much more reliably identifies the main galaxy than does the center of mass. We calculate the evolving rest-frame position of the central galaxy using the bulk motion of the star and dark matter particles within 3 proper kpc of the defined center. Finally, we keep the orientation of the axes of the galaxy-centered reference frame $(x, y, z)$ parallel to the original grid of the box $(x', y', z')$.  That is, the orientation of the reference frame is kept fixed even as the orientation of the disk changes. This choice is ideal because it allows us to track the {\emph{absolute}} evolution of the orientation of angular momentum---i.e., with respect to the same reference orientation---in the simulation across timesteps. In the galaxy-centered frame, the proper position and velocity vectors of the cells and particles are simple translations from the box frame: $\vec{r} = \vec{r'} - \vec{r}_{\mathrm{galaxy}}$ and $\vec{v} = \vec{v'} - \vec{v}_{\mathrm{galaxy}}$. We re-establish the galaxy-centered reference frame for each time output of each simulation.

\begin{deluxetable*}{lcc|c|c|c|c|c|c|c}
\tablecaption{The radial profiles of the specific angular momenta $j$ of the dark matter and gas inside the virial volume of the FOGGIE halos are reported from $z\sim2$ to \zzero. The median $j$, as reported here, is calculated for the Blizzard, Squall, Maelstrom, and Tempest FOGGIE halos for each redshift listed. The units of $j$ in this table are 10$^3$ pkpc$\cdot$ km/s. \label{table:j_rvir}}
\tablehead{&
}
\startdata
    R/R$_{\mathrm{vir}}$ & & 0 -- 0.02 &
    0.02 -- 0.04 &
    0.05 -- 0.1 &
    0.1 -- 0.2 &
    0.2 -- 0.4 &
    0.4 -- 0.6 &
    0.6 -- 0.8 &
    0.8 -- 1\\
    \hline
    \\
\multicolumn{10}{c}{\hspace{3cm}\bf{ {Dark Matter}}}\\
\\
$z<0.25$      & & 0.05 & 0.11 & 0.21 & 0.50 & 0.80 & 1.14 & 1.27 & 1.67 \\
$0.25<z<0.75$ & & 0.05 & 0.11 & 0.23 & 0.42 & 0.70 & 0.96 & 1.03 & 1.18 \\
$0.75<z<1.25$ & & 0.03 & 0.07 & 0.19 & 0.31 & 0.56 & 0.77 & 0.97 & 1.29 \\
$1.25<z<1.75$ & & 0.03 & 0.05 & 0.12 & 0.22 & 0.37 & 0.52 & 0.75 & 0.88 \\
$1.75<z<2.25$ & & 0.03 & 0.05 & 0.14 & 0.18 & 0.28 & 0.46 & 0.59 & 0.68 \\
\\
\multicolumn{10}{c}{\hspace{3cm}\bf{ {Cold Gas}} ($\mathrm{T}<10^4\,\mathrm{K}$)}\\
\\
$z<0.25$      & & 0.81 & 1.83 & 3.40 & 4.34 & 5.40 & 6.37 & 8.08 & 10.37 \\
$0.25<z<0.75$ & & 0.70 & 1.51 & 2.58 & 3.72 & 4.90 & 6.31 & 8.20 & 9.64 \\
$0.75<z<1.25$ & & 0.47 & 1.12 & 2.08 & 2.92 & 4.02 & 4.41 & 4.61 & 5.73 \\
$1.25<z<1.75$ & & 0.36 & 0.82 & 1.31 & 1.60 & 2.12 & 1.89 & 2.47 & 3.23 \\
$1.75<z<2.25$ & & 0.27 & 0.60 & 1.13 & 1.30 & 1.61 & 1.32 & 1.80 & 3.10 \\
\\
\multicolumn{10}{c}{\hspace{3cm}\bf{ {Warm Gas}} ($10^4<\mathrm{T}<10^5\,\mathrm{K}$)}\\
\\
$z<0.25$      & & 0.54 & 1.38 & 2.58 & 3.43 & 4.44 & 5.07 & 4.69 & 4.64  \\
$0.25<z<0.75$ & & 0.44 & 1.09 & 1.92 & 3.08 & 4.38 & 4.37 & 4.15 & 3.28 \\
$0.75<z<1.25$ & & 0.34 & 0.79 & 1.44 & 2.35 & 3.36 & 3.19 & 2.36 & 2.09 \\
$1.25<z<1.75$ & & 0.25 & 0.54 & 0.87 & 1.32 & 1.74 & 1.69 & 1.53 & 1.53 \\
$1.75<z<2.25$ & & 0.19 & 0.41 & 0.74 & 0.99 & 1.23 & 1.11 & 0.93 & 0.87 \\
\\
\multicolumn{10}{c}{\hspace{3cm}\bf{ {Warm-Hot Gas}} ($10^5<\mathrm{T}<10^6\,\mathrm{K}$)}\\
\\
$z<0.25$      & & 0.49 & 1.18 & 2.06 & 2.62 & 3.55 & 3.73 & 2.87 & 2.42  \\
$0.25<z<0.75$ & & 0.38 & 0.93 & 1.42 & 2.26 & 3.12 & 2.85 & 2.43 & 2.20 \\
$0.75<z<1.25$ & & 0.30 & 0.67 & 1.17 & 1.83 & 2.43 & 2.31 & 2.16 & 1.89 \\
$1.25<z<1.75$ & & 0.21 & 0.43 & 0.71 & 1.06 & 1.60 & 1.83 & 1.61 & 1.42 \\
$1.75<z<2.25$ & & 0.16 & 0.33 & 0.54 & 0.64 & 0.77 & 0.84 & 0.84 & 0.92 \\
\\
\multicolumn{10}{c}{\hspace{3cm}\bf{ {Hot Gas}} ($\mathrm{T}>10^6\,\mathrm{K}$)}\\
\\
$z<0.25$      & & 0.25 & 0.63 & 1.22 & 1.69 & 2.34 & 2.21 & 3.43 & 4.89  \\
$0.25<z<0.75$ & & 0.20 & 0.52 & 0.86 & 1.33 & 1.79 & 1.75 & 2.65 & 4.45 \\
$0.75<z<1.25$ & & 0.17 & 0.41 & 0.72 & 1.06 & 1.36 & 1.77 & 2.74 & 3.75 \\
$1.25<z<1.75$ & & 0.12 & 0.27 & 0.47 & 0.69 & 0.89 & 1.19 & 1.88 & 3.11 \\
$1.75<z<2.25$ & & 0.10 & 0.20 & 0.37 & 0.48 & 0.75 & 1.09 & 1.82 & 2.24 \\
\\
\enddata
\end{deluxetable*}

With respect to the galaxy-centered frame, the angular momentum $\vec{L}$ of each parcel of mass in the simulation is defined by:

\begin{equation} \label{L_eq}
\begin{split}
\vec{L} & = m\cdot(\vec{r}\times\vec{v})=(L_{x}, L_{y}, L_{z}),
\end{split}
\end{equation}

\noindent Without an external force (e.g., gravitational interactions, collisions, gas pressures, viscous stress forces) acting on it, the angular momentum vector of a parcel would be constant. Moreover, in a closed and non-expanding system, the sum of the angular momentum vectors of all of the system's parcels is a conserved quantity both in the total ($\sum \vec{L}$) and along each axis ($\sum L_x$, $\sum L_y$, $\sum L_z$). Of course, the parcels of mass in a halo do experience external forces and galaxies and their halos are not closed systems. By measuring how the distribution of the vectors in the halo varies with time and among the components of mass, we can study how angular momentum is exchanged and transported around the halo. By tracking how the sum of the vectors evolves, we can study how angular momentum is gained by and lost from the system. 

\begin{figure*}[!ht]
\centering
\includegraphics[width=\textwidth]{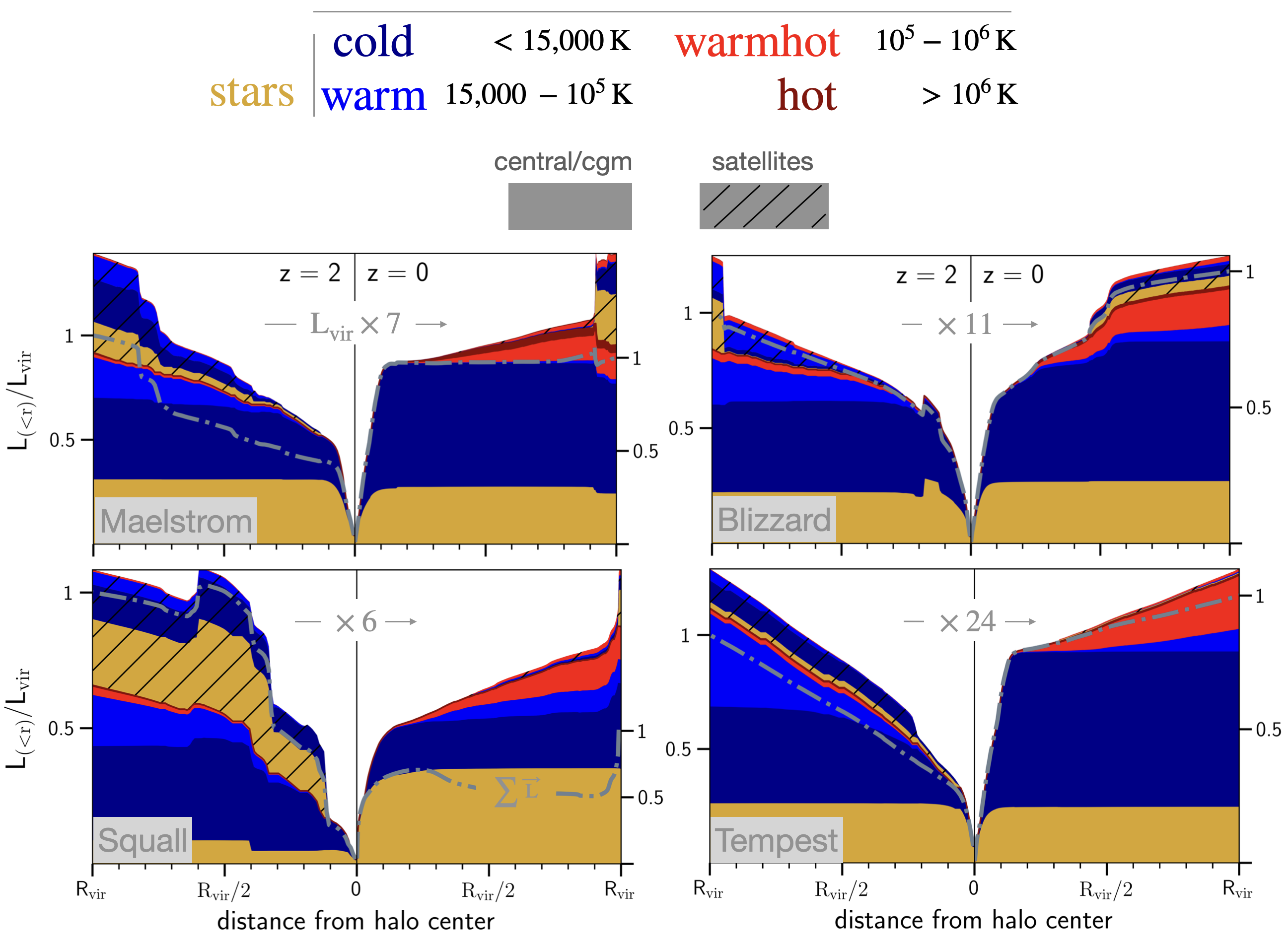}
\caption{The cumulative angular momentum of the baryons within the virial volume of four simulated Milky Way-like halos from the FOGGIE suite are shown at $z=2$ (left side of each panel) and at $z=0$ (right side of each panel). The colors indicate the baryonic component and are labeled at the top. The hatched regions indicate angular momentum associated with satellite galaxies. The cumulative total angular momentum of all components ($\sum\vec{L}$) is shown with the grey line. The total momentum is less than the absolute sum of the individual components because some components have angular momenta that are misaligned with each other. In all four halos at both redshifts, more than half of the baryonic angular momentum resides outside of $\sim$R$_{\mathrm{vir}}$/5---associated with circumgalactic gas and baryons bound to satellite galaxies. The total angular momentum of the baryons within R$_{\mathrm{vir}}$ increases by a factor of $\times7-24$ from $z=2$ to the present. At both redshifts, the bulk ($\gtrsim\,80\%$) of the angular momentum is associated with the stars and cold gas contained in the central and satellite galaxies and the warm gas in the CGM. \vspace{0.4cm}} \label{fig:cumulative_momentum_allhalos}
\end{figure*}

Throughout this paper, we will discuss the net angular momentum vector of the halo. At times, this will be shown using a cumulative radial profile (see e.g., Figure \ref{fig:cumulative_momentum_allhalos}). We note here that the magnitude of the net angular momentum vector is necessarily less than or equal to the sum of the magnitudes of the individual angular momentum vectors ($\left|\vec {L}_{\mathrm{net}}\right| = \left|\sum \vec{L}_i \right| \leq \sum\left|\vec{L}_i\right|$). This results from the fact that individual angular momentum vectors that are severely misaligned with each other may cancel when added together. For similar reasons, the cumulative angular momentum profile is allowed to decrease with radius --- i.e., if an annulus has angular momentum that is strongly misaligned with the existing cumulative vector. This is not true for cumulative radial profiles of typical positive scalar quantities, e.g., mass and energy. This effect is clear in the cumulative radial profiles shown in Figure \ref{fig:cumulative_momentum_allhalos} (discussed more later).

The specific angular momentum (angular momentum per unit mass) $\vec{j}$ of each parcel of mass in the simulation is defined by:
\begin{equation} \label{L_eq}
\begin{split}
\vec{j} & = \vec{L}/m  = \vec{r}\times\vec{v}=(j_{x}, j_{y}, j_{z}),
\end{split}
\end{equation}
where
\begin{equation} \label{L_eq_i}
\begin{split}
j_x\,&=\,y\dot{z} - z\dot{y}, \\
j_y\,&=\,-x\dot{z} + z\dot{x}, \\
j_z\,&=\,x\dot{y} - y\dot{x}, \\
\end{split}
\end{equation}
and ($x$, $y$, $z$), ($\dot{x}$, $\dot{y}$, $\dot{z}$) and ($j_x$, $j_y$, $j_z$) are the components of the proper position, velocity and specific angular momentum in the $x$-, $y$-, and $z$-axis of the galaxy-centered reference frame, respectively. 

\subsection{Visualizing the Distribution of the Orientations of Angular Momentum Vectors in the Halo: \philthel}\label{subsec:philthel}

One of the goals of this paper is to study how the {\emph{distribution of angular momentum vectors}} in the halo varies as a function of time, position in the halo, and component of mass. To do that, we need a way to visualize a distribution of many 3D-vectors without losing information on their magnitude or orientation. To that end, we look to the spherical coordinate system. 

Each parcel of mass in the simulation has a defined angular momentum vector, as described above. We measure the relative orientation of that vector with respect to the x-, y-, z- reference frame using $\phi$ and $\theta$ of a spherical coordinate system defined,
\begin{eqnarray}\label{phil_thel}
\theta_L&=&\arctan\left(\frac{L_x}{L_y}\right),\mathrm{~and}\\
\phi_L&=&\arccos\left(\frac{L_z}{L}\right),
\end{eqnarray}
where the subscript $L$ indicates that \phil{} and \thel{} are measurements of the orientation of a parcel's {\emph{angular momentum vector}}---and not the parcel's position vector. The coordinates \phil{} and \thel{} are defined such that angular momentum vectors lying parallel to the $-z$, $+z$, $-y$, $+y$, $-x$, and $+x$ axes of the frame will have (\phil{}, \thel{}) values of ($0^{\circ}$, $180^{\circ}$), ($0^{\circ}$, $0^{\circ}$), ($-90^{\circ}$, $90^{\circ}$), ($90^{\circ}$, $90^{\circ}$), ($180^{\circ}$, $90^{\circ}$), and ($0^{\circ}$, $90^{\circ}$), respectively.

That brings us to the three quantities of interest: the direction of the parcel's angular momentum vector described by (\phil{}, \thel{}) and the magnitude of the angular momentum $|L| = \sqrt{L_x^2 + L_y^2 + L_z^2}$. With those three quantities, the full 3D information of each vector is preserved. Note that we did not reduce the number of terms needed to describe the vector ($N = 3$): we only recast them into quantities that can be more intuitively captured on a 2D graphic. 

We introduce a visualization that considers how the total angular momentum $|L| $ is distributed in the 2D parameter space of \phil{} and \thel{}. Each point in the \philthel{} plane represents a unique orientation of the angular momentum vector. Parcels of mass with aligned angular momentum vectors will occupy the same space in \philthel{}. The distribution of $|L|$ in this plane indicates how much angular momentum is carried at each orientation. Throughout this paper, we use the \philthel{} space to study how the distribution of angular momenta in the halo evolves in magnitude and direction with time. In this space, ordered rotational motion around the center of the halo (i.e., a disk, where the angular momentum vectors of many parcels are aligned) occupy a narrow region. On the other hand, a wide distribution in \philthel{} indicates motions that are incoherent and turbulent. 

In later sections, we calculate \philthel{} of the net angular momentum vector for different classifications of the mass in the halo (described later; e.g., young stars in the disk, cold gas inflow, outflowing gas) and study its time evolution. By definition, the net angular momentum occupies a single location in the \philthel{} plane.

\section{The Evolution of the Magnitude of the Angular Momentum}\label{sec:AM_magnitude}

In this section, we examine how angular momenta (total and specific) is distributed in the FOGGIE halos as a function of radial distance from the halo center and among the mass components. We consider only the magnitude of the angular momentum vector here, and discuss its orientation in the next section. We briefly summarize the main conclusions of this section below, and refer to the subsections for full details.
 
First, in \S\ref{sec:total_AM_baryons} and Figure \ref{fig:cumulative_momentum_allhalos} we examine the cumulative baryonic angular momentum as a function of radial distance and baryonic component for four of the FOGGIE halos at \ztwo{} and \zzero. These four halos are those that were run to \zzero{} at the time of this analysis. At both redshifts, we find that the majority ($>50\%$) of the angular momentum of the baryons lies in the outer halo ($\gtrsim R_{\mathrm{vir}}/5$).  This outer halo angular momentum is split (with a proportion that changes between \ztwo{} and \zzero{}) between the gas and stellar mass associated with satellite galaxies and the circumgalactic medium.

\begin{figure*}[!ht]
\centering
\includegraphics[width=\textwidth]{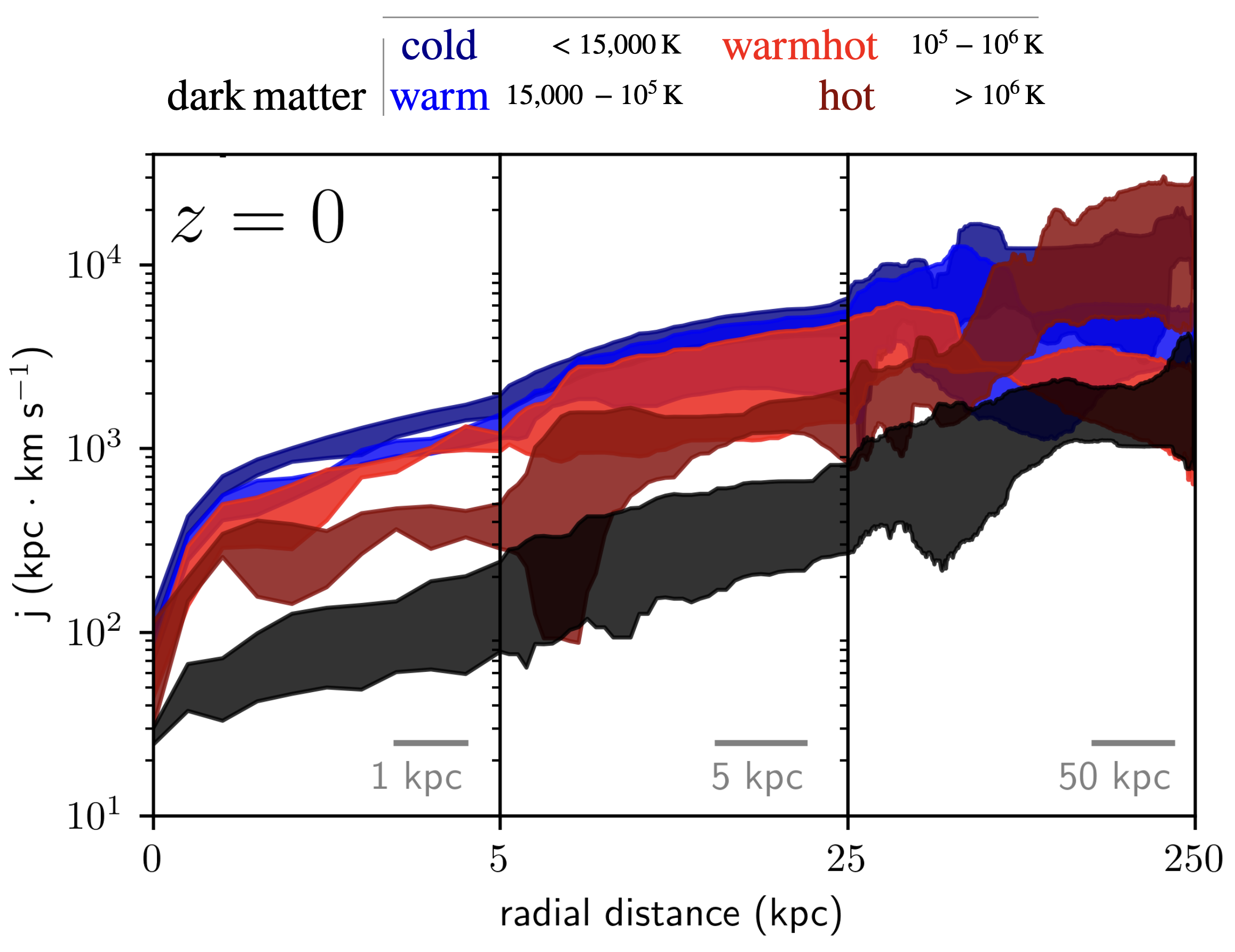}
\caption{The 16$^{\mathrm{th}}$--84$^{\mathrm{th}}$ percentiles of the radial profiles of the specific angular momentum for the dark matter (black) and gas (colors) for the FOGGIE halos at $z = 0$ is shown. The gas profiles are split by gas temperature. The x-axis is broken into three rough sections: the inner disk ($r = 1$--5 kpc), the outer disk ($r = 5$--25 kpc), and the circumgalactic medium ($r = 25$--250 kpc). The scale of each section is indicated with the grey bar. \vspace{0.4cm} \label{fig:z0_halos}}
\end{figure*}

Next, in \S\ref{sec:specific_AM} and Figure \ref{fig:z0_halos}, we quantify the specific angular momentum profiles of the dark matter and baryonic gas for the FOGGIE halos at \zzero{}. We find that the gas in the halo carries a higher specific angular momentum ($j$) than that of the dark matter. This is the case for all gas temperature phases and at all radial distances. Moreover, we find that the $j$ of the gas has a strong temperature dependence. On average, colder gas carries a higher $j$ at all radial distances. The population-averaged radial profiles of the dark matter and gas are provided in Table \ref{table:j_rvir} at various redshifts, normalized by R$_{\mathrm{vir}}$.

\begin{figure*}
\centering
\includegraphics[width=\textwidth]{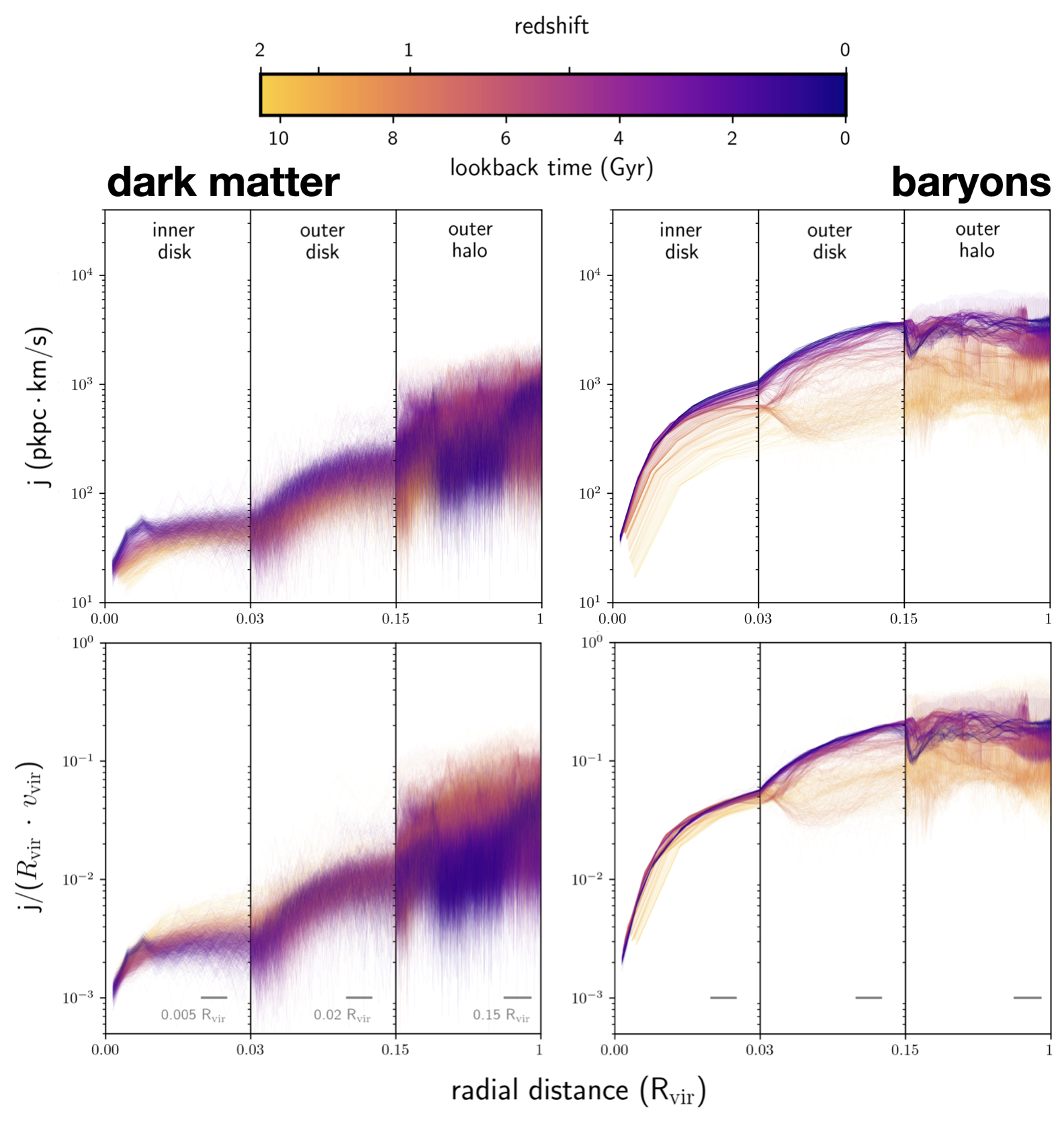}
\caption{The time evolution of the radial profile of the specific angular momentum $j(r)$ of the dark matter (left) and baryons (right) of the Tempest halo is shown. The top panels show the quantity $j$ as-is and the bottom panels are normalized by the growth in the size and virial velocity of the halo in cosmic time ($j$/R$_{\mathrm{vir}}\cdot v_{vir}$). The color-coding indicates the cosmic time---yellow is $z=2$ and dark purple is $z=0$. The radial distance is normalized by the virial radius ($\mathrm{R}_{\mathrm{vir}}$) at each redshift. The horizontal spatial scale increases from left to right across the three sub-panels, and the scale is shown in the lower left panel. In general, and shown here for Tempest, the FOGGIE halos undergo strong evolution in the baryonic angular momentum in the outer disk and CGM. In contrast, the inner disk and dark matter halo minimally evolve.} %The dashed grey line indicates the average specific angular momentum of the dark matter inside the virial radius ($<\mathrm{j}_{_{\mathrm{DM, vir}}}>$). } 
\label{fig:Tempest_jprof_dm_baryons}
\end{figure*}

\begin{figure*}
\centering
\includegraphics[width=\textwidth]{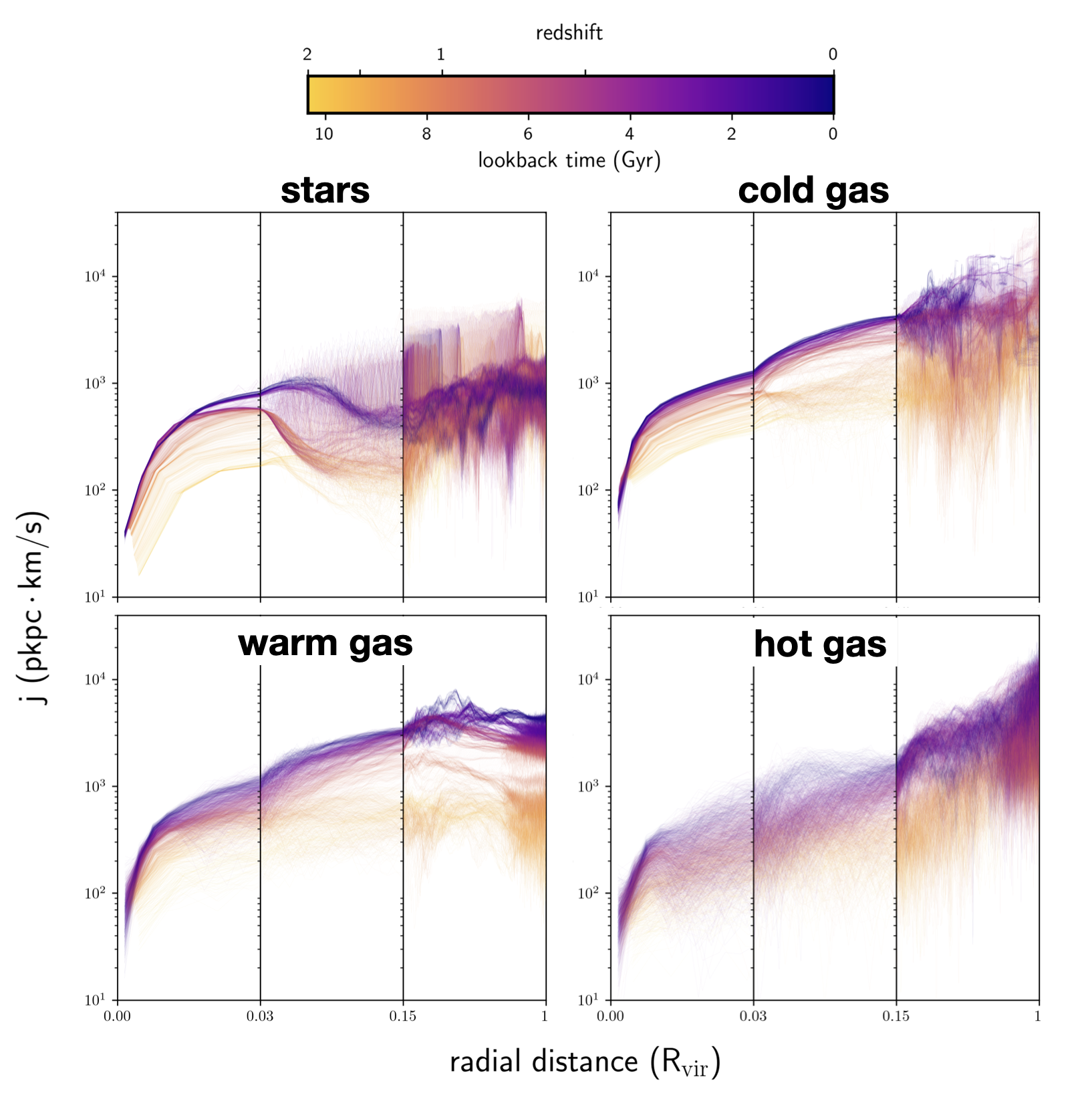}
\caption{The same as the top right panel of Figure \ref{fig:Tempest_jprof_dm_baryons}, but now distinguishing by the baryonic components---stars, cold gas ($<15,000\,\mathrm{K}$), warm gas (15,000 K - 10$^5$ K), and hot gas ($>\,10^6$ K). From \ztwo{} to \zzero, there is significant evolution in the specific angular momentum $j$ of the cold and warm gas in the outer disk, increasing by nearly an order of magnitude over that period. There is a similar degree of evolution for those components in the outer halo. In contrast, there is only mild evolution of $j$ for the hot gas on both the disk and halo scales.} \label{fig:Tempest_jprof_baryonic_components}
\end{figure*}

Then, in Figures \ref{fig:Tempest_jprof_dm_baryons} and \ref{fig:Tempest_jprof_baryonic_components}, we examine in detail the evolution of the radial profile of $j$ for the dark and baryonic matter of a single halo (Tempest). In Tempest, we show that the $j$ of the cold and warm components of both the central galaxy and the CGM generally increase together with time, while those of the dark and hot halo minimally change on average. Qualitatively, these conclusions are generally true for the other FOGGIE halos as well.

\begin{figure*}[ht]
\centering
\includegraphics[width=0.85\textwidth]{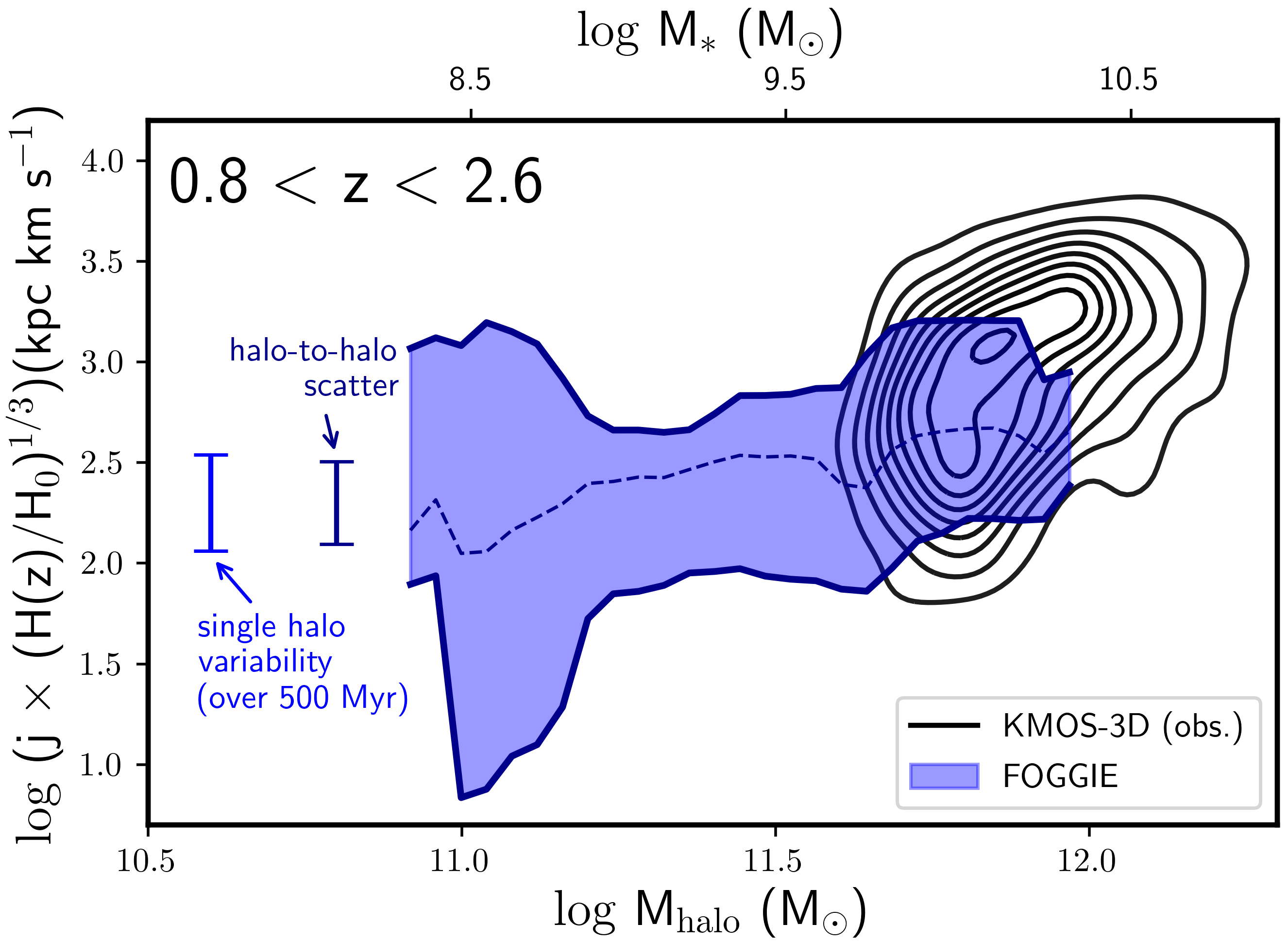}
\caption{The angular momentum of the star-forming components (young stars; age $\leq$10 Myr) of the central galaxies in FOGGIE (blue shade) are compared with observations of the star-forming gas in disk galaxies from the KMOS-3D survey (black contour; \citealt{burkert16}) over the redshift range $0.8<\mathrm{z}<2.6$. The simulation measurements include every snapshot (separated by $\sim$5 Myr) of the six simulations over this redshift period. The relation between stellar mass (top x-axis) and halo mass (bottom x-axis) is taken from \citet{behroozi19}, and is used to estimate halo masses for the observations to compare with the simulations. The average variability of a single simulated galaxy over a timescale of 500 Myr is indicated with the light blue error bar on the left. This ``single halo'' term is one of the two dominant sources of scatter in the angular momentum versus stellar mass relation in FOGGIE. The second source of scatter is the mean difference from halo-to-halo---a ``halo population'' term. The standard deviation of the average of the six halos over this redshift range is shown with the dark blue error bar on the right. Both of these terms contribute to the observed width of the FOGGIE distribution. \vspace{0.4cm}} \label{fig:observational_comparison}
\end{figure*}

Finally, in \S\ref{sec:obs_specific_AM} and Figure \ref{fig:observational_comparison} we compare the specific angular momentum of the central disk galaxies of the full suite of FOGGIE halos with observations of star-forming galaxies over $0.8<z<2.6$. We find good agreement in the limited range of halo mass where the samples overlap ($\log M_{\mathrm{halo}}/\mathrm{M}_{\odot}] = 11.5-12.5$) and provide predictions for the mean and variance of Milky-Way like halos at lower masses and higher redshifts. We show that the observed scatter in $j$ for the FOGGIE halos reflects in roughly equal parts: (1) systematic differences in the mean angular momenta of the halos with one another, and (2) the variability of the individual halos in time.

\begin{comment}
\begin{figure*}
\centering
\includegraphics[width=\textwidth]{./figures/Tempest_jprof_dm_baryons_normed.png}
\caption{Same as Figure \ref{fig:Tempest_jprof_dm_baryons}, but normalized by the evolving size ($R_{\mathrm{vir}}$) and virial circular velocity ($v_{\mathrm{vir}}$) of the parent halo.} \label{fig:Tempest_jprof_dm_baryons_normed}
\end{figure*}
\end{comment}

\subsection{The Total Angular Momentum of the Baryons}\label{sec:total_AM_baryons}

In Figure \ref{fig:cumulative_momentum_allhalos}, we show the radial cumulative angular momentum of the baryons inside four of the FOGGIE halos at \ztwo{} and at \zzero. The profiles are split by the baryonic component and normalized by the total angular momentum inside the virial radius R$_{\mathrm{vir}}$ at each redshift. The total vector sum of the angular momentum is indicated by the dashed grey line. The total is lower than the stack of the individual components because a portion of the angular momentum cancels between components.

At \ztwo{}, the majority of the baryonic angular momentum inside $R_\mathrm{vir}$ resides in the outer halo ($\gtrsim R_{\mathrm{vir}}/5$)---mostly in warm circumgalactic gas, and the stars and cold interstellar gas associated with satellite galaxies. As noted above, cold gas (for our definition; $<15,000$\,K) and stars are generally only found in and around the galaxies in the FOGGIE halos. 

Galaxies carry angular momentum in two forms, (1) through their orbital motion through the halo, and (2) through their internal rotation. Since we adopt a rest-frame in which the central galaxy is at rest, the central galaxy carries only rotational angular momentum. Satellite galaxies carry both orbital and rotational angular momentum\footnote{The total angular momentum does not depend on the chosen reference frame. Orbital and rotational angular momentum are interchanged in different reference frames, without changing the total.}. These two forms of angular momentum can be exchanged with one another (e.g., as galaxies merge or through long-range gravitational interactions) and they both contribute to the total shown in Figure \ref{fig:cumulative_momentum_allhalos}. This is true not only for galaxies, but for any definable structure in the halo that has tangential motion around the central galaxy (e.g., clumps, turbulence)---they all carry angular momentum through both their bulk orbital motions through the halo and their internal relative motions. For purely isotropic turbulence, the random orientations of convective motions throughout the halo will cancel out. We will see examples of angular momentum in the form of small-scale turbulence in hot CGM gas in  \S\ref{sec:AM_orientation}. For these reasons, the reader should not interpret the angular momentum shown in Figure \ref{fig:cumulative_momentum_allhalos} as that of simple and well-ordered rotation around the center of the halo.

From $r=0$ to $\sim$R$_{\mathrm{vir}}$/10, the cumulative profiles in Figure \ref{fig:cumulative_momentum_allhalos} sharply rise. This is true in all four halos and at both redshifts. This rise is associated with the rotation of the stars and cold gas in the central disk. Outside of R$_{\mathrm{vir}}$/10, there are several discrete increments and decrements in the total cumulative angular momentum. These jumps are generally associated with the orbital and rotational angular momenta of the satellite galaxies, shown by the hatched colors. The increments are common in the \ztwo{} snapshot when the halos contain several $10^5$--$10^{10}$ M$_{\odot}$ satellites (\citealt{simons20}). By \zzero{}, nearly all of the stellar and cold gas angular momenta in the four halos is contained within $r<R_{\mathrm{vir}}/5$, confined to the central galaxy and its stellar halo.

In all four halos, the volume-filling warm-hot gas ($T=10^{5-6}$ K) accounts for $<5$\% of the total angular momentum at \ztwo{} but up to $20$\% of the total at \zzero. The hot component of gas ($>10^6$ K) carries only trace amounts of the total angular momentum at both redshifts. The star-formation feedback routine implemented in this generation of the FOGGIE simulations launches primarily hot outflows. The majority of the hot gas in the halos originates in these outflows. These outflows generally have high radial velocities and low tangential velocities with respect to the central galaxy. Moreover, this diffuse hot component comprises only a small fraction ($<1$\%) of the baryonic mass in the halo.  Both of these factors lead to the relatively small amount of angular momentum carried by the hot gas component. The fact that the hot component carries such a trace amount of angular momentum suggests that the FOGGIE outflows remove an inconsequential amount of angular momentum from galaxies. In a later section (\S\ref{sec:outflowing}), we verify that conclusion.

From \ztwo{} to \zzero, the cumulative profiles generally maintain their shape. Qualitatively, the ratio of angular momentum in the outer halo, inner halo, and central galaxy are similar at both times. At a surface level, this suggests that the baryonic angular momentum associated with the central galaxy grows in lockstep with the baryonic angular momentum in the halo. 

The half-angular momentum radius (where L$_{(<r)}/$L$_{\mathrm{vir}} = 0.5$) is roughly constant in R/R$_{\mathrm{vir}}$ from \ztwo{} to \zzero. At both redshifts, more than half of the total angular momentum in the halos resides outside of the proximity of the central galaxies ($\gtrsim\,$R$_{\mathrm{vir}}$/5). At \ztwo{}, the angular momentum in the outer halo is mostly carried by satellites and cold + warm circumgalactic gas ($T<10^5$). At \zzero{}, it is mostly carried by warm + warmhot circumgalactic gas ($15,000\,<\,T\,<10^6$ K).

From \ztwo{} to \zzero, the total baryonic angular momentum inside $R_{\mathrm{vir}}$ increases by a factor of $\times 10$--20 across the four halos. We attribute this to three factors. The first two are the mass and size growth of the halo: both the true mass growth from accretion and the pseudo-growth from the evolving and ever-growing virial volume, as defined. The virial masses ($M_{vir}$) of the four halos shown in Figure \ref{fig:cumulative_momentum_allhalos} increase by factors of $\times2.4$--5 from \ztwo{} to \zzero{} (see Table \ref{table:angmom_rvir}).

If the growth in the total baryonic angular momentum inside $R_{\mathrm{vir}}$ were to follow that expected for the dark matter halo, then we should expect for it to evolve as $L_{\mathrm{vir}}\propto R_{\mathrm{vir}}\times M_{\mathrm{vir}} \times v_{\mathrm{vir}}\propto M_{\mathrm{vir}}^{5/3} \times (1+z)^{-1/2}$. For the four halos in Figure \ref{fig:cumulative_momentum_allhalos}, the quantity $R_{\mathrm{vir}}\times M_{\mathrm{vir}} \times v_{\mathrm{vir}}$ increases by factors of $\times6-21$ across the halos (Table \ref{table:angmom_rvir}). While this range is similar to the ranges of the true growths ($\times6-24$), there is a poor one-to-one correspondence for a given halo. This indicates that the halo growth is not alone governing the growth in the total angular momentum.

The third reason for the growth in $L_{\mathrm{vir}}$, and what will be part of the focus of the next sub-section, is that the gas specific angular momentum at fixed radius generally increases in all of the FOGGIE halos over this time period.

\subsection{The Specific Angular Momentum of the Baryons and Dark Matter}\label{sec:specific_AM}

We now turn our attention to the time evolution of the magnitude of the specific angular momentum $j$ (i.e., the angular momentum per unit mass).

%%% Figure 2 %%%

In Figure \ref{fig:z0_halos}, we show the composite $j(r)$ radial profiles of the dark matter and baryonic gas mass for four of the FOGGIE halos at \zzero{} --- Blizzard, Squall, Maelstrom, and Tempest. The vertical span of the profiles indicate the running 16$^{\mathrm{th}}$-84$^{\mathrm{th}}$ percentile of the halos. The running percentiles are calculated using 250 pc annular bins for each halo, and then collated across the four halos.

Qualitatively, we find that the composite $j(r)$ profiles at \zzero{} are generally larger for:

\begin{enumerate}
\item all temperatures of gas relative to the dark matter at fixed radii,
\item cooler gas relative to hotter gas at fixed radii, and 
\item all mass components at larger radii.
\end{enumerate}

\noindent For instance, $j(r)$ of the coldest gas is roughly a factor of $\times10$ higher than that of the dark matter at all radii. The factor is smaller when comparing this coldest gas to the warmer gas bins. In Table \ref{table:j_rvir}, we provide the population-median $j(r)$ radial profile of the same four halos included in Figure \ref{fig:z0_halos}. These profiles are normalized by the virial radius and contain multiple bins spanning $0\,<\,z\,<\,2.25$. At all of the redshifts studied, the three rules enumerated above hold.

There is one notable exception to these rules. In the outer halo, we find that the hottest gas ($>\,10^6\,\mathrm{K}$) carries higher or similar $j$ than the rest of the gas at fixed radius. This is true at \zzero{} (Figure \ref{fig:z0_halos}), but also up to \ztwo{} (Table \ref{table:j_rvir}). This high $j$ hot gas may have a number of origins including: hot winds from star-formation in the satellite population inheriting the high $j$ of their satellite's orbit, or tangential ram pressure induced on the hot halo from the orbits of the satellites or from cooler material inflowing from the intergalactic medium. While the existence of this hot high-$j$ gas is noteworthy, this material represents a negligible fraction of the total mass \citep{lochhaas23} and angular momentum (Table \ref{fig:cumulative_momentum_allhalos}) of the halos at these redshifts ($0\,<\,z\,<\,2$).

%%% Figure 3 %%%

In Figure \ref{fig:Tempest_jprof_dm_baryons} we show the time evolution of the $j(r)$ profile of the dark matter (left) and baryons (right) in the Tempest halo from \ztwo{} to \zzero{}. We split the radial distance shown on the horizontal axis into three bins normalized by the virial radius at each redshift. These choices {\emph{roughly}} mark the edges of the inner disk, the outer disk, and the outer halo of Tempest across the redshift range shown. The three sections are shown with increasing linear spatial scales. In the top two panels, we show the raw $j$ profile. In the bottom two panels, we show $j$ normalized by the product of the virial radius and virial velocity of Tempest at each redshift. By comparing both, we can isolate the classically-expected evolution due to the growth of the halo ($\propto R_{\mathrm{vir}}\cdot v_{\mathrm{vir}}$) from any excess growth in $j$.

In all three regions of the halo, we find relatively little ($\lesssim\times2$) evolution in the raw specific angular momentum ($j$) of the dark matter over the $\sim 10$\,Gyr of cosmic time shown. When normalized by the growth of the halo ($\mathrm{R}_{\mathrm{vir}} \cdot \mathrm{v}_{\mathrm{vir}}$), this evolution is suppressed (bottom panel). The dark matter profile roughly follows a $j(r)\propto r$ scaling at all redshifts. This scaling is expected from tidal torque theory and is produced in N-body simulations of purely dark halos \citep{1987ApJ...319..575B, bullock01}. The linear scaling with $r$ is consistent with a dark halo with a flat rotation curve---e.g., a flat circular velocity curve and a constant velocity dispersion with radius.

In contrast, we find strong evolution for the baryons in {\emph{both}} $j$ and $j/(\mathrm{R}_{\mathrm{vir}} \cdot \mathrm{v}_{\mathrm{vir}}$). All three regions of the halo show an increase in $j$ of $\sim\times5-10$ increase over this time. When normalized by $R_{\mathrm{vir}}\cdot v_{\mathrm{vir}}$, the evolution in the inner portion of the halo (the ``inner disk") is suppressed but the outer disk and outer halo still exhibit a large growth. We note here that the majority of star formation in the FOGGIE centrals occurs inside 2 kpc (see also \citealp{wright24,Acharyya24}). At $0.03 \,\mathrm{R}_{\mathrm{vir}}$, the specific angular momentum of the baryons is roughly equal to the average specific angular momentum of the dark halo. At face value, this is consistent with one of the key assertions in the classic model of disk galaxy formation in $\Lambda$CDM---that the $j$ of the disk and the dark halo are linked.

However, importantly, the specific angular momentum of the innermost portion of the disk is not representative of that of the rest of the halo. The outer disk and CGM carry the bulk ($>80$\%) of the baryonic angular momentum in the halo (Figure \ref{fig:cumulative_momentum_allhalos}). To restate, in these regions we do find strong evolution of $j/R_{\mathrm{vir}}\cdot v_{\mathrm{vir}}$---which increases on average by a factor of $\sim 5$ from \ztwo{} to \zzero{}(Figure \ref{fig:Tempest_jprof_dm_baryons}).

%%% Figure 4 %%%

In Figure \ref{fig:Tempest_jprof_baryonic_components}, we further explore the evolution in the baryonic specific angular momentum $j$ of Tempest. We show the time evolution of the raw $j$ split into four components---stars, cold gas, warm gas, and hot gas. The definition of the gas bins are the same as in Figure \ref{fig:cumulative_momentum_allhalos}. 

We find significant evolution in $j$ in the outer disk and CGM for the cold ($T\,<\,15,000$ K) and warm ($15,000\,<\,T\,<\,10^5$ K) gas, reflecting that found for the total baryons in Figure \ref{fig:Tempest_jprof_dm_baryons}. The hot gas, on the other hand, shows mild to no evolution in these regions of the halo. The stars also display evolution in $j$ in the inner and outer disk, but not in the outer halo. The stellar $j$ in the outer halo reflects the stellar component of the changing satellite population orbiting around Tempest. This stellar component contributes only a small fraction of the total halo angular momentum at both \ztwo{} and \zzero{} (Figure \ref{fig:cumulative_momentum_allhalos}). We conclude that it is the evolution of the $j$ of the cooler material ($<10^5$ K, ``cold" and ``warm" gas) in the CGM that establishes the strong evolution in $j$ for the baryons.

\subsection{Comparison with Observations}\label{sec:obs_specific_AM}

We now compare the distribution of $j$ from the full suite of FOGGIE simulations against the real galaxy population at $0.8<z<2.6$.

In Figure \ref{fig:observational_comparison}, we show the relation between the halo mass and the specific angular momentum $j$ of the star-forming components (young stars; age $<10$ Myr) of the central galaxies in the six FOGGIE simulations over $0.8<z<2.6$. The observational dataset of $j$ measurements were carried out \citet{burkert16} as a part of the KMOS-3D integral field galaxy survey \citep{wisnioski15, wisnioski19}. We normalize the $j$ of the simulations by the evolving Hubble parameter, $H(z)$, to compare with the measurements provided in \citet{burkert16}. We translate the stellar masses of the observational sample into their equivalent dark matter halo masses using the stellar mass-halo mass correspondence provided by \citet{behroozi19}. That correspondence is shown by the dual axes in Figure \ref{fig:observational_comparison}. We caution that the simulation swath shows the time evolution of only six halos. We are not fully sampling the distribution of properties that halos at these masses and redshifts possess---i.e., spin, environment, and accretion histories.

Over these redshifts, the central galaxies in FOGGIE generally span lower halo masses than the observational sample. In the small mass range where they overlap ($11.7< \log M_{\mathrm{halo}}/M_{\odot}<12.0)$, we find general agreement in both the population mean and scatter. In the simulations, we recover a mild dependence between specific angular momentum and halo mass. Specifically, the central galaxies have higher specific angular momentum when their halos are more massive. This is mainly driven by the specific angular momentum evolution shown in Figure \ref{fig:Tempest_jprof_baryonic_components}. The $j$ of the cold star-forming material increases with decreasing redshift---indicating that the growth in the baryonic angular momentum of the simulated halos outpaces their growth in mass.

We use the simulations to investigate the source of scatter in the $j-M$ relation at fixed mass. At fixed mass, the $1\sigma$ scatter of $j$ of the simulated FOGGIE population is $\sim 0.5$--1 dex. We attribute this scatter to two sources: systematic differences in the average $j$ of the six halos at fixed mass and the time variability of a single halo. The former is a measure of the mean variance of the population, while the latter is a measure of the stochastic evolution of a single halo.

First, we quantify how the variability of a single halo contributes to the scatter in the $j$-$M$ relation. To do that, we calculate the variance in $j$ of each halo around their mean evolution over a time baseline of 500 Myr. The light blue bar in Figure \ref{fig:observational_comparison} shows the measured variability averaged across the six halos. 

Next, we quantify how the differences in the mean evolution of the halos contributes to the scatter. To do that, we calculate the running standard deviation of the average of the six halos in bins of halo mass. The dark blue uncertainty bar shows this contribution.

We find that the single halo variability contributes slightly more to the observed scatter of the distribution than does the mean differences between the halos. Observations of $j$ in star-forming galaxies in this mass range (from e.g., {\emph{JWST}} spectroscopy) could measure the observed scatter in this low mass population at these redshifts. The FOGGIE simulations indicate that this scatter will carry information on both (1) the variability of single halos and (2) differences between halos---in roughly equal parts.

\begin{figure*}
\centering
\includegraphics[width=\textwidth]{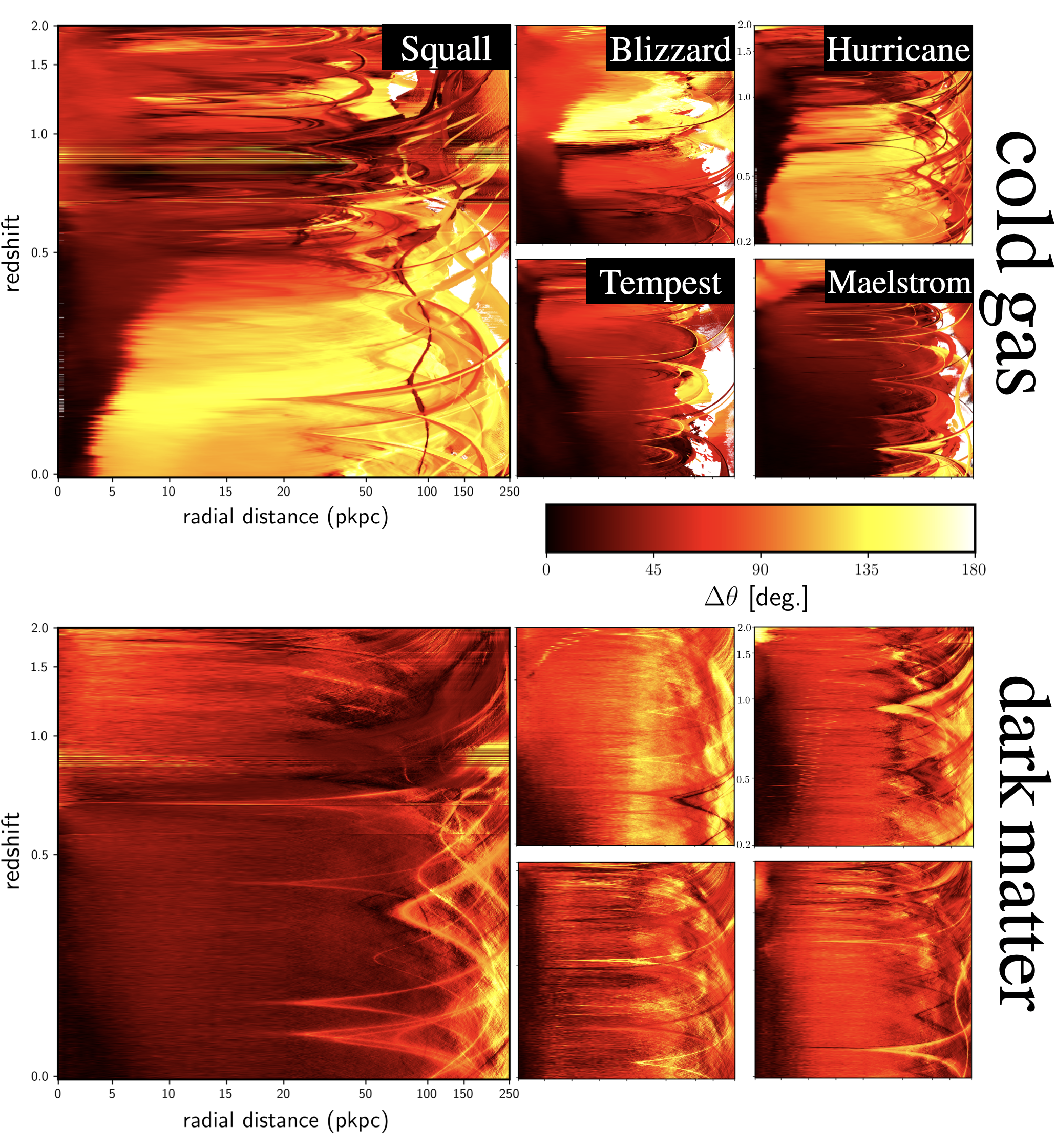}
\caption{With respect to the central stellar disk at \zzero{}, the relative orientations $\Delta\theta$ of the angular momentum of the halo's cold gas ($<\mathrm{15,000}\,\mathrm{K}$; top) and dark matter (bottom) are shown as a function of redshift and radial distance. The color shows the degree of alignment, with black indicating total alignment and white-yellow indicating total anti-alignment. The $z=0$ orientation of the cold central disk is generally only established at late times---after $z=2$ in all five of the FOGGIE halos shown. The orientation of the central disk is generally not well aligned with cold gas beyond 20 kpc in all halos. In the inner region of all five halos, the angular momentum of the dark matter is well-aligned with the disk. However, in the outer regions of the halos the dark matter is largely misaligned with the central disk---up to $\sim$45--90 degrees. Taken together, the central disk and inner dark halo generally co-establish an orientation at late times that is disconnected from the the circumgalactic cold gas and outer dark halo. } \label{fig:R-Z-theta}
\end{figure*}

\begin{figure*}[t!]
\centering
\includegraphics[width=\textwidth]{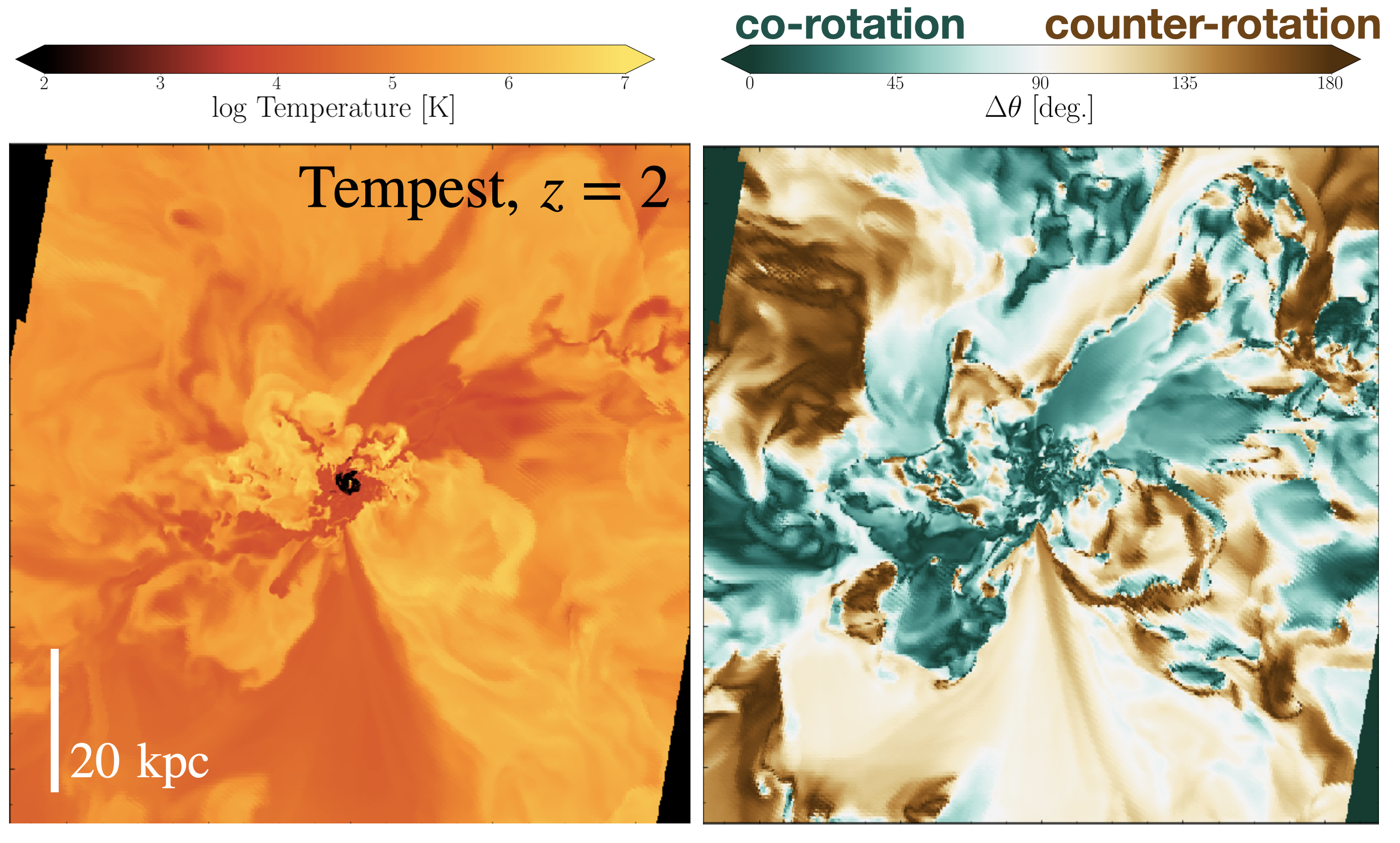}
\includegraphics[width=\textwidth]{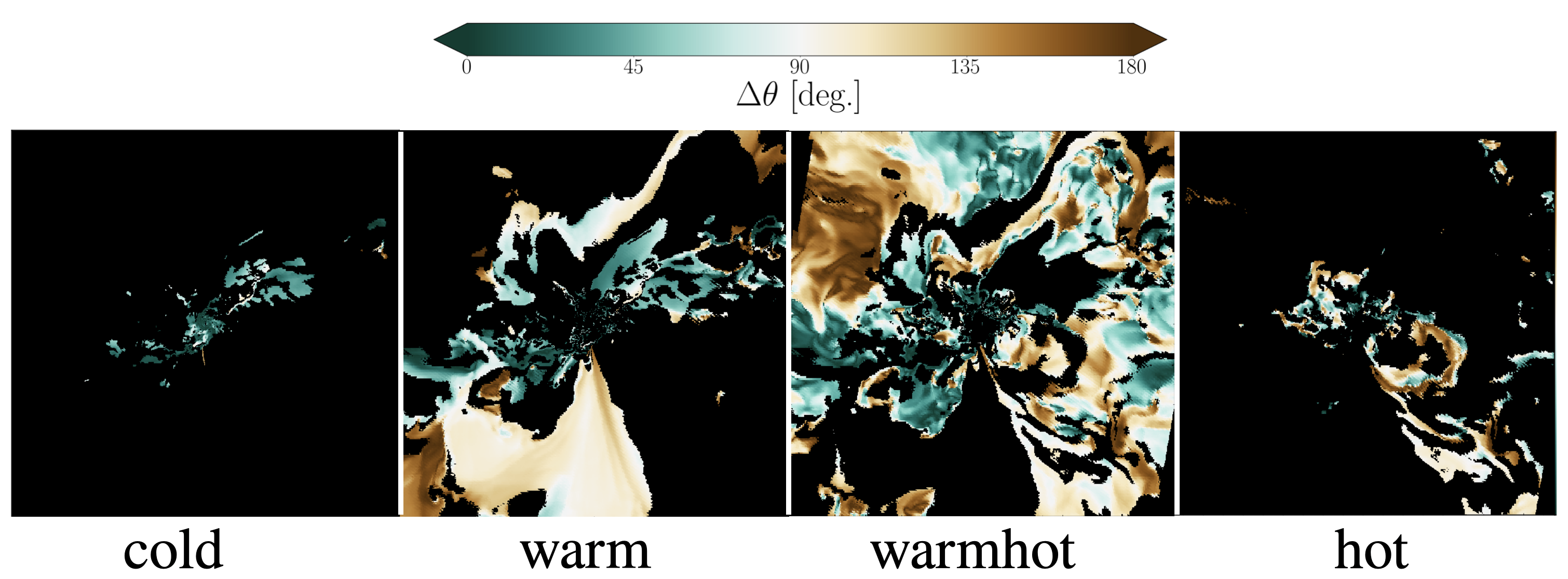}
\caption{The temperature (left) and orientation of the angular momentum (right) of a slice through the circumgalactic medium of the Tempest halo at $z=2$ is shown. The slice lies in the plane of the central disk---i.e., looking down on the face of the disk as defined by its net angular momentum vector. The bottom panels show the orientation of the angular momentum split by temperature.}\label{fig:Tempest_z2slice}
\end{figure*}

\begin{figure*}[t!]

\centering
\includegraphics[width=\textwidth]{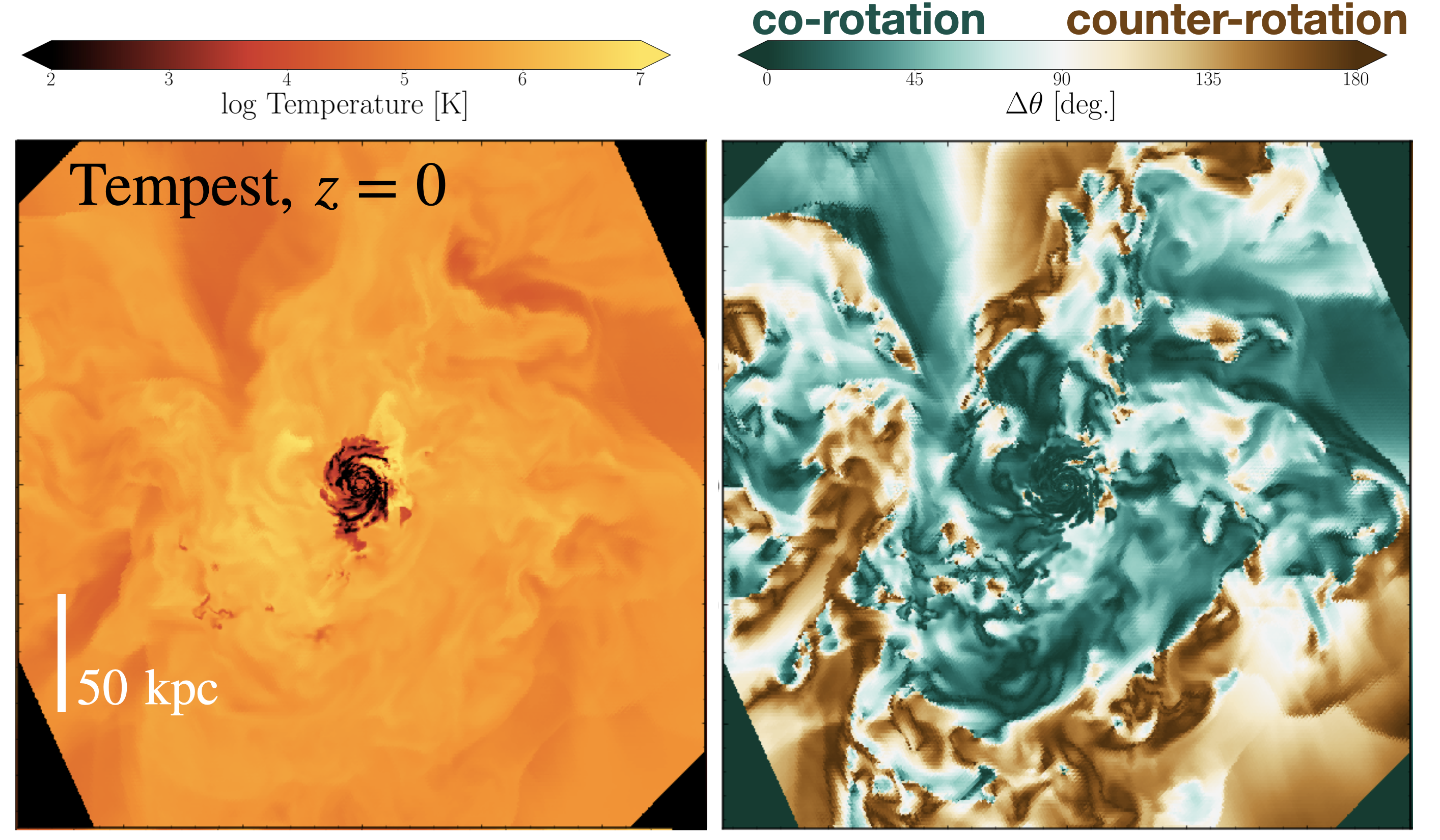}
\includegraphics[width=\textwidth]{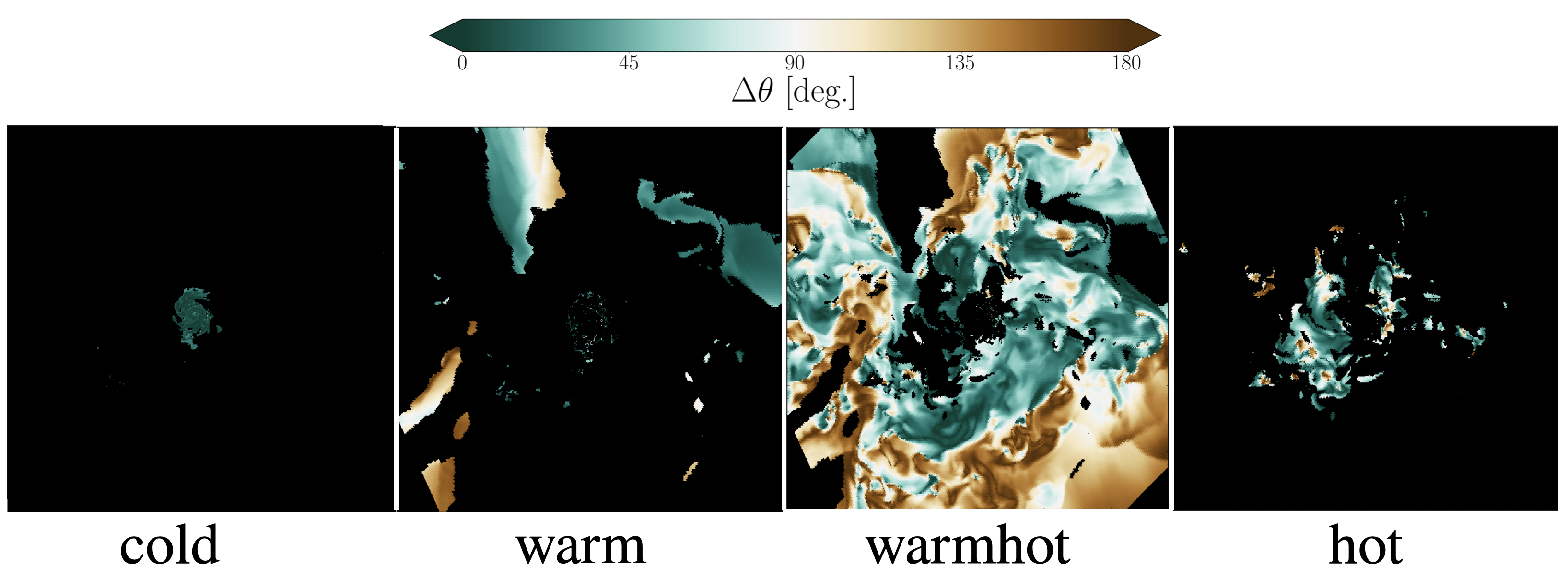}
\caption{The same as Figure \ref{fig:Tempest_z2slice}, but for the Tempest halo at $z=0$.\vspace{0.4cm}}\label{fig:Tempest_z0slice}

\end{figure*}

\begin{figure*}
\centering
\includegraphics[width=\textwidth]{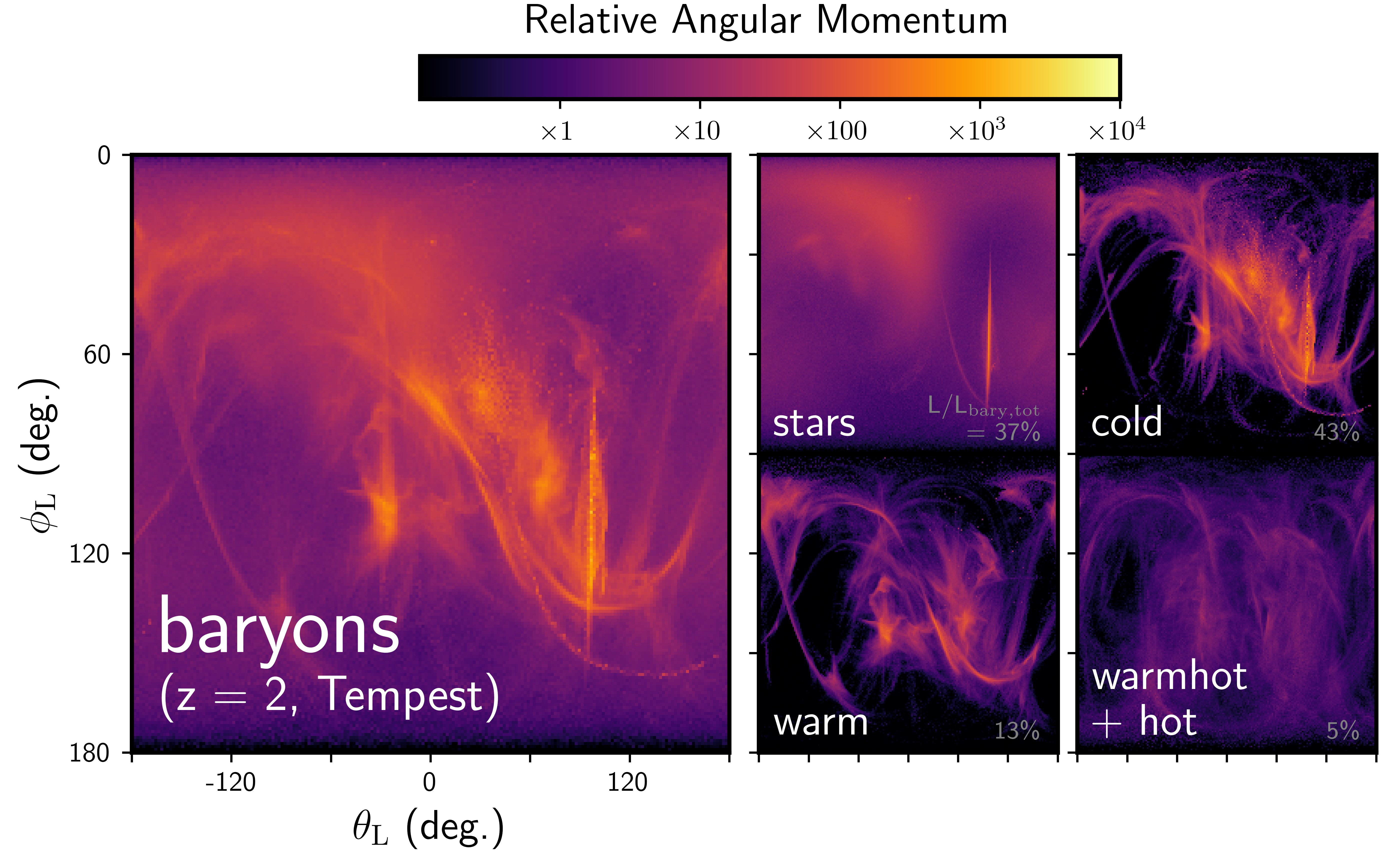}
\includegraphics[width=\textwidth]{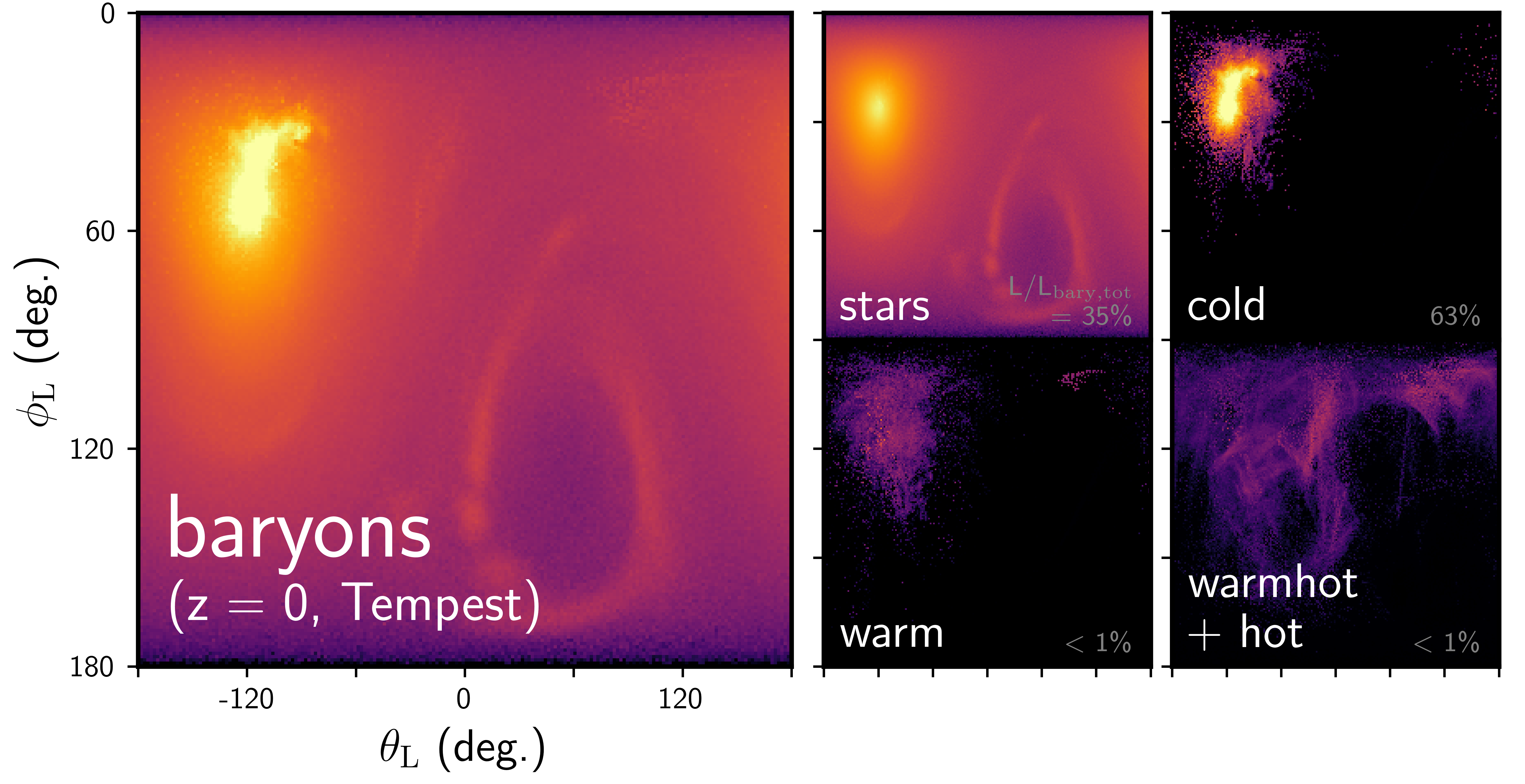}
\caption{The distribution of the orientation of the angular momentum in the inner 25 pkpc of the Tempest halo at \ztwo{} (top) and \zzero{} (bottom) is shown. All of the baryons are shown in the left panel, and the various baryonic components are shown in the right panels. In the \philthel{} plane, coherent rotation around the halo center appears as a cluster of angular momentum around a given position. Rotation that is not centered on the halo center (e.g., the internal rotation of a satellite disk galaxy, a spinning clump, or an eddy in the CGM gas) appears as a sinuisoidal wave. Incoherent and non-rotational motions (e.g., turbulence, the dispersion-supported stellar halo)---where the angular momentum vectors point in random directions---span a wide distribution in this plane.} \label{fig:phil-thel-tempest}
\end{figure*}

%%%%%%%%%%%%%%%%%%%%%%%%

\section{The Orientation of Angular Momentum}\label{sec:AM_orientation}

In this section, we turn our attention towards the distribution of the {\emph{orientation}} of the angular momentum inside the FOGGIE halos. We briefly summarize the main conclusions here, and refer to the subsections for details.

In \S\ref{sec:dtheta_dr_dz} and Figure \ref{fig:R-Z-theta} we examine how the orientations of the angular momenta ($\vec{L}$) of the cold gas and dark matter in each of the FOGGIE halos: (1) evolve in time, (2) vary as a function of the radial distance from the halo center and (3) relate to the orientation of the central galaxy at \zzero{}. We find that the \zzero{} orientations of the FOGGIE central galaxies are only established at late times in all of the halos studied (after \ztwo{}). At most cosmic times after \ztwo{}, the orientation of the cold gas $\vec{L}$ in the central disk is generally (but not always) misaligned with that of the circumgalactic cold gas in the rest of the halo. Moreover, beyond the inner few kpc, the orientation of the dark matter $\vec{L}$ is generally misaligned with {\emph{both}} that of the central disk and that of the cold circumgalactic gas. We describe the $z\,<\,2$ accretion history of each halo to provide context to their unique evolution in the orientation of their angular momentum.

In \S\ref{sec:theta_distribution_tempest} and Figures \ref{fig:Tempest_z2slice} and \ref{fig:Tempest_z0slice}, we cut slices through the Tempest halo to study how the orientation of the angular momentum of the circumgalactic gas varies as a function of position in the halo and the temperature. We find that the orientation is {\emph{highly variable}} in space, varying by up to the maximum $180^{\circ}$ difference over kpc-scales. The orientation of the warmer gas ($>\,10^5\,\mathrm{K}$) varies more than the colder gas, and is also more poorly correlated with the orientation of the central galaxy than the colder gas. In Figure \ref{fig:phil-thel-tempest}, we explore this further by visualizing the time evolution of the distribution of the 3D orientation of the baryonic angular momentum for the inner region of Tempest. At \ztwo{}, we find that the angular momentum of each baryonic component (stars and gas of various temperatures) generally spans a wide range of 3D orientations---indicating that the distribution of angular momentum is largely disorganized. By \zzero{}, when the rates of galaxy merging and cosmological accretion have significantly reduced, the vast majority ($>99\%$) of the baryonic angular momentum in the inner halo is well-aligned with the orientation of the central stellar disk.

\subsection{Angular Momentum Orientation as a Function of Radial Distance and Redshift}\label{sec:dtheta_dr_dz}
In Figure \ref{fig:R-Z-theta}, we show the redshift evolution of the orientation of the angular momentum vector in radial shells centered on the central galaxy. In the top panel, we show the cold gas ($T<$ 15,000 K). In the bottom panel, we show the dark matter. In each shell at each redshift, we measure the difference between the orientation of the net angular momentum of each component with that of the central stellar disk at $z=0$ ($z=0.2$ for Hurricane). This difference angle is defined as $\Delta \theta$. Since all of the orientations are measured with respect to the {\emph{same reference orientation}}, the different portions of this diagram can be safely compared. That is, in the same halo one can interpret differences between redshifts, radial bins, and mass components as true differences in the absolute orientation. 

Satellite galaxies oscillate in this diagram as their orbits bring them towards and away from the center of the halo. These oscillatory features appear as distinguished, curved tracks in Figure \ref{fig:R-Z-theta}. The alignment between the net angular momentum of the satellite galaxy (including its orbit and internal rotation)  and the central disk can be inferred from the colors of these features on the plot. In all five halos shown and all of the redshifts considered, the angular momentum orientations of the satellites vary widely and they are generally---but not always---poorly aligned with the angular momentum orientation of the central disk galaxy. 

Each of our halos present a case study in angular momentum acquisition. In that spirit, we now briefly explain some of the nuances in the accretion history of each halo at $z\,<\,2$ that drive the orientation of their \zzero{} angular momentum. We connect back with the visual features seen in Figure \ref{fig:R-Z-theta}, where relevant. Below, the ratios of the virial masses of the satellites and central (main parent halo) are provided in the shorthand $M_{\mathrm{central}}:M_{\mathrm{satellite}}$.

\begin{itemize}

\item In the {\bf{Squall halo}}, the angular momentum of the central disk is quickly ($<50$ Myr) re-oriented following a 2:1 major merger at $z\sim0.7$. This event is evident in Figure \ref{fig:R-Z-theta}, where the orientation in the center of the halo (at the disk scale) abruptly changes from $\Delta\theta \sim\,70^{\circ}$ to $\sim0^{\circ}$ --- with 0$^{\circ}$ defined as pure alignment with the orientation of the final \zzero{} disk. Shortly after this event, a minor satellite (4.5\,:\,1 mass ratio) is stripped of its gas as it passes through the inner halo. This stripped gas is counter-rotating with respect to the central disk, and is distributed around the CGM over $\sim200\,\mathrm{Myr}$. Together, the pair of these events lead to: (1) the re-formation of the central disk of Squall and (2) the development of a CGM whose angular momentum orientation is highly misaligned with that of that central disk.

\item In the {\bf{Tempest}} halo, a warp develops in the outer gaseous disk at $z\,<\,1$. This warp coincides with the orientation of the angular momentum of the dominant cool filament feeding the disk at $z\,=\,0.5-2$. At late times ($z\,<\,0.3$), this filament dissipates, but the warp persists in the outer-most portions of the Tempest disk to \zzero{}. By \zzero{}, the majority of the Tempest disk (by volume and mass) has a single coherent angular momentum orientation. This direction is aligned with the inner CGM of Tempest but is misaligned with the warm-hot circumgalactic gas in its outer halo. We more fully explore the evolving orientation of the Tempest halos in Figures \ref{fig:Tempest_z2slice}, \ref{fig:Tempest_z0slice}, and \ref{fig:phil-thel-tempest}, and in the next subsection.

\item The {\bf{Blizzard}} halo forms a gaseous polar ring at $z\,=\,1$, which is rotating in a plane that is nearly perpendicular to the central disk ($\sim90^{\circ}$). In Figure \ref{fig:R-Z-theta}, the polar ring is apparent as the contrast between the orange color in the center of the halo and the outer bright yellow color. The ring is aligned with the net orientation of the angular momentum of the cold gas in the halo to $R\sim200$ kpc. At $z\sim0.7$, a minor satellite (60:1 mass ratio) makes a pass with the Blizzard central galaxy. This first pass destroys the ring while also abruptly quenching the star-formation of the central galaxy. Following this pass, there is a transitory period ($\Delta t\sim1\,\mathrm{Gyr}$) where the gas associated with the interstellar medium of the satellite remains outside of the Squall central disk. The eventual accretion of this gas establishes the \zzero{} orientation of the central disk. The final orientation is roughly identical to that of the original orbit of the satellite.

\item The {\bf{Hurricane}} halo forms a gaseous polar ring around $z\sim0.5$. The angular momentum of the circumgalactic gas is organized at the orientation of this polar ring preceding its formation. This ring is formed as a result of a pair of mergers at z $\sim$ 0.5 that are orbiting with the same angular momentum orientation as the CGM. Stars begin to form out of the ring around $z\sim0.3$.  By z$<$0.3, the orientation of the CGM angular momentum is $90-180^\circ$ misaligned with that of the central disk and with that of the dark matter halo.

\item The {\bf{Maelstrom}} halo has a relatively quiet merger history. It experiences its last {\emph{major}} merger at $z\sim4$. At $z\sim1.3$, it experiences a gas-rich merger with a minor satellite (13:1 mass ratio). This satellite carries an orbital angular momentum that is $\sim$180$^{\circ}$ misaligned with the existing disk. In response to this merger, the central disk of Maelstrom is reoriented by $\Delta 90^{\circ}$.

\end{itemize}

In general, we find that the $z=0$ orientation of the central disks in the FOGGIE halos are established only at late cosmic times. 
 In the Squall and Blizzard halos, the central disk orientation is established at $z\sim 0.5$. In Hurricane, Tempest and Maelstrom, it is established at $z\sim 1.5$. In short, the central disk at $z=0$ has effectively lost memory of its orientation at $z=2$ in all of the halos studied. By $z=0$, the orientation of the cold gas, dark matter, and stars in the center of the central galaxy ($r<3$ kpc) are well aligned. As indicated above, in many cases it is the late time mergers and interactions with the satellite population that a major role in re-shaping the orientation of angular momentum in the halo and of the central disk (via mergers).

Except for Maelstrom, the orientations of the inner regions of the halos are generally misaligned with the outer regions---including both the dark matter and cold circumgalactic gas. Together, these components make up the bulk of the (total and baryonic) mass and angular momentum in the halo.

At most times below \ztwo{} and in all of the halos except Maelstrom, the cold central gas disk is surrounded by misaligned cold material. Often, this cold material is associated with an outer secondary disk. These secondary disks are created through filamentary accretion, which is often misaligned from the central disk galaxy (C.\ Lochhaas et al. in prep). In some halos (e.g., Squall and Hurricane) the misalignment is large enough that the secondary-disk can be considered counter-rotating ($\Delta\theta>90^{\circ}$), while in others the misalignment is small ($\Delta\theta<45^{\circ}$). These misaligned secondary disks are generally non-star forming and metal poor \citep{Acharyya24}. 

In the bottom panel of Figure \ref{fig:R-Z-theta}, we show the relative orientation of the angular momentum of the dark matter. Near the central galaxy (r $\lesssim$ 5 kpc), the dark matter angular momentum is generally well-aligned with that of the disk. However, in the vast majority of the volume of the halo, there is strong misalignment with {\emph{both}} the central disk and the cold circumgalactic gas (top panels). 

In sum, Figure \ref{fig:R-Z-theta} suggests three important conclusions:

\begin{enumerate}
    \item the present-day orientation of the cold gas disks in FOGGIE are generally only established at late times ($z<1$), 
    \item the angular momentum of the stellar and cold gas disks are generally `disconnected' from that of the dark matter halo at all times, and
    \item an individual galaxy merger can play the dominant role in defining the \zzero{} orientation of the central galaxy.
\end{enumerate}

\subsection{The Distribution of the Orientations of the Angular Momentum Vector in the Halo}\label{sec:theta_distribution_tempest}

In Figures \ref{fig:Tempest_z2slice} and \ref{fig:Tempest_z0slice}, we examine slices through the Tempest halo in the plane of the central galaxy---i.e., the ``face-on'' view---at $z=2$ and $z=0$, respectively. The face-on view is defined to run parallel to the net angular momentum of the central gaseous disk at each redshift. In the top left panel, we show the temperature of the gas. In the top right panel, we show the orientation of the angular momentum of the gas relative to that of the central disk. In the bottom panels we show the relative orientation split into different bins of gas temperature, following the definitions in \S\ref{sec:measurements} and Figure \ref{fig:cumulative_momentum_allhalos}.

At both redshifts and in all temperature bins, we find spatial variability in the orientation of the angular momentum orientation in the halo. In some cases, the orientation varies abruptly (by up to 180$^{\circ}$) over kpc-scales. 
In general, we find that the coldest material in the Tempest halo is aligned with the disk rotation at both redshifts. Although not shown, we note that this is also the case for the Maelstrom halo. However, it is not the case for the Squall, Blizzard and Hurricane halos (see Figure \ref{fig:R-Z-theta}).

In contrast, at \ztwo{} the angular momentum orientation of the warm (15,000--$10^5$\,K), warm-hot ($10^5$--$10^6$\,K) and hot ($>10^6$ K) gas are generally all strongly misaligned with the central disk. These phases also exhibit higher spatial variability than the cold gas. By \zzero{}, most of the gas in the inner halo of Tempest (at $r<50$\,kpc) is co-rotating with the same angular momentum orientation as the central galaxy.

\begin{figure*}
\centering
\includegraphics[width=\textwidth]{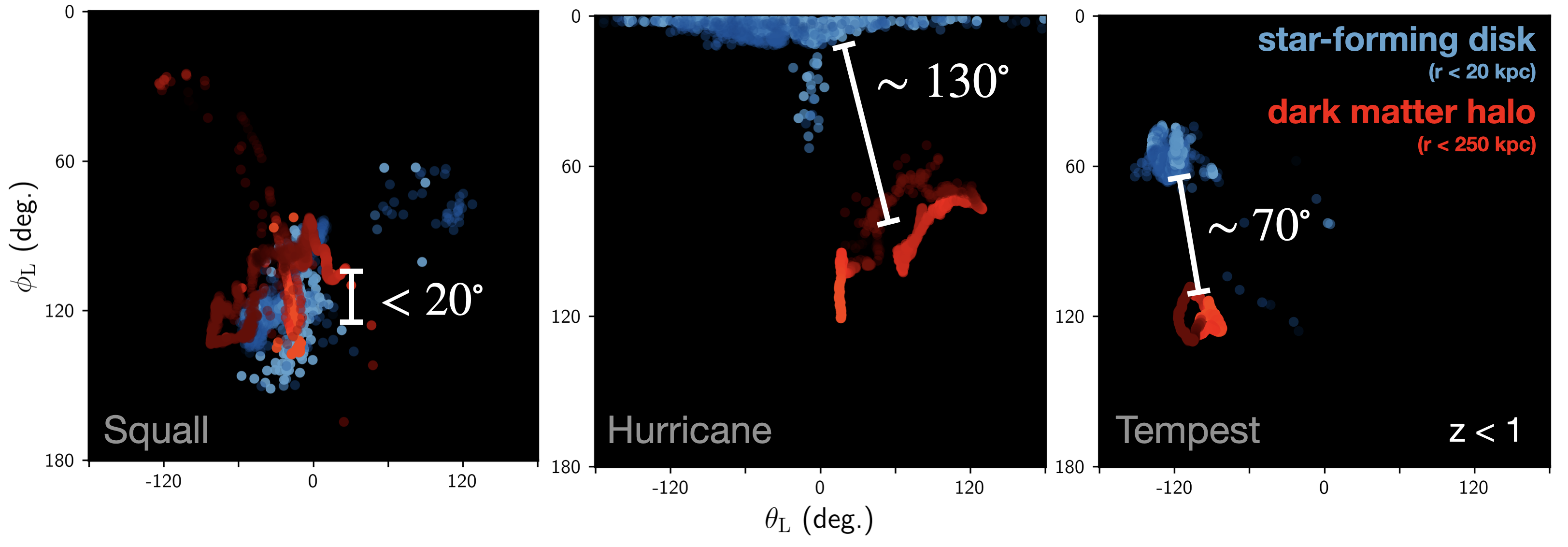}
\caption{The orientation of the net angular momentum of the dark matter halo (red) and the star-forming disk (blue) is shown for three halos at $z<1$. Each dot represents an independent snapshot of the simulations ($\Delta\mathrm{t}\sim 5\,\mathrm{Myr}$). The brightness of the points codes the total angular momentum in that component at each snapshot---brighter points contain more angular momentum. The star-forming components of the central disks in FOGGIE are often misaligned with the dark matter halo.\vspace{0.4cm}} \label{fig:sf_versus_dm_phil-thel}
\end{figure*}

\begin{figure*}
\centering
\includegraphics[width=0.9\textwidth]{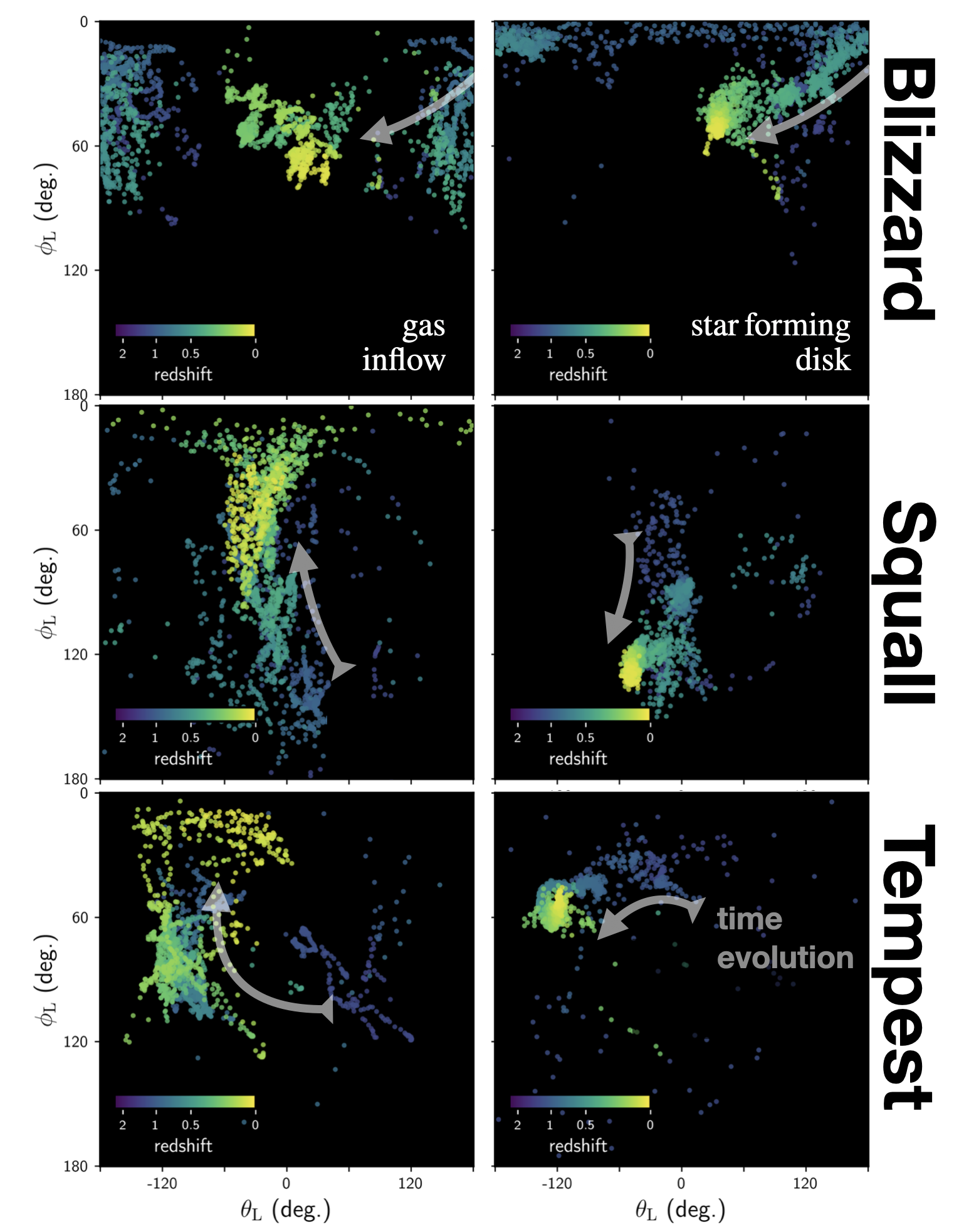}
\caption{The evolution of the orientation of the angular momentum of gas that is inflowing onto (left), and the star-forming disk of (right), the FOGGIE central galaxies is shown from \ztwo{} to \zzero{}. The orientation is described in terms of the spherical coordinates \philthel{}. The quantities \phil{} and \thel{} are defined with respect to the fixed simulation box. Inflowing gas is shown on the left, and the star-forming disk is shown on the right. Inflowing gas is defined as gas within 20 pkpc of the central and with radial velocities $v_{\mathrm{r}}<-100~ \mathrm{km}\,\mathrm{s}^{-1}$. The star-forming disk is defined using the young star-particles ($<10$\,Myr) within 20 pkpc of the center of the central galaxy. Each dot represents a single snapshot in time, and the redshift of the snapshot is color-coded. The qualitative time evolution of each panel is indicated with a grey arrow.} \label{fig:phil_thel_inflow_disk}
\end{figure*}

In Figure \ref{fig:phil-thel-tempest}, we use the \philthel{} plane (see \S\ref{subsec:philthel}) to examine the distribution of the orientation of the baryonic angular momentum in the inner region of the Tempest halo ($r<25$\,kpc) at \ztwo{} (top panel) and \zzero{} (bottom panel). The angular momentum of all of the baryons (stars and gas) is shown in the large panel on the left. The distributions for the different components of the baryons are shown on the right. The fraction of the total angular momentum in each component is listed in the lower right of each panel. The color-coding indicates the amount of relative angular momentum and is shown with a log scale.

Figure \ref{fig:phil-thel-tempest} is the complete description of the {\emph{distribution}} of the angular momentum {\emph{vectors}} around Tempest. This set of panels shows not only which components of the baryons carry the bulk angular momentum, but also how that angular momentum is distributed among all possible 3D orientations. In the \philthel{} plane, material that is rotating coherently around the halo center (e.g., the central disk) will appear as a well-localized cluster of angular momentum at a given position. Rotation that is not centered on the halo center (e.g., the internal rotation of a satellite disk galaxy, a clump with rotation, or a turbulent eddy in the CGM gas) appears as a sinusoidal wave\footnote{The length of the wave is set by the object's internal rotation velocity and its position in the plane is set by its orbital velocity. Those details are beyond the scope of this paper, but we emphasize that the sinusoidal waves indicate {\emph{real angular momentum}} in the form of localized spin.}. Finally, incoherent and non-rotational motions (e.g., turbulence and stars in the dispersion-supported stellar halo) span a wide distribution in this plane.

Figure \ref{fig:phil-thel-tempest} indicates that at \ztwo{} the angular momentum of the baryons in the inner halo of Tempest is largely incoherent. In all four baryonic components, the angular momentum is distributed over a wide range of \phil{} and \thel{}. This suggests that the majority of the baryonic angular momentum (which would occupy a narrow region in \philthel{} if in coherent rotation) is {\emph{not}} in a well-established rotating disk, nor is it in a coherently rotating sphere.

By \zzero{}, the \philthel{} distribution is markedly changed. The stars, cold gas, and warm gas occupy a small locus in \philthel{}. This locus is that of the central disk, which by this late time has developed a coherent rotation-supported disk. The angular momenta of the warm-hot + hot material ($>10^{5}\,\mathrm{K}$) span a wider range in \philthel{} than the cooler gas. This material indicates little preference for the orientation of the disk. This warmer material is almost exclusively heated by the outflows driven by the thermal feedback scheme included in the simulations to model SNe feedback. Note that this material comprises a negligible portion of the total angular momentum in the system ($<1\%$).

In Figure \ref{fig:sf_versus_dm_phil-thel}, we show the {\emph{net orientation}} of the angular momenta of the dark matter halo (red) and the star-forming (SF) central disk (blue) of the Squall, Hurricane, and Tempest FOGGIE halos for every simulation timestep (separated $\Delta\,t\sim 5\mathrm{Myr}$) below $z\sim1$. The ``star-forming disk" is measured using the young stars in the central galaxy ($<10\,$Myr), tracing recent star-formation. At these later cosmic times, the \zzero{} orientation of both the star-forming disks and the dark matter halo are generally well established. This is evident in how localized the points are in Figure \ref{fig:sf_versus_dm_phil-thel}.

In Hurricane and Tempest, the orientation of the SF disk and dark matter halo are poorly associated at \zzero{}. Compared to their respective SF disks, the Hurricane dark halo is $\sim130^{\circ}$ offset and the Tempest halo $\sim70^{\circ}$ offset. In contrast, the SF disk and dark matter halo are well-aligned ($<30^{\circ}$) for the Squall and Maelstrom (the latter is not shown in Fig. \ref{fig:sf_versus_dm_phil-thel}). As stated above, the Squall central galaxy contains two disks that are strongly misaligned with one another. The inner of these disks is star-forming, while the outer is comprised only of cold non star-forming gas. It is suggestive that the orientation of Squall's inner star-forming disk maintains a close association with the orientation of the dark matter halo, while the same is not true for the outer gas disk (Figure \ref{fig:z0_halos}) which is reoriented in response to a minor satellite merger (see \S \ref{sec:dtheta_dr_dz}).

In short, Figures \ref{fig:Tempest_z0slice}-\ref{fig:sf_versus_dm_phil-thel} suggest that (across the FOGGIE suite) the angular momentum orientation exhibits high spatial, temporal, and component-by-component variability. This variability is {\emph{not expected}} in the classic model of angular momentum evolution in which the baryonic and dark matter mass experience handcuffed evolution in both the magnitude and orientation of the vector {$\vec{L}$}.

\begin{figure*}
\centering
\includegraphics[width=\textwidth]{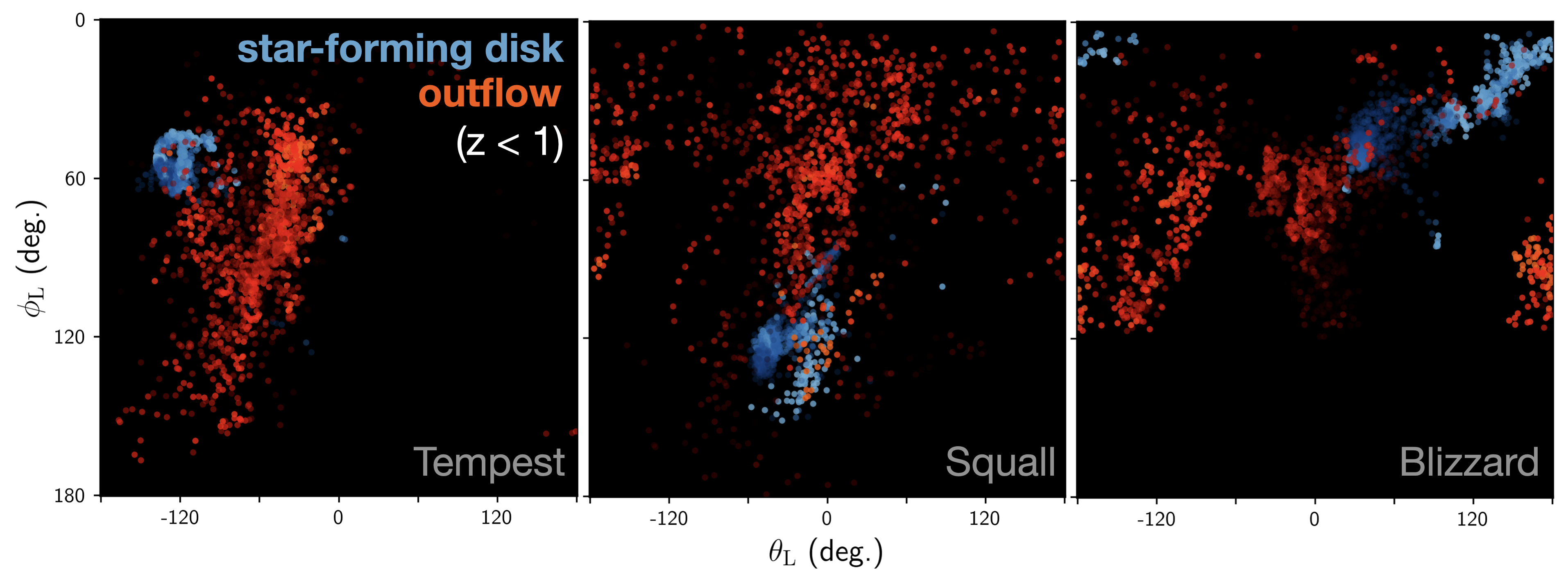}
\caption{The absolute orientation of the outflowing gas (red) and the star-forming disk (blue) is shown for each time output of three FOGGIE halos at $z<1$. The outflows generally remove angular momentum that is misaligned with the star-forming disk. Each point represents one time output of the simulation. The brightness of the points codes the total angular momentum in that component at each snapshot.} \label{fig:thel_phil_outflow}
\end{figure*}

% relabel y-axis as "angular momentum depletion time from outflows"

\begin{figure*}
\centering
\includegraphics[width=\textwidth]{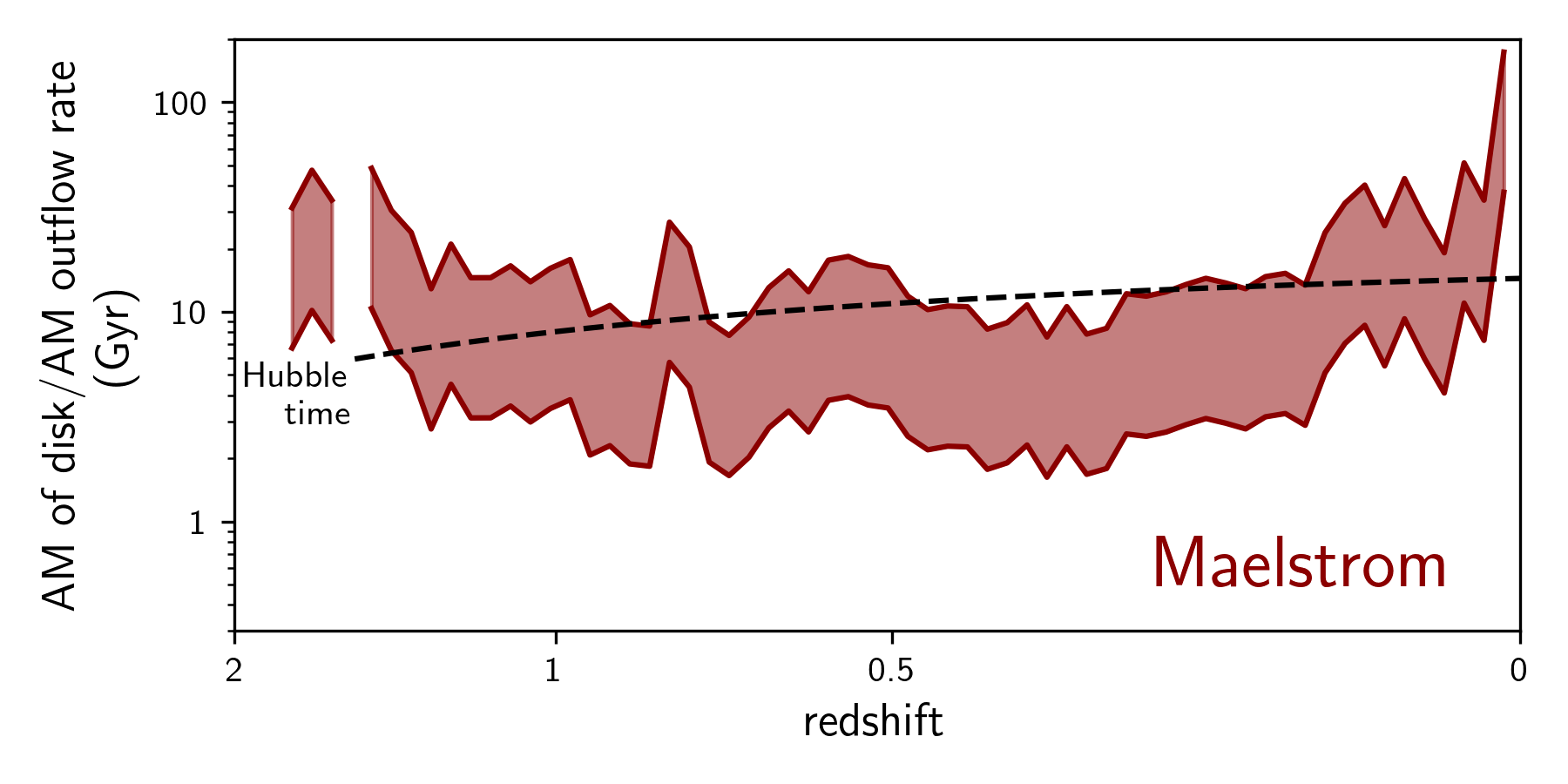}
\caption{The timescale that it would take the outflowing gas to fully deplete the disk of its angular momentum is shown from \ztwo{} to \zzero{} for the Maelstrom central galaxy. This ratio is calculated considering the relative orientation of the angular momenta of each component. This calculation neglects replenishment of angular momentum from inflows and gravitational torques. The upper boundary of the shaded region shows an outflow moving at 150 km s$^{-1}$ and the lower boundary shows 700 km s$^{-1}$. Faster outflows remove angular momentum at quicker rates (leading to lower depletion times). The FOGGIE outflows generally fall between these velocity boundaries. This figure shows that the SF-driven outflows that are generated in FOGGIE {\emph{do not}} remove a significant amount of angular momentum from galaxies. Assuming no replenishment, it would take several billions of years for the outflowing material to deplete the disk of angular momentum---comparable to the Hubble time. The Hubble time is indicated with the dashed black line.}\label{fig:outflow_timescale}
\end{figure*}

\section{The Angular Momentum of Inflowing and Outflowing Gas}\label{sec:inflowing_outflowing}

In this section, we examine how angular momentum is added to and removed from the FOGGIE centrals through inflows (\S\ref{sec:inflowing}) and outflows (\S\ref{sec:outflowing}), respectively. We show how the orientation of the inflowing (Figure \ref{fig:phil_thel_inflow_disk}) and outflowing gas (Figure \ref{fig:thel_phil_outflow}) relate to the orientation of the star-forming disks over time, and quantify the negligible role that the FOGGIE outflows serve in the angular momentum evolution of the FOGGIE centrals (Figure \ref{fig:outflow_timescale}).

\subsection{Inflowing Gas and the Re-orientation of the Star-Forming Disk}\label{sec:inflowing}

In the left panels of Figure \ref{fig:phil_thel_inflow_disk}, we show the orientation (using the \philthel{} plane) of the {\emph{net}} angular momentum of the gas that is inflowing onto the central galaxies in three of the FOGGIE halos: Blizzard, Squall, and Tempest. In the right panels, we show the orientation of the central star-forming disk. To reiterate, each position in the \philthel{} plane corresponds to a unique orientation of the angular momentum vector in 3D space. Every simulation output from \ztwo{} to \zzero{} is shown, with the redshift of the outputs indicated. The outputs are separated in time by $\sim 5$\,Myr.

For each simulation output, we define gas inflow as that which has a radial velocity $v_{R}<-100$ km s$^{-1}$ and is within 20 pkpc of the central galaxy. These criteria select the gas that will contact the central disk within $\sim 200$\,Myr. This selection captures gas that originated in the filamentary cool cosmological accretion ($\sim10^{5}$ K metal-poor gas from the cosmic web; C.\ Lochhaas et al. in prep) as well as inflowing clumps of dense gas (R.\ Augustin et al. in prep).
It occasionally (moreso at $z>1$) includes the gas that is bound to satellite galaxies that pass nearby the central galaxy or that are in the process of merging. The radial velocity criterion rejects the turbulent gas that comprises most of the CGM by-volume ($\sigma_{r, \rm turb.}\sim 40$ km s$^{-1}$; \citealt{lochhaas21})---i.e., gas which can have random and transient negative radial velocities but which is not truly inflowing. The star-forming disk is defined as the collection of star particles younger than 10 Myr (i.e., tracing recent star-formation) within 20 pkpc of the center of the central galaxy. 

Figure \ref{fig:phil_thel_inflow_disk} shows that, in each of the three halos, the net orientation of both the star-forming disk and the inflowing gas evolve significantly from $z\sim2$ to $z\sim0$. This mirrors the conclusions we drew from Figure \ref{fig:R-Z-theta}, which shows that there is large evolution in the {\emph{relative orientation}} of the angular momentum of the cold gas of the eventual \zzero{} disk over this period. Now, Figure \ref{fig:phil_thel_inflow_disk} more explicitly shows that the {\emph{absolute}} orientation of the \zzero{} star-forming disks are not established until late times---generally after $z=1$. Arrows are included to highlight the qualitative time evolution in both panels.

In the Blizzard halo, the star-forming disk generally tracks the evolving direction of the gas inflow. The orientation of the \zzero{} star-forming disk is established at $z<0.5$ with the same orientation as the gas inflow at $z<0.5$. This indicates that the net angular momentum direction is governed by the direction of the gas inflow direction at these times.
As the orientation of the gas inflows shift with time, the star-forming disks do too.
By contrast, in the Squall halo the star-forming disk is detached from the evolution of the gas inflow. We note that Squall maintains a secondary misaligned disk for most of the redshifts considered here (Figure \ref{fig:R-Z-theta}). It is onto this secondary disk (which is not forming stars) that the inflowing mass and angular momentum deposits.
It is generally true in all six halos that the star-forming central disk re-orients over time from \ztwo{} to \zzero{}. The same is not true for the dark matter halo, whose orientation is relatively constant with time (Figure \ref{fig:R-Z-theta}).

\subsection{Outflowing Gas}\label{sec:outflowing}

In Figures \ref{fig:thel_phil_outflow} and \ref{fig:outflow_timescale}, we examine how angular momentum is carried away from the FOGGIE central galaxies through gas outflows. In the current generation of the simulations, outflows are exclusively driven by thermal feedback from star-formation and supernovae \citep{cen_ostriker_06}; see \S\,2.4 of \citet{wright24} for more detail. Black holes and associated AGN feedback are not included.

We select outflowing gas as that which has a radial velocity $v_{R}>+100$ km s$^{-1}$ and which is located between a radial distance of 10 and 35 kpc from the center of the central galaxy. As in the previous subsection, the radial velocity criterion rejects gas that is turbulent ($\sigma_{r,\,\mathrm{turb}}\sim 40$ km s$^{-1}$; \citealt{lochhaas21}) but not outflowing. The outflowing material generally span speeds of $v_{R}\sim100$--700 km s$^{-1}$, but the majority ($>95$\%) of the mass in the outflows have velocities $v_{R}<350$ km s$^{-1}$ \citep{lochhaas21}. The star-forming disk is defined as before.

In Figure \ref{fig:thel_phil_outflow}, we show the distribution of the {\emph{net orientation}} of the angular momentum of both the outflowing material and the star-forming disk from $z=1$ to $z=0$. Every simulation output (which are separated by $\Delta\,t\sim$ 5 Myr) below $z=1$ is shown for the Tempest, Squall, and Blizzard halos. The conclusions are similar for the other three halos. Figure \ref{fig:thel_phil_outflow} shows that the outflows generally carry angular momentum that {\emph{is strongly misaligned}} with the star-forming disk. To understand why that is the case, we refer back to the bottom panel of Figure \ref{fig:phil-thel-tempest}---which shows the full distribution of angular momentum split by phase for Tempest. In general, the central galaxies in FOGGIE are surrounded by a corona of ``warm-hot + hot'' ($>10^5$ K) gas that maintains a wide distribution in angular momentum. The majority of the angular momentum stored in this phase of the inner CGM is poorly aligned with that of the star-forming central disk. As the outflowing gas crashes into this corona, it acts to lift this misaligned material from the central galaxy.

In Figure \ref{fig:outflow_timescale}, we calculate the ratio of (1) the magnitude of the angular momentum of the disk and (2) the rate of change of the magnitude due to gas removal from outflows. In doing so, we consider the vector nature of the outflowing gas --- which can either increase or decrease the magnitude of the disk depending on the relative alignment of the material being removed. We use this ratio to quantify the time it would take for the outflowing gas to fully deplete the disk of angular momentum at its current rate. It can be thought of as an ``angular momentum depletion time".

We show this quantity for the Maelstrom halo, and note that the general conclusions below are similar for the other halos. At a given snapshot in the simulation, the outflowing gas generally spans a distribution of speeds ($\sim 150$--700 km s$^{-1}$; \citealt{lochhaas21}). To estimate the rate that angular momentum is carried away from the disk, we measure the total angular momentum associated with the material defined as outflowing (see above) and calculate the flux rate assuming edge case scenarios: (1) all of the material is moving in a slow outflow (150 km s$^{-1}$), and (2) all of the material is moving in a fast outflow (700 km s$^{-1}$). These bounds are indicated with the shaded swath.

Figure \ref{fig:outflow_timescale} indicates that the outflows carry away only a small fraction of the total angular momentum of the Maelstrom central at all times from \ztwo{} to \zzero{}. The time it would take to carry away all of the angular momentum ranges from 2 to 100 Gyr. This conclusion is similar for the other halos and is mostly related to the thermal feedback scheme implemented in the simulation---which launches mostly low density gas on purely radial orbits (i.e., gas with relatively low angular momentum per volume).

\section{Discussion}\label{sec:discussion}
In this section, we briefly summarize the results of the paper and place them within the context of existing observations and theoretical frameworks, where relevant. We begin with an overview of the classic model of disk galaxy formation, and modern updates, in \S\ref{sec:models}.
In \S\ref{sec:disc_AM_magnitude} and \S\ref{sec:disc_AM_orientation} we discuss the evolution of the magnitude and orientation, respectively, of the angular momentum of the FOGGIE central galaxies and their halos. Finally, in \S\ref{sec:disc_AM_inflow_outflow} we discuss the connection between the evolution of the angular momentum of the FOGGIE central galaxies and their inflowing and outflowing gas.

\subsection{Classic model of disk galaxy formation}\label{sec:models}
First, to establish context and contrast with our results,  we briefly review the classic model of disk galaxy formation in a $\Lambda$CDM cosmology (e.g., \citealt{Fall80, Mo98}). 

In this picture, it is presumed that the present-day spin characteristics of galaxies are seeded early in their formation. Baryons and dark matter acquire angular momentum in the early universe through tidal interactions with nearby structure \citep{Hoyle51, peebles69, white84}. As the baryons cool into the centers of their parent dark matter halos, it is perhaps reasonable to expect that they would conserve the magnitude and orientation of their specific angular momentum vector \citep{Fall80, Mo98}. The magnitude of the acquired angular momentum can be expressed with respect to the available dynamical support of the halo via the ``spin parameter'', $\lambda$, where 
\begin{equation}
    \lambda\propto \frac{j\,E^{1/2}}{M^{3/2}}\propto \frac{j}{R_{\mathrm{virial}}v_{\mathrm{virial}}}
\end{equation}\label{eq:spin_eq}

\noindent which relates the specific angular momentum of a given component of mass (e.g., dark matter or baryonic matter), $j$, to that of a circular orbit at the virial radius of the halo, $R_{\mathrm{virial}}\times v_{\mathrm{virial}}$ \citep{peebles69, bullock01}. If $\lambda$ is constant in cosmic time, then the specific angular momentum of the baryons and dark matter should {\emph{both}} increase as the halo grows in mass and size ($j\propto\,H(z)^{-1/3}$; see e.g., \citealt{burkert16}). Commonly, it is assumed that $j_{\mathrm{DM}} = j_{\mathrm{baryons}}$ --- on the notion that both components are initially torqued by the same tidal field and will acquire the same or similar initial spins. 

Together, these assumptions impose that the net specific angular momentum {\emph{vector}} of the baryons should mirror that of the dark matter halo in which they cooled (in magnitude and direction) at all times---regardless of the fact that the dark matter is dissipationless (in energy) while the baryonic gas is not.

In short, this classic picture rests on two key points: (1) galaxies retain memory of their initial spin with a specific angular momentum vector that is fixed in direction and evolves in magnitude alongside the growth of the halo and (2) the evolution of the specific angular momentum of the baryons is, at all times, handcuffed to that of the dark matter halo. The observable predictions from this model e.g., the sizes and angular momenta of disk galaxies, can successfully match the average observed properties of today's disk galaxy population \citep{1983IAUS..100..391F, Dalcanton97, Mo98, RF12, RF13}.

However, over the past two decades and also including the results presented in this paper, a more complicated picture of the angular momentum evolution of galaxies has begun to emerge, mostly through the efforts of hydrodynamic simulations of galaxy formation in the context of a $\Lambda$CDM cosmology \citep[e.g.,][]{sharma05, bailin05, Brook12, Kassin12a, stewart13, ubler14, welker14, danovich15, genel15, Lagos17, EB18, Oppenheimer18, LiJ22, Hafen22, trapp22, stern24, trapp24, Liu24}. As discussed below, these simulations broadly suggest that the acquisition of angular momentum in galaxies is much more complex and stochastic than is suggested by the classic analytic picture above. There are several processes that act to exchange angular momentum between galaxies and their surroundings, many of which are empirically (or likely) prevalent at early cosmic times. These include fresh accretion from the intergalactic medium, recycled accretion from the circumgalactic medium, major and minor mergers, galaxy outflows, and gravitational torques between galaxies, their circumgalactic gas, and their dark matter halos.

We now summarize our results and then briefly, and incompletely, discuss how they fit in with these classic and emerging frameworks for the evolution of angular momentum on galaxy and halo scales.

\begin{figure*}[ht]
\centering
\includegraphics[width=1.0\textwidth]{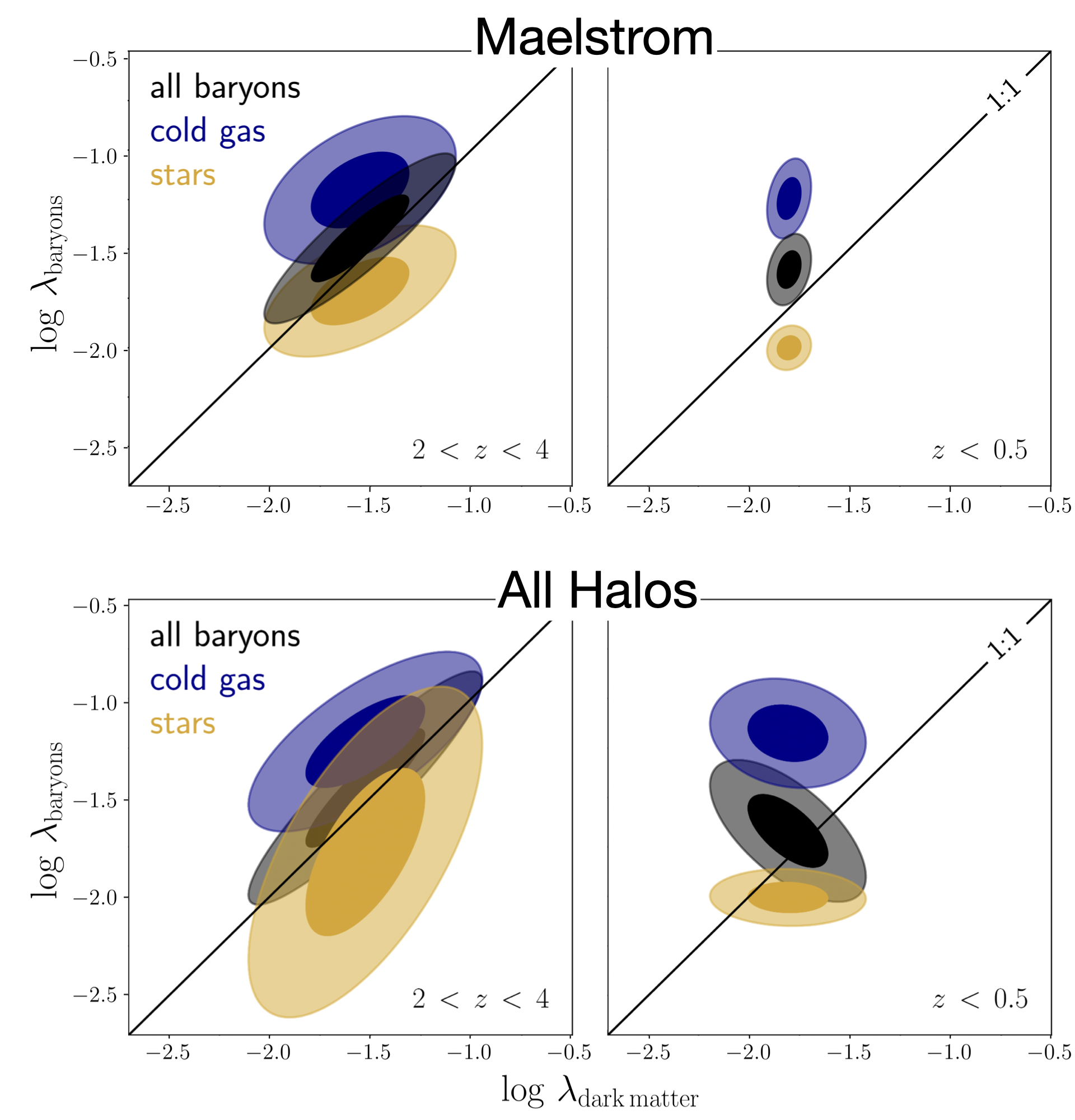}
\caption{The distribution of the spin parameter $\lambda$  \citep{bullock01} of the baryons and dark matter is shown. The Maelstrom halo is shown at top and the full collection of FOGGIE halos is shown at bottom. In both rows, the left panel shows early cosmic times ($2\,<\,z\,<\,4$) and the right panel shows late cosmic times ($z\,<\,0.5$). At early cosmic times the evolving spin of the full collection of baryons in the halo (black) is well-correlated with that of the dark matter: the distribution of simulation snapshots in the $\lambda_{\mathrm{baryon}}-\lambda_{\mathrm{DM}}$ plane is close to the 1:1 line. At late cosmic times, this correlation is lost. The cold gas (blue) maintains a higher spin than the dark matter halo, while the stars (yellow) maintain a lower spin.\vspace{0.4cm}} \label{fig:spin_comparison}
\end{figure*}

\subsection{The Evolution of the Magnitude of the Angular Momentum}\label{sec:disc_AM_magnitude}

First, in \S\,\ref{sec:AM_magnitude}, we measure the evolution of the cumulative profile of the angular momentum of the baryonic mass in the FOGGIE halos from \ztwo{} to \zzero{} ($L_{\mathrm{bary}}$; Figure \ref{fig:cumulative_momentum_allhalos}) sorted by radial distance and split by the component of baryonic mass. At both redshifts, a significant fraction of the total $L_{\mathrm{bary}}$ resides in the halo well-outside the central galaxies ($\sim50$--80\% depending on the halo and redshift). The angular momentum in the outer halo is split in roughly equal portions between: the `warm' 15,000--10$^6$ K circumgalactic gas and the stars and cold interstellar gas of the satellite galaxy population. Each satellite will carry angular momentum through both its internal rotation and its orbit through the halo. The remaining angular momentum in the halo is associated with the rotation of the cold gas and stars in the central disk. 

The fact that the bulk of the angular momentum resides outside the central galaxy across cosmic times (for the MW-halo masses simulated here, at least) presents an obvious but nonetheless important point. To model how the spin of the central galaxies evolve, we must both understand how the angular momentum budget on the halo scale evolves and how and when that angular momentum is transferred down to the scale of the disk.

From \ztwo{} to \zzero{} and across the halos, we find that the magnitude of the total baryonic angular momentum in the halo $L_{\mathrm{bary, vir}} = |\sum_{R_{\mathrm{vir}}} \vec{L}_{\mathrm{bary}} (r)|$ increases by $\times 6 - 24$ (see Table \ref{table:angmom_rvir}). We compare these factors against the classic expectation that $L_{\mathrm{bary, vir}}$  should simply increase in lockstep with the evolving size, mass, and circular velocity of the halo with time, i.e., $L_{\mathrm{bary, vir}} (z)\propto\,R_{\mathrm{vir}} (z)\times M_{\mathrm{vir}}(z)\times v_{\mathrm{vir}} (z)$. From \ztwo{} to \zzero{}, we indeed find that the quantity $R_{\mathrm{vir}}\times M_{\mathrm{vir}} \times v_{\mathrm{vir}}$ increases by a similar range of factors ($\times 6 - 21$) across the halos (Table \ref{table:angmom_rvir}). In other words, the assumption that the baryonic angular momentum is handcuffed to the growth of the halo well-reflects the evolution {\emph{averaged over the full FOGGIE galaxy population\footnote{Keep in mind: this is a statement over only four halos, selected to have a Milky Way-like merger history.}}}.
%, as is often assumed considering a classic analytic assumption of a constant spin parameter $\lambda$.

However, importantly, we find that the one-to-one correspondence between the expected growth and actual growth is {\emph{poor}} for any given halo (Table \ref{table:angmom_rvir}). Only one of the halos, Maelstrom, is roughly consistent with its classic expectation.  The other three halos have growths in $L_{\mathrm{bary, vir}}$ that are $\times2-3$ discrepant with their expectation. For context, the Maelstrom halo experiences its last major merger at $z\sim4$. At late cosmic times $z\,<\,3$, the mass growth of Maelstrom on both the galaxy and halo scale is dominated primarily by a few stable and long-lived filaments and secondarily by a number of minor mergers (with mass ratios lower than 1:10).

We conclude that the growth in the baryonic angular momentum of a halo is not uniquely governed by its mass and size growth alone. The actual growth depends sensitively on the details of the kinematics of the accreted baryons, namely the orbits and frequencies of galaxy mergers and the orientations and impact parameters of the warm filaments of gas that are feeding the halo. 

As an example, Squall experiences a major merger at $z\sim0.75$ with a massive satellite whose orbital angular momentum is $\sim180^{\circ}$ misaligned with the existing angular momentum of the halo. In Figure \ref{fig:R-Z-theta}, this satellite appears as a distinct yellow streak in the dark matter panel for Squall at this redshift. The addition of this retrograde satellite into the halo acts to {\emph{reduce}} the net angular momentum of Squall, while still {\emph{increasing}} its mass and virial radius, in turn steepening Squall's gravitational potential and so increasing $v_{\mathrm{vir}}$. By \zzero{}, the true baryonic angular momentum in the Squall halo is a factor of $\sim3.5\times$ lower than that predicted by the halo growth---a direct result of this single retrograde merger. 

This highlights a first shortcoming of any analytic prescription that ties the angular momentum growth of galaxies to the mass growth of their halos: it will not capture the important variance in the accretion kinematics for individual systems.

In parallel, we also find strong differences in the evolution of the specific angular momentum $j$ of the baryons and the dark matter (Figures \ref{fig:z0_halos} and \ref{fig:Tempest_jprof_dm_baryons}). Among the baryons, $j$ is generally larger for cooler gas and for gas at larger radii. At both \zzero{} and \ztwo{}, the quantity $j$ is generally larger for the baryons than it is for the dark matter, 3--10$\times$ at fixed radius, depending on the temperature of the gas (see also e.g., \citealt{Kimm11, 2015MNRAS.449.2087D,Oppenheimer18, elbadry18} for similar reports at these halo masses). We show that the evolution in $j$ is above-and-beyond what is expected from the growth of the halo and steepening of its potential well ($\propto R_{\mathrm{vir}}\cdot v_{\mathrm{vir}}$; bottom panels of Fig. \ref{fig:Tempest_jprof_dm_baryons}).

A re-framing of this result is that the spin parameter $\lambda$ (Eq \ref{eq:spin_eq}, \citealt{peebles69}, \citealt{bullock01}) can (and does in the FOGGIE halos) both evolve with time and vary between the dark and baryonic components of matter. To explore this, in Figure \ref{fig:spin_comparison} we compare the distribution of the \citet{bullock01} spin parameter evaluated at $R_{\mathrm{vir}}$ ($\lambda=J_{\mathrm{vir}}/(\sqrt{2}M_{\mathrm{vir}}v_{\mathrm{vir}}R_{\mathrm{vir}})$) of the baryons and dark matter for the Maelstrom halo (top row) and the full suite of FOGGIE halos (bottom row) over two windows in cosmic time: one early ($2\,<\,z\,<\,4$) and one late ($z\,<\,0.5$). We subdivide the baryons into cold gas and stars --- which together contain the majority of the angular momentum in the halo at both redshifts (see e.g., Figure \ref{fig:cumulative_momentum_allhalos}). We first note that there are wide distributions of the spin parameter in both components --- both for the collection of halos (bottom panel) and for Maelstrom alone. This indicates that there are two contributions to the scatter in $\lambda$: (i) halo-to-halo differences and (ii) true time evolution over the windows examined ($\sim1.8$ Gyr). 

At high redshift (left panel), the distribution for the baryons $\lambda_{\mathrm{baryons}}$ is well-correlated with that of the dark matter $\lambda_{\mathrm{DM}}$.  By low redshifts (right panel), the link between the dark matter and the baryonic matter is lost: the distributions are no longer sloped along the 1:1 line. This is consistent with the results of \citet{2019MNRAS.488.4801J}, which arrived at a similar conclusion using a larger population of simulated galaxies (N $\sim$ 150) from two simulation suites. \citet{2019MNRAS.488.4801J} report that while the global distributions of the spin parameters of the baryonic components of simulated galaxies are similar to their dark matter halos, there is effectively zero correlation between the two parameters on a galaxy-by-galaxy basis. In FOGGIE, we find that the discrepancy between $\lambda_{\mathrm{DM}}$ and $\lambda_{\mathrm{baryons}}$ for individual systems in large part reflects the stochastic nature of the kinematics of the baryonic matter accreted onto the halo (from satellites and warm filaments). 

To understand when and where the baryons develop high $j$, we dissect the $j$ evolution of each baryonic component as a function of radial distance and cosmic time, separately, for the Tempest halo (Figure \ref{fig:Tempest_jprof_dm_baryons}, \ref{fig:Tempest_jprof_baryonic_components}) --- generalizing for the other halos. We show that the radial profile of $j$ for the dark matter in the Tempest halo undergoes little evolution from \ztwo{} to \zzero{}. Moreover, it follows the $j(r)\propto r$ scaling expected from tidal torque theory \citep{1987ApJ...319..575B, bullock01}. By contrast, we find a strong evolution in the $j$ profile of the baryonic matter in the outer disk and halo (i.e., CGM). Most of this growth can be attributed to the increase in $j$ of the warm and cold gas (i.e., $T<10^5\,\mathrm{K}$; Figure \ref{fig:Tempest_jprof_baryonic_components}).

\subsection{The Evolution of the Orientation of the Angular Momentum Orientation of the Disk and CGM} \label{sec:disc_AM_orientation}

In \S\ref{sec:AM_orientation}, we turn our attention towards the distribution and time evolution of the {\emph{orientation}} of the angular momentum vector --- examining how it varies within an individual halo, across the components of mass, and with cosmic time. 

In Figure \ref{fig:R-Z-theta} and Figure \ref{fig:sf_versus_dm_phil-thel}, we examine the evolution of the absolute orientation of the cold gas and dark matter as a function of radial position and redshift from $z\sim 2$ to $z\sim 0$. There are two key results that we highlight:

\begin{enumerate}[label=(\roman*)]
    \item The angular momentum orientation of the \zzero{} cold gas disk generally only develops at late cosmic times --- after $z\sim1.5$ in {\emph{all of the FOGGIE halos}}.
    
    \item The angular momentum orientation of the central disks are largely disconnected from that of {\emph{both}} their dark matter host halo and their circumgalactic gas after \ztwo. 
\end{enumerate}

These results are each in independent conflict with the traditional disk formation model, which suggest  that the spin properties of galaxies are established early and persist linked to those of their host halos. Instead, we find that the angular momentum vector of the FOGGIE centrals develops late, influenced primarily by the specific kinematic properties of the late time accretion onto the central disk.

The alignment between the orientation of the angular momentum vector of the dark matter and the baryonic disk has been studied in other zoom-in (e.g., \citealt{2010MNRAS.405..274H, 2019MNRAS.488.4801J, 2023ApJ...958...44B}) and large box \citep{Sheng23, Rodriguez24} simulation suites of halos that are similar in mass to the FOGGIE suite. Misalignments up to 90$^\circ$ have been reported (although see \citealt{Sheng23}). Across the FOGGIE suite, the orientation of the outer dark matter halos are relatively constant from \ztwo{} to \zzero{} --- in general, misaligned  with the central disk by $75-90^{\circ}$. However, the {\emph{inner}} portion of the dark matter halo is generally well-aligned with the central disk across these same redshifts.

The late settling of the angular momentum orientation of the FOGGIE disks echoes observations of the ionized cold gas disks in galaxies, which show that regular rotation (rotation to velocity dispersion ratio $> 3$) develops in galaxies at late times (e.g., \citealt{kassin12, simons17}). The kinematics of the ionized star-forming gas have now been mapped for several thousand Milky Way-mass galaxies at \zzero{} (e.g., \citealt{bundy21, croom21, sanchez23}) and several hundred Milky Way progenitors to \ztwo{} (e.g., \citealt{wisnioski15, wisnioski19, girard20, tiley21}). For the majority ($>85\%$) of these systems both in the local universe \citep{Jin16, Ristea22} and at \ztwo{} \citep{wisnioski15}, the direction of the rotation axis of the ionized gas disk is generally found to be well-aligned with the minor axis of the continuum tracing the stars. This indicates that the angular momentum of the star-forming cold gas disk and the stellar disk are aligned. However, fewer observational constraints are available for the relative alignment of the disk and its gas in the circumgalactic medium. 

Observationally, the line-of-sight kinematics of the circumgalactic gas surrounding external galaxies can be measured through the intervening absorption features they will imprint on the spectra of distant quasars (e.g., \citealt{Ho17, augustin18, Lopez18, lochhaas19, Zabl19, Martin19, Lopez20, lehner21}). 

At high redshift ($1\,<\,z\,<\,2$), observations of \mgii\ absorption (associated with cool $\sim10^4$ K photoionized gas around galaxies) often reveal signs of rotation in the inner circumgalactic medium with a rotation axis that is generally, but not always, aligned with that of the central disk \citep{Zabl19, Martin19}. In the Illustris TNG50 simulations, \citet{DeFelippis21} showed that Mg II is a reliable tracer of the cooler circumgalactic gas, and that the specific angular momentum of the Mg II gas is reflective of the full population of gas in the CGM.

On the other hand, Lyman-$\alpha$ (also tracing cool gas in the halo) that has been imaged in emission around high redshift galaxies generally does {\emph{not}} show signatures of rotation \citep{2016ApJ...831...39B, 2019ApJS..245...23C}. It is important to note that Lyman-$\alpha$ is a resonant line and is highly susceptible to scattering. The Lyman-$\alpha$ photons that eventually escape the halo will preferentially be those that last scattered off of a H atom in the extreme blue-shifted and red-shifted tails of the velocity distribution of of the CGM. For this reason, Lyman-$\alpha$ is a more complicated, and less reliable, tracer of the kinematics of the halo.

In Figures \ref{fig:Tempest_z2slice} and \ref{fig:Tempest_z0slice}, we use the Tempest halo to illustrate the high spatial variability of the angular momentum vector in the FOGGIE halos. We find that the angular momentum orientation can vary by up to $180^{\circ}$ over small distances ($\sim$tens of kpc). This is in direct conflict with any model that describes the circumgalactic medium as a slowly rotating spheroid with a single angular momentum orientation. At \ztwo{}, the orientation of the angular momentum exhibits high spatial variability throughout the full virial volume of the halos. By \zzero{}, the angular momentum of the inner regions of the halos generally develop a coherent direction while the outer regions still exhibit high variability.

\subsection{Angular Momentum Exchanged Through Halo-Scale Gas Flows}\label{sec:disc_AM_inflow_outflow}

In \S\,\ref{sec:inflowing_outflowing}, we examine the time evolution of the angular momentum that is being carried onto and away from the FOGGIE centrals by inflowing and outflowing gas, respectively. We then relate the evolving character of those flows of angular momentum to the evolution of the angular momentum of the central disks. 

\vspace{0.2cm}
\noindent{\bf{Inflowing Gas:}} In Figure \ref{fig:phil_thel_inflow_disk} we use the \philthel{} plane to track the co-evolution of the 3-D orientation of the net angular momentum of both the star-forming disk and the cool gas inflow in the FOGGIE halos. In all of the halos, we find that the 3-D orientations of the net angular momentum vectors of both the star-forming disk and the inflowing gas vary significantly from \ztwo{} to \zzero{}. Specifically, we measure net changes in the orientation that range between $\sim45^{\circ}$--$180^{\circ}$ across the halos over that period.

The role that inflows play in the angular momentum evolution of Milky Way-like galaxies has been studied across a number of different simulation suites (e.g., \citealt{Lagos17, Stern21, Hafen22}). 
\citet{Hafen22} and \citet{Stern21} studied how angular momentum is acquired in $z\sim0$ Milky Way-mass galaxies in the FIRE  and FIRE-2 simulation suites, respectively. In FIRE-2, the formation of physically thin and regularly-rotating disks is precipitated by the virialization of the inner CGM \citep{Stern21, Yu23}, and a transition in the dominant mode of accretion from a supersonic ``cold mode accretion'' provided by intergalactic filaments to a more gentle ``cooling flow accretion'' of gas cooling from the hot virialized CGM \citep{Stern21, Hafen22, stern24, hopkins23}. For FIRE's Milky Way-mass halos, this transition occurs at late times ($z<1$). \citet{Hafen22} showed that the cooling flows generally flatten into the plane of the disk as they cool---and that the fraction of the angular momentum that is aligned with the existing disk orientation governs the mass fraction of the thin disk. It is important to note that the smooth gas accretion in FOGGIE is dominated by warm filamentary inflows across all halos and at all cosmic times (C.\ Lochhaas et al., in preparation)---in contrast with the hot CGM cooling flow accretion seen in FIRE-2 (\citealp{Stern21}; see also \citealp{stern24}).

%In the Eagle simulations, \citet{Lagos17} studied how angular momentum is added to systems through inflowing gas. [more here]. We emphasize that the Milky Way population studied in this paper represents only a subset of the population studied in large-box simulations such as EAGLE \citep{Lagos17}, Illustris/Illustris-TNG \citep{genel15, RG22}, and New Horizons \citep{welker14}, which can examine mass and redshift trends independently albeit at more coarse mass, spatial, and temporal resolution.

%---combine all of the below into their own paragraph
%Compared to the Eagle simulations (Lagos+)...

%Compared to the Illustris simulations (need reference+)...

%Compared to the New Horizons simulations (Welker+)...

%In semi-analytic models (Dave+)...

\vspace{0.2cm}
\noindent{\bf{Outflowing Gas:}} We now turn our attention to the angular momentum carried away from the FOGGIE centrals by outflows. We emphasize that our conclusions are largely statements on the specific thermal SNe stellar feedback model implemented in FOGGIE and should not be taken as generalized conclusions for this galaxy population.

In Figure \ref{fig:outflow_timescale}, we show that the rate at which the FOGGIE outflows remove angular momentum from the central disks is small and negligible compared to the angular momentum content of the disks. We measure the time that it would take for the outflows to fully deplete the disk of its angular momentum---at the rate they are doing so at each point in time. We conclude that the ``depletion time'' of the angular momentum is large in all halos and at all cosmic times. It ranges from $\sim 5$ Gyr to several factors of the Hubble time. The inefficiency of the outflows in removing angular momentum is due to two factors: (1) the outflows generally carry away angular momentum that is misaligned with the net orientation of the disk (Figure \ref{fig:thel_phil_outflow}) and (2) the outflows are generally diffuse and carry relatively little mass (see Fig. 5 of \citealt{lochhaas21}).

It has been noted by several previous works that the re-accretion of outflowing material plays an important role in the angular momentum evolution of galaxies (so called ``fountain flows"; e.g., \citealt{Brook12, ubler14, Christensen16, DeFelippis17, Grand19, Semenov23a, Semenov23b}). For instance, using the Illustris-TNG50 simulation \citet{Semenov23a, Semenov23b} showed that at the time of disk formation in the Milky Way-like population, the specific angular momentum $j$ of the inflowing and outflowing gas have similar distributions in $j$. This suggests that the accretion onto these systems is dominated by that which has been ejected and re-accreted.

In general, low angular momentum material carried away by the outflow can also mix and inherit high angular momentum from the CGM , and can also be torqued to higher angular momentum by the dark halo. If this material eventually re-joins the central galaxy, it will act as a conduit for angular momentum to exchange between the circumgalactic gas, the dark matter halo and the galaxy.

We note that we can not track the history of individual gas parcels, and so we can not directly interrogate this scenario. However, the majority of the accretion onto the disks at $z<2$ is dominated by intergalactic cold + warm flows, and not by the smooth accretion from the warm-hot circumgalactic medium.

\section{Summary}\label{sec:conclusions}

We investigate the co-evolution of the angular momenta of Milky Way-like galaxies, their circumgalactic gas, and their dark matter halos using zoom-in simulations from the Figuring Out Gas \& Galaxies in Enzo (FOGGIE) simulation suite. In each halo, we measure how the magnitude and orientation of the angular momentum varies with the position in the halo, between the components of mass, and as a function of cosmic time. Our conclusions are as follows.

\begin{enumerate}
\item From $z\sim2$ to today, and in general across the simulated halos, the specific angular momentum of the central galaxies and the cool gas in their circumgalactic media ($T<10^5\,\mathrm{K}$) increase in concert. Over that same period, the specific angular momentum of the hot ($>\,10^6\,\mathrm{K}$) and dark components of the halo minimally change. The growth in the baryonic angular momentum is larger than that expected from the mass and size growth of the halo alone.

\item By $z\sim 1$, the central galaxies have generally lost association with the angular momentum of their dark halo---both in magnitude and orientation. We find a wide distribution of angular momentum orientations in the halo, varying by up to $180^{\circ}$ over small ($\sim$tens of kpc) scales and between the different components of mass. 

\item The {\emph{net}} angular momenta of the galaxy, circumgalactic gas, and dark matter halo are generally misaligned with one another at all cosmic times. The present-day orientation of the central galaxies are only established at late times (after $z=1$), after the rates of cosmic accretion and merging sufficiently decline---and the disks are able to settle and stabilize their orientation.

\item Across our halos and the cosmic time studied, the SNe-driven outflows in FOGGIE remove a negligible fraction of the angular momentum of the central disks.

\item We provide predictions for the population mean and variance of angular momentum in low mass galaxies at high redshift---a population that is just now being explored with the near-infrared spectrographs on the \textit{JWST} observatory. We predict that the observed population scatter at those masses in large part reflect the short-timescale behavior of individual galaxies.
\end{enumerate}

In sum, the results presented in this paper highlight a complex story for the acquisition and re-distribution of angular momentum in Milky Way-like systems and their halos. The evolution of the dark and baryonic matter are largely decoupled at late times, as is that of the matter in the central disk and the halo. Perhaps the most striking implication of our findings is that present-day Milky Way-like galaxies retain little to no memory of their angular momentum from the early universe. This suggests that their angular momentum identity is shaped primarily by late-time processes rather than their initial conditions.

\section*{Acknowledgments}
RCS appreciates support from a Giacconi Fellowship at the Space Telescope Science Institute. RA, CL, AA, and MSP were supported by NASA via an Astrophysics Theory Program grant 80NSSC18K1105. AA's efforts for this work were additionally supported by NSF-AST 1910414 and HST AR \#16151. KHC was supported by HST AR \#15052. RA's efforts for this work were additionally supported by HST AR \#15012 and HST GO \#16730.  RA and CL also acknowledge financial support from the STScI Director’s Discretionary Research Fund (DDRF). RA also acknowledges funding by the European Research Council through ERC-AdG SPECMAP-CGM, GA 101020943. BWO acknowledges support from NSF grants \#1908109 and \#2106575, NASA ATP grants NNX15AP39G and 80NSSC18K1105, and NASA TCAN grant 80NSSC21K1053. JT and ACW acknowledge support from the \textit{Nancy Grace Roman Space Telescope} Project, under the Milky Way Science Investigation Team.   RA's efforts for this work were additionally supported by HST GO \#16730. RA acknowledges funding by the European Research Council through ERC-AdG SPECMAP-CGM, GA 101020943. CL's efforts for this work were additionally supported by NASA through the NASA Hubble Fellowship grant \#HF2-51538 awarded by the Space Telescope Science Institute, which is operated by the Association of Universities for Research in Astronomy, Inc., for NASA, under contract NAS5-26555.

Computations described in this work were performed using the publicly-available \textsc{Enzo} code (\href{http://enzo-project.org}{http://enzo-project.org}), which is the product of a collaborative effort of many independent scientists from numerous institutions around the world. Their commitment to open science has helped make this work possible. The python packages {\sc matplotlib} \citep{hunter2007}, {\sc numpy} \citep{walt2011numpy}, {\sc rockstar} \citep{Behroozi2013a}, {\sc tangos} \citep{pontzen2018}, \textsc{scipy} \citep{scipy2020}, {\sc yt} \citep{ytpaper}, and {\sc Astropy} \citep{astropy2013,astropy2018,astropy2022} were all used in parts of this analysis or in products used by this paper. 

Resources supporting this work were provided by the NASA High-End Computing (HEC) Program through the NASA Advanced Supercomputing (NAS) Division at Ames Research Center and were sponsored by NASA's Science Mission Directorate; we are grateful for the superb user-support provided by NAS. Resources were also provided by the Blue Waters sustained-petascale computing project, which is supported by the NSF (award number ACI-1238993 and ACI-1514580) and the state of Illinois. Blue Waters is a joint effort of the University of Illinois at Urbana-Champaign and its NCSA. Computations described in this work were performed using the publicly-available Enzo code, which is the product of a collaborative effort of many independent scientists from numerous institutions around the world.

\bibliography{angular_momentum}{}

\begin{thebibliography}{}
\expandafter\ifx\csname natexlab\endcsname\relax\def\natexlab#1{#1}\fi
\providecommand{\url}[1]{\href{#1}{#1}}
\providecommand{\dodoi}[1]{doi:~\href{http://doi.org/#1}{\nolinkurl{#1}}}
\providecommand{\doeprint}[1]{\href{http://ascl.net/#1}{\nolinkurl{http://ascl.net/#1}}}
\providecommand{\doarXiv}[1]{\href{https://arxiv.org/abs/#1}{\nolinkurl{https://arxiv.org/abs/#1}}}

\bibitem[{{Acharyya} {et~al.}(2024){Acharyya}, {Peeples}, {Tumlinson}, {Shea},
  {Lochhaas}, {Wright}, {Simons}, {Augustin}, {Smith}, \& {Hyeonmin
  Lee}}]{Acharyya24}
{Acharyya}, A., {Peeples}, M.~S., {Tumlinson}, J., {et~al.} 2024, arXiv
  e-prints, arXiv:2404.06613, \dodoi{10.48550/arXiv.2404.06613}

\bibitem[{{Astropy Collaboration} {et~al.}(2013){Astropy Collaboration},
  {Robitaille}, {Tollerud}, {Greenfield}, {Droettboom}, {Bray}, {Aldcroft},
  {Davis}, {Ginsburg}, {Price-Whelan}, {Kerzendorf}, {Conley}, {Crighton},
  {Barbary}, {Muna}, {Ferguson}, {Grollier}, {Parikh}, {Nair}, {Unther},
  {Deil}, {Woillez}, {Conseil}, {Kramer}, {Turner}, {Singer}, {Fox}, {Weaver},
  {Zabalza}, {Edwards}, {Azalee Bostroem}, {Burke}, {Casey}, {Crawford},
  {Dencheva}, {Ely}, {Jenness}, {Labrie}, {Lim}, {Pierfederici}, {Pontzen},
  {Ptak}, {Refsdal}, {Servillat}, \& {Streicher}}]{astropy2013}
{Astropy Collaboration}, {Robitaille}, T.~P., {Tollerud}, E.~J., {et~al.} 2013,
  \aap, 558, A33, \dodoi{10.1051/0004-6361/201322068}

\bibitem[{{Astropy Collaboration} {et~al.}(2018){Astropy Collaboration},
  {Price-Whelan}, {Sip{\H{o}}cz}, {G{\"u}nther}, {Lim}, {Crawford}, {Conseil},
  {Shupe}, {Craig}, {Dencheva}, {Ginsburg}, {VanderPlas}, {Bradley},
  {P{\'e}rez-Su{\'a}rez}, {de Val-Borro}, {Aldcroft}, {Cruz}, {Robitaille},
  {Tollerud}, {Ardelean}, {Babej}, {Bach}, {Bachetti}, {Bakanov}, {Bamford},
  {Barentsen}, {Barmby}, {Baumbach}, {Berry}, {Biscani}, {Boquien}, {Bostroem},
  {Bouma}, {Brammer}, {Bray}, {Breytenbach}, {Buddelmeijer}, {Burke},
  {Calderone}, {Cano Rodr{\'\i}guez}, {Cara}, {Cardoso}, {Cheedella}, {Copin},
  {Corrales}, {Crichton}, {D'Avella}, {Deil}, {Depagne}, {Dietrich}, {Donath},
  {Droettboom}, {Earl}, {Erben}, {Fabbro}, {Ferreira}, {Finethy}, {Fox},
  {Garrison}, {Gibbons}, {Goldstein}, {Gommers}, {Greco}, {Greenfield},
  {Groener}, {Grollier}, {Hagen}, {Hirst}, {Homeier}, {Horton}, {Hosseinzadeh},
  {Hu}, {Hunkeler}, {Ivezi{\'c}}, {Jain}, {Jenness}, {Kanarek}, {Kendrew},
  {Kern}, {Kerzendorf}, {Khvalko}, {King}, {Kirkby}, {Kulkarni}, {Kumar},
  {Lee}, {Lenz}, {Littlefair}, {Ma}, {Macleod}, {Mastropietro}, {McCully},
  {Montagnac}, {Morris}, {Mueller}, {Mumford}, {Muna}, {Murphy}, {Nelson},
  {Nguyen}, {Ninan}, {N{\"o}the}, {Ogaz}, {Oh}, {Parejko}, {Parley}, {Pascual},
  {Patil}, {Patil}, {Plunkett}, {Prochaska}, {Rastogi}, {Reddy Janga},
  {Sabater}, {Sakurikar}, {Seifert}, {Sherbert}, {Sherwood-Taylor}, {Shih},
  {Sick}, {Silbiger}, {Singanamalla}, {Singer}, {Sladen}, {Sooley},
  {Sornarajah}, {Streicher}, {Teuben}, {Thomas}, {Tremblay}, {Turner},
  {Terr{\'o}n}, {van Kerkwijk}, {de la Vega}, {Watkins}, {Weaver}, {Whitmore},
  {Woillez}, {Zabalza}, \& {Astropy Contributors}}]{astropy2018}
{Astropy Collaboration}, {Price-Whelan}, A.~M., {Sip{\H{o}}cz}, B.~M., {et~al.}
  2018, \aj, 156, 123, \dodoi{10.3847/1538-3881/aabc4f}

\bibitem[{{Astropy Collaboration} {et~al.}(2022){Astropy Collaboration},
  {Price-Whelan}, {Lim}, {Earl}, {Starkman}, {Bradley}, {Shupe}, {Patil},
  {Corrales}, {Brasseur}, {N{\"o}the}, {Donath}, {Tollerud}, {Morris},
  {Ginsburg}, {Vaher}, {Weaver}, {Tocknell}, {Jamieson}, {van Kerkwijk},
  {Robitaille}, {Merry}, {Bachetti}, {G{\"u}nther}, {Aldcroft},
  {Alvarado-Montes}, {Archibald}, {B{\'o}di}, {Bapat}, {Barentsen},
  {Baz{\'a}n}, {Biswas}, {Boquien}, {Burke}, {Cara}, {Cara}, {Conroy},
  {Conseil}, {Craig}, {Cross}, {Cruz}, {D'Eugenio}, {Dencheva}, {Devillepoix},
  {Dietrich}, {Eigenbrot}, {Erben}, {Ferreira}, {Foreman-Mackey}, {Fox},
  {Freij}, {Garg}, {Geda}, {Glattly}, {Gondhalekar}, {Gordon}, {Grant},
  {Greenfield}, {Groener}, {Guest}, {Gurovich}, {Handberg}, {Hart},
  {Hatfield-Dodds}, {Homeier}, {Hosseinzadeh}, {Jenness}, {Jones}, {Joseph},
  {Kalmbach}, {Karamehmetoglu}, {Ka{\l}uszy{\'n}ski}, {Kelley}, {Kern},
  {Kerzendorf}, {Koch}, {Kulumani}, {Lee}, {Ly}, {Ma}, {MacBride}, {Maljaars},
  {Muna}, {Murphy}, {Norman}, {O'Steen}, {Oman}, {Pacifici}, {Pascual},
  {Pascual-Granado}, {Patil}, {Perren}, {Pickering}, {Rastogi}, {Roulston},
  {Ryan}, {Rykoff}, {Sabater}, {Sakurikar}, {Salgado}, {Sanghi}, {Saunders},
  {Savchenko}, {Schwardt}, {Seifert-Eckert}, {Shih}, {Jain}, {Shukla}, {Sick},
  {Simpson}, {Singanamalla}, {Singer}, {Singhal}, {Sinha}, {Sip{\H{o}}cz},
  {Spitler}, {Stansby}, {Streicher}, {{\v{S}}umak}, {Swinbank}, {Taranu},
  {Tewary}, {Tremblay}, {Val-Borro}, {Van Kooten}, {Vasovi{\'c}}, {Verma}, {de
  Miranda Cardoso}, {Williams}, {Wilson}, {Winkel}, {Wood-Vasey}, {Xue},
  {Yoachim}, {Zhang}, {Zonca}, \& {Astropy Project Contributors}}]{astropy2022}
{Astropy Collaboration}, {Price-Whelan}, A.~M., {Lim}, P.~L., {et~al.} 2022,
  \apj, 935, 167, \dodoi{10.3847/1538-4357/ac7c74}

\bibitem[{{Augustin} {et~al.}(2018){Augustin}, {P{\'e}roux}, {M{\o}ller},
  {Kulkarni}, {Rahmani}, {Milliard}, {Pieri}, {York}, {Vladilo}, {Aller}, \&
  {Zwaan}}]{augustin18}
{Augustin}, R., {P{\'e}roux}, C., {M{\o}ller}, P., {et~al.} 2018, \mnras, 478,
  3120, \dodoi{10.1093/mnras/sty1287}

\bibitem[{{Bailin} \& {Steinmetz}(2005)}]{bailin05}
{Bailin}, J., \& {Steinmetz}, M. 2005, \apj, 627, 647, \dodoi{10.1086/430397}

\bibitem[{{Baptista} {et~al.}(2023){Baptista}, {Sanderson}, {Huber}, {Wetzel},
  {Sameie}, {Boylan-Kolchin}, {Bailin}, {Hopkins}, {Faucher-Giguere},
  {Chakrabarti}, {Vargya}, {Panithanpaisal}, {Arora}, \&
  {Cunningham}}]{2023ApJ...958...44B}
{Baptista}, J., {Sanderson}, R., {Huber}, D., {et~al.} 2023, \apj, 958, 44,
  \dodoi{10.3847/1538-4357/acea79}

\bibitem[{{Barnes} \& {Efstathiou}(1987)}]{1987ApJ...319..575B}
{Barnes}, J., \& {Efstathiou}, G. 1987, \apj, 319, 575, \dodoi{10.1086/165480}

\bibitem[{{Behroozi} {et~al.}(2019){Behroozi}, {Wechsler}, {Hearin}, \&
  {Conroy}}]{behroozi19}
{Behroozi}, P., {Wechsler}, R.~H., {Hearin}, A.~P., \& {Conroy}, C. 2019,
  \mnras, 488, 3143, \dodoi{10.1093/mnras/stz1182}

\bibitem[{{Behroozi} {et~al.}(2013){Behroozi}, {Wechsler}, \&
  {Wu}}]{Behroozi2013a}
{Behroozi}, P.~S., {Wechsler}, R.~H., \& {Wu}, H.-Y. 2013, \apj, 762, 109,
  \dodoi{10.1088/0004-637X/762/2/109}

\bibitem[{{Bird} {et~al.}(2021){Bird}, {Loebman}, {Weinberg}, {Brooks},
  {Quinn}, \& {Christensen}}]{Bird21}
{Bird}, J.~C., {Loebman}, S.~R., {Weinberg}, D.~H., {et~al.} 2021, \mnras, 503,
  1815, \dodoi{10.1093/mnras/stab289}

\bibitem[{{Bland-Hawthorn} \& {Gerhard}(2016)}]{2016ARA&A..54..529B}
{Bland-Hawthorn}, J., \& {Gerhard}, O. 2016, \araa, 54, 529,
  \dodoi{10.1146/annurev-astro-081915-023441}

\bibitem[{{Borisova} {et~al.}(2016){Borisova}, {Cantalupo}, {Lilly}, {Marino},
  {Gallego}, {Bacon}, {Blaizot}, {Bouch{\'e}}, {Brinchmann}, {Carollo},
  {Caruana}, {Finley}, {Herenz}, {Richard}, {Schaye}, {Straka}, {Turner},
  {Urrutia}, {Verhamme}, \& {Wisotzki}}]{2016ApJ...831...39B}
{Borisova}, E., {Cantalupo}, S., {Lilly}, S.~J., {et~al.} 2016, \apj, 831, 39,
  \dodoi{10.3847/0004-637X/831/1/39}

\bibitem[{{Brook} {et~al.}(2012){Brook}, {Stinson}, {Gibson}, {Ro{\v{s}}kar},
  {Wadsley}, \& {Quinn}}]{Brook12}
{Brook}, C.~B., {Stinson}, G., {Gibson}, B.~K., {et~al.} 2012, \mnras, 419,
  771, \dodoi{10.1111/j.1365-2966.2011.19740.x}

\bibitem[{{Brooks} {et~al.}(2009){Brooks}, {Governato}, {Quinn}, {Brook}, \&
  {Wadsley}}]{Brooks09}
{Brooks}, A.~M., {Governato}, F., {Quinn}, T., {Brook}, C.~B., \& {Wadsley}, J.
  2009, \apj, 694, 396, \dodoi{10.1088/0004-637X/694/1/396}

\bibitem[{{Brummel-Smith} {et~al.}(2019){Brummel-Smith}, {Bryan}, {Butsky},
  {Corlies}, {Emerick}, {Forbes}, {Fujimoto}, {Goldbaum}, {Grete}, {Hummels},
  {Kim}, {Koh}, {Li}, {Li}, {Li}, {OShea}, {Peeples}, {Regan}, {Salem},
  {Schmidt}, {Simpson}, {Smith}, {Tumlinson}, {Turk}, {Wise}, {Abel},
  {Bordner}, {Cen}, {Collins}, {Crosby}, {Edelmann}, {Hahn}, {Harkness},
  {Harper-Clark}, {Kong}, {Kritsuk}, {Kuhlen}, {Larrue}, {Lee}, {Meece},
  {Norman}, {Oishi}, {Paschos}, {Peruta}, {Razoumov}, {Reynolds}, {Silvia},
  {Skillman}, {Skory}, {So}, {Tasker}, {Wagner}, {Wang}, {Xu}, \&
  {Zhao}}]{brummel-smith19}
{Brummel-Smith}, C., {Bryan}, G., {Butsky}, I., {et~al.} 2019, The Journal of
  Open Source Software, 4, 1636, \dodoi{10.21105/joss.01636}

\bibitem[{{Bryan} \& {Norman}(1998)}]{bryan98}
{Bryan}, G.~L., \& {Norman}, M.~L. 1998, \apj, 495, 80, \dodoi{10.1086/305262}

\bibitem[{{Bryan} {et~al.}(2014){Bryan}, {Norman}, {O'Shea}, {Abel}, {Wise},
  {Turk}, {Reynolds}, {Collins}, {Wang}, {Skillman}, {Smith}, {Harkness},
  {Bordner}, {Kim}, {Kuhlen}, {Xu}, {Goldbaum}, {Hummels}, {Kritsuk}, {Tasker},
  {Skory}, {Simpson}, {Hahn}, {Oishi}, {So}, {Zhao}, {Cen}, {Li}, \& {Enzo
  Collaboration}}]{bryan14}
{Bryan}, G.~L., {Norman}, M.~L., {O'Shea}, B.~W., {et~al.} 2014, \apjs, 211,
  19, \dodoi{10.1088/0067-0049/211/2/19}

\bibitem[{{Bullock} {et~al.}(2001){Bullock}, {Dekel}, {Kolatt}, {Kravtsov},
  {Klypin}, {Porciani}, \& {Primack}}]{bullock01}
{Bullock}, J.~S., {Dekel}, A., {Kolatt}, T.~S., {et~al.} 2001, \apj, 555, 240,
  \dodoi{10.1086/321477}

\bibitem[{{Bundy} {et~al.}(2015){Bundy}, {Bershady}, {Law}, {Yan}, {Drory},
  {MacDonald}, {Wake}, {Cherinka}, {S{\'a}nchez-Gallego}, {Weijmans}, {Thomas},
  {Tremonti}, {Masters}, {Coccato}, {Diamond-Stanic}, {Arag{\'o}n-Salamanca},
  {Avila-Reese}, {Badenes}, {Falc{\'o}n-Barroso}, {Belfiore}, {Bizyaev},
  {Blanc}, {Bland-Hawthorn}, {Blanton}, {Brownstein}, {Byler}, {Cappellari},
  {Conroy}, {Dutton}, {Emsellem}, {Etherington}, {Frinchaboy}, {Fu}, {Gunn},
  {Harding}, {Johnston}, {Kauffmann}, {Kinemuchi}, {Klaene}, {Knapen},
  {Leauthaud}, {Li}, {Lin}, {Maiolino}, {Malanushenko}, {Malanushenko}, {Mao},
  {Maraston}, {McDermid}, {Merrifield}, {Nichol}, {Oravetz}, {Pan}, {Parejko},
  {Sanchez}, {Schlegel}, {Simmons}, {Steele}, {Steinmetz}, {Thanjavur},
  {Thompson}, {Tinker}, {van den Bosch}, {Westfall}, {Wilkinson}, {Wright},
  {Xiao}, \& {Zhang}}]{bundy21}
{Bundy}, K., {Bershady}, M.~A., {Law}, D.~R., {et~al.} 2015, \apj, 798, 7,
  \dodoi{10.1088/0004-637X/798/1/7}

\bibitem[{{Burkert} {et~al.}(2016){Burkert}, {F{\"o}rster Schreiber}, {Genzel},
  {Lang}, {Tacconi}, {Wisnioski}, {Wuyts}, {Bandara}, {Beifiori}, {Bender},
  {Brammer}, {Chan}, {Davies}, {Dekel}, {Fabricius}, {Fossati}, {Kulkarni},
  {Lutz}, {Mendel}, {Momcheva}, {Nelson}, {Naab}, {Renzini}, {Saglia},
  {Sharples}, {Sternberg}, {Wilman}, \& {Wuyts}}]{burkert16}
{Burkert}, A., {F{\"o}rster Schreiber}, N.~M., {Genzel}, R., {et~al.} 2016,
  \apj, 826, 214, \dodoi{10.3847/0004-637X/826/2/214}

\bibitem[{{Cai} {et~al.}(2019){Cai}, {Cantalupo}, {Prochaska}, {Arrigoni
  Battaia}, {Burchett}, {Li}, {Chisholm}, {Bundy}, \&
  {Hennawi}}]{2019ApJS..245...23C}
{Cai}, Z., {Cantalupo}, S., {Prochaska}, J.~X., {et~al.} 2019, \apjs, 245, 23,
  \dodoi{10.3847/1538-4365/ab4796}

\bibitem[{{Cen} \& {Ostriker}(2006)}]{cen_ostriker_06}
{Cen}, R., \& {Ostriker}, J.~P. 2006, \apj, 650, 560, \dodoi{10.1086/506505}

\bibitem[{{Christensen} {et~al.}(2016){Christensen}, {Dav{\'e}}, {Governato},
  {Pontzen}, {Brooks}, {Munshi}, {Quinn}, \& {Wadsley}}]{Christensen16}
{Christensen}, C.~R., {Dav{\'e}}, R., {Governato}, F., {et~al.} 2016, \apj,
  824, 57, \dodoi{10.3847/0004-637X/824/1/57}

\bibitem[{{Corlies} {et~al.}(2020){Corlies}, {Peeples}, {Tumlinson}, {O'Shea},
  {Lehner}, {Howk}, {O'Meara}, \& {Smith}}]{corlies20}
{Corlies}, L., {Peeples}, M.~S., {Tumlinson}, J., {et~al.} 2020, \apj, 896,
  125, \dodoi{10.3847/1538-4357/ab9310}

\bibitem[{{Croom} {et~al.}(2021){Croom}, {Owers}, {Scott}, {Poetrodjojo},
  {Groves}, {van de Sande}, {Barone}, {Cortese}, {D'Eugenio}, {Bland-Hawthorn},
  {Bryant}, {Oh}, {Brough}, {Agostino}, {Casura}, {Catinella}, {Colless},
  {Cecil}, {Davies}, {Drinkwater}, {Driver}, {Ferreras}, {Foster},
  {Fraser-McKelvie}, {Lawrence}, {Leslie}, {Liske}, {L{\'o}pez-S{\'a}nchez},
  {Lorente}, {McElroy}, {Medling}, {Obreschkow}, {Richards}, {Sharp}, {Sweet},
  {Taranu}, {Taylor}, {Tescari}, {Thomas}, {Tocknell}, \& {Vaughan}}]{croom21}
{Croom}, S.~M., {Owers}, M.~S., {Scott}, N., {et~al.} 2021, \mnras, 505, 991,
  \dodoi{10.1093/mnras/stab229}

\bibitem[{{Dalcanton} {et~al.}(1997){Dalcanton}, {Spergel}, \&
  {Summers}}]{Dalcanton97}
{Dalcanton}, J.~J., {Spergel}, D.~N., \& {Summers}, F.~J. 1997, \apj, 482, 659,
  \dodoi{10.1086/304182}

\bibitem[{{Danovich} {et~al.}(2015{\natexlab{a}}){Danovich}, {Dekel}, {Hahn},
  {Ceverino}, \& {Primack}}]{2015MNRAS.449.2087D}
{Danovich}, M., {Dekel}, A., {Hahn}, O., {Ceverino}, D., \& {Primack}, J.
  2015{\natexlab{a}}, \mnras, 449, 2087, \dodoi{10.1093/mnras/stv270}

\bibitem[{{Danovich} {et~al.}(2015{\natexlab{b}}){Danovich}, {Dekel}, {Hahn},
  {Ceverino}, \& {Primack}}]{danovich15}
---. 2015{\natexlab{b}}, \mnras, 449, 2087, \dodoi{10.1093/mnras/stv270}

\bibitem[{{DeFelippis} {et~al.}(2021){DeFelippis}, {Bouch{\'e}}, {Genel},
  {Bryan}, {Nelson}, {Marinacci}, \& {Hernquist}}]{DeFelippis21}
{DeFelippis}, D., {Bouch{\'e}}, N.~F., {Genel}, S., {et~al.} 2021, \apj, 923,
  56, \dodoi{10.3847/1538-4357/ac2cbf}

\bibitem[{{DeFelippis} {et~al.}(2017){DeFelippis}, {Genel}, {Bryan}, \&
  {Fall}}]{DeFelippis17}
{DeFelippis}, D., {Genel}, S., {Bryan}, G.~L., \& {Fall}, S.~M. 2017, \apj,
  841, 16, \dodoi{10.3847/1538-4357/aa6dfc}

\bibitem[{{Dekel} {et~al.}(2009{\natexlab{a}}){Dekel}, {Sari}, \&
  {Ceverino}}]{2009ApJ...703..785D}
{Dekel}, A., {Sari}, R., \& {Ceverino}, D. 2009{\natexlab{a}}, \apj, 703, 785,
  \dodoi{10.1088/0004-637X/703/1/785}

\bibitem[{{Dekel} {et~al.}(2009{\natexlab{b}}){Dekel}, {Birnboim}, {Engel},
  {Freundlich}, {Goerdt}, {Mumcuoglu}, {Neistein}, {Pichon}, {Teyssier}, \&
  {Zinger}}]{dekel09}
{Dekel}, A., {Birnboim}, Y., {Engel}, G., {et~al.} 2009{\natexlab{b}}, \nat,
  457, 451, \dodoi{10.1038/nature07648}

\bibitem[{{El-Badry} {et~al.}(2018{\natexlab{a}}){El-Badry}, {Bradford},
  {Quataert}, {Geha}, {Boylan-Kolchin}, {Weisz}, {Wetzel}, {Hopkins}, {Chan},
  {Fitts}, {Kere{\v{s}}}, \& {Faucher-Gigu{\`e}re}}]{2018MNRAS.477.1536E}
{El-Badry}, K., {Bradford}, J., {Quataert}, E., {et~al.} 2018{\natexlab{a}},
  \mnras, 477, 1536, \dodoi{10.1093/mnras/sty730}

\bibitem[{{El-Badry} {et~al.}(2018{\natexlab{b}}){El-Badry}, {Quataert},
  {Wetzel}, {Hopkins}, {Weisz}, {Chan}, {Fitts}, {Boylan-Kolchin},
  {Kere{\v{s}}}, {Faucher-Gigu{\`e}re}, \& {Garrison-Kimmel}}]{EB18}
{El-Badry}, K., {Quataert}, E., {Wetzel}, A., {et~al.} 2018{\natexlab{b}},
  \mnras, 473, 1930, \dodoi{10.1093/mnras/stx2482}

\bibitem[{{El-Badry} {et~al.}(2018{\natexlab{c}}){El-Badry}, {Quataert},
  {Wetzel}, {Hopkins}, {Weisz}, {Chan}, {Fitts}, {Boylan-Kolchin},
  {Kere{\v{s}}}, {Faucher-Gigu{\`e}re}, \& {Garrison-Kimmel}}]{elbadry18}
---. 2018{\natexlab{c}}, \mnras, 473, 1930, \dodoi{10.1093/mnras/stx2482}

\bibitem[{{Fall}(1983)}]{1983IAUS..100..391F}
{Fall}, S.~M. 1983, in Internal Kinematics and Dynamics of Galaxies, ed.
  E.~{Athanassoula}, Vol. 100, 391--398

\bibitem[{{Fall} \& {Efstathiou}(1980)}]{Fall80}
{Fall}, S.~M., \& {Efstathiou}, G. 1980, \mnras, 193, 189,
  \dodoi{10.1093/mnras/193.2.189}

\bibitem[{{Fall} \& {Romanowsky}(2013)}]{RF13}
{Fall}, S.~M., \& {Romanowsky}, A.~J. 2013, \apjl, 769, L26,
  \dodoi{10.1088/2041-8205/769/2/L26}

\bibitem[{{Garrison-Kimmel} {et~al.}(2018){Garrison-Kimmel}, {Hopkins},
  {Wetzel}, {El-Badry}, {Sanderson}, {Bullock}, {Ma}, {van de Voort}, {Hafen},
  {Faucher-Gigu{\`e}re}, {Hayward}, {Quataert}, {Kere{\v{s}}}, \&
  {Boylan-Kolchin}}]{GK18}
{Garrison-Kimmel}, S., {Hopkins}, P.~F., {Wetzel}, A., {et~al.} 2018, \mnras,
  481, 4133, \dodoi{10.1093/mnras/sty2513}

\bibitem[{{Genel} {et~al.}(2015){Genel}, {Fall}, {Hernquist}, {Vogelsberger},
  {Snyder}, {Rodriguez-Gomez}, {Sijacki}, \& {Springel}}]{genel15}
{Genel}, S., {Fall}, S.~M., {Hernquist}, L., {et~al.} 2015, \apjl, 804, L40,
  \dodoi{10.1088/2041-8205/804/2/L40}

\bibitem[{{Girard} {et~al.}(2020){Girard}, {Mason}, {Fontana},
  {Dessauges-Zavadsky}, {Morishita}, {Amor{\'\i}n}, {Fisher}, {Jones},
  {Schaerer}, {Schmidt}, {Treu}, \& {Vulcani}}]{girard20}
{Girard}, M., {Mason}, C.~A., {Fontana}, A., {et~al.} 2020, \mnras, 497, 173,
  \dodoi{10.1093/mnras/staa1907}

\bibitem[{{Grand} {et~al.}(2024){Grand}, {Fragkoudi}, {G{\'o}mez}, {Jenkins},
  {Marinacci}, {Pakmor}, \& {Springel}}]{Grand24}
{Grand}, R. J.~J., {Fragkoudi}, F., {G{\'o}mez}, F.~A., {et~al.} 2024, \mnras,
  \dodoi{10.1093/mnras/stae1598}

\bibitem[{{Grand} {et~al.}(2017){Grand}, {G{\'o}mez}, {Marinacci}, {Pakmor},
  {Springel}, {Campbell}, {Frenk}, {Jenkins}, \& {White}}]{2017MNRAS.467..179G}
{Grand}, R. J.~J., {G{\'o}mez}, F.~A., {Marinacci}, F., {et~al.} 2017, \mnras,
  467, 179, \dodoi{10.1093/mnras/stx071}

\bibitem[{{Grand} {et~al.}(2019){Grand}, {van de Voort}, {Zjupa}, {Fragkoudi},
  {G{\'o}mez}, {Kauffmann}, {Marinacci}, {Pakmor}, {Springel}, \&
  {White}}]{Grand19}
{Grand}, R. J.~J., {van de Voort}, F., {Zjupa}, J., {et~al.} 2019, \mnras, 490,
  4786, \dodoi{10.1093/mnras/stz2928}

\bibitem[{{Hafen} {et~al.}(2022){Hafen}, {Stern}, {Bullock}, {Gurvich}, {Yu},
  {Faucher-Gigu{\`e}re}, {Fielding}, {Angl{\'e}s-Alc{\'a}zar}, {Quataert},
  {Wetzel}, {Starkenburg}, {Boylan-Kolchin}, {Moreno}, {Feldmann}, {El-Badry},
  {Chan}, {Trapp}, {Kere{\v{s}}}, \& {Hopkins}}]{Hafen22}
{Hafen}, Z., {Stern}, J., {Bullock}, J., {et~al.} 2022, \mnras, 514, 5056,
  \dodoi{10.1093/mnras/stac1603}

\bibitem[{{Hahn} {et~al.}(2010){Hahn}, {Teyssier}, \&
  {Carollo}}]{2010MNRAS.405..274H}
{Hahn}, O., {Teyssier}, R., \& {Carollo}, C.~M. 2010, \mnras, 405, 274,
  \dodoi{10.1111/j.1365-2966.2010.16494.x}

\bibitem[{{Helmi} {et~al.}(2018){Helmi}, {Babusiaux}, {Koppelman}, {Massari},
  {Veljanoski}, \& {Brown}}]{2018Natur.563...85H}
{Helmi}, A., {Babusiaux}, C., {Koppelman}, H.~H., {et~al.} 2018, \nat, 563, 85,
  \dodoi{10.1038/s41586-018-0625-x}

\bibitem[{{Ho} {et~al.}(2017){Ho}, {Martin}, {Kacprzak}, \& {Churchill}}]{Ho17}
{Ho}, S.~H., {Martin}, C.~L., {Kacprzak}, G.~G., \& {Churchill}, C.~W. 2017,
  \apj, 835, 267, \dodoi{10.3847/1538-4357/835/2/267}

\bibitem[{{Hodges-Kluck} {et~al.}(2016){Hodges-Kluck}, {Miller}, \&
  {Bregman}}]{2016ApJ...822...21H}
{Hodges-Kluck}, E.~J., {Miller}, M.~J., \& {Bregman}, J.~N. 2016, \apj, 822,
  21, \dodoi{10.3847/0004-637X/822/1/21}

\bibitem[{{Hopkins} {et~al.}(2023){Hopkins}, {Gurvich}, {Shen}, {Hafen},
  {Grudi{\'c}}, {Kurinchi-Vendhan}, {Hayward}, {Jiang}, {Orr}, {Wetzel},
  {Kere{\v{s}}}, {Stern}, {Faucher-Gigu{\`e}re}, {Bullock}, {Wheeler},
  {El-Badry}, {Loebman}, {Moreno}, {Boylan-Kolchin}, \& {Quataert}}]{hopkins23}
{Hopkins}, P.~F., {Gurvich}, A.~B., {Shen}, X., {et~al.} 2023, \mnras, 525,
  2241, \dodoi{10.1093/mnras/stad1902}

\bibitem[{{Hoyle}(1951)}]{Hoyle51}
{Hoyle}, F. 1951, in Problems of Cosmical Aerodynamics, 195

\bibitem[{{Hummels} {et~al.}(2019){Hummels}, {Smith}, {Hopkins}, {O'Shea},
  {Silvia}, {Werk}, {Lehner}, {Wise}, {Collins}, \& {Butsky}}]{hummels19}
{Hummels}, C.~B., {Smith}, B.~D., {Hopkins}, P.~F., {et~al.} 2019, \apj, 882,
  156, \dodoi{10.3847/1538-4357/ab378f}

\bibitem[{{Hunter}(2007)}]{hunter2007}
{Hunter}, J.~D. 2007, Computing in Science and Engineering, 9, 90,
  \dodoi{10.1109/MCSE.2007.55}

\bibitem[{{Jiang} {et~al.}(2019){Jiang}, {Dekel}, {Kneller}, {Lapiner},
  {Ceverino}, {Primack}, {Faber}, {Macci{\`o}}, {Dutton}, {Genel}, \&
  {Somerville}}]{2019MNRAS.488.4801J}
{Jiang}, F., {Dekel}, A., {Kneller}, O., {et~al.} 2019, \mnras, 488, 4801,
  \dodoi{10.1093/mnras/stz1952}

\bibitem[{{Jin} {et~al.}(2016){Jin}, {Chen}, {Shi}, {Tremonti}, {Bershady},
  {Merrifield}, {Emsellem}, {Fu}, {Wake}, {Bundy}, {Lin}, {Argudo-Fernandez},
  {Huang}, {Stark}, {Storchi-Bergmann}, {Bizyaev}, {Brownstein}, {Chisholm},
  {Guo}, {Hao}, {Hu}, {Li}, {Li}, {Masters}, {Malanushenko}, {Pan}, {Riffel},
  {Roman-Lopes}, {Simmons}, {Thomas}, {Wang}, {Westfall}, \& {Yan}}]{Jin16}
{Jin}, Y., {Chen}, Y., {Shi}, Y., {et~al.} 2016, \mnras, 463, 913,
  \dodoi{10.1093/mnras/stw2055}

\bibitem[{{Kassin} {et~al.}(2014){Kassin}, {Brooks}, {Governato}, {Weiner}, \&
  {Gardner}}]{kassin14}
{Kassin}, S.~A., {Brooks}, A., {Governato}, F., {Weiner}, B.~J., \& {Gardner},
  J.~P. 2014, \apj, 790, 89, \dodoi{10.1088/0004-637X/790/2/89}

\bibitem[{{Kassin} {et~al.}(2012{\natexlab{a}}){Kassin}, {Devriendt}, {Fall},
  {de Jong}, {Allgood}, \& {Primack}}]{Kassin12a}
{Kassin}, S.~A., {Devriendt}, J., {Fall}, S.~M., {et~al.} 2012{\natexlab{a}},
  \mnras, 424, 502, \dodoi{10.1111/j.1365-2966.2012.21219.x}

\bibitem[{{Kassin} {et~al.}(2012{\natexlab{b}}){Kassin}, {Weiner}, {Faber},
  {Gardner}, {Willmer}, {Coil}, {Cooper}, {Devriendt}, {Dutton},
  {Guhathakurta}, {Koo}, {Metevier}, {Noeske}, \& {Primack}}]{kassin12}
{Kassin}, S.~A., {Weiner}, B.~J., {Faber}, S.~M., {et~al.} 2012{\natexlab{b}},
  \apj, 758, 106, \dodoi{10.1088/0004-637X/758/2/106}

\bibitem[{{Kere{\v{s}}} {et~al.}(2009){Kere{\v{s}}}, {Katz}, {Fardal},
  {Dav{\'e}}, \& {Weinberg}}]{keres09}
{Kere{\v{s}}}, D., {Katz}, N., {Fardal}, M., {Dav{\'e}}, R., \& {Weinberg},
  D.~H. 2009, \mnras, 395, 160, \dodoi{10.1111/j.1365-2966.2009.14541.x}

\bibitem[{{Kere{\v{s}}} {et~al.}(2005){Kere{\v{s}}}, {Katz}, {Weinberg}, \&
  {Dav{\'e}}}]{keres05}
{Kere{\v{s}}}, D., {Katz}, N., {Weinberg}, D.~H., \& {Dav{\'e}}, R. 2005,
  \mnras, 363, 2, \dodoi{10.1111/j.1365-2966.2005.09451.x}

\bibitem[{{Kimm} {et~al.}(2011){Kimm}, {Devriendt}, {Slyz}, {Pichon}, {Kassin},
  \& {Dubois}}]{Kimm11}
{Kimm}, T., {Devriendt}, J., {Slyz}, A., {et~al.} 2011, arXiv e-prints,
  arXiv:1106.0538.
\newblock \doarXiv{1106.0538}

\bibitem[{{Kopenhafer} {et~al.}(2023){Kopenhafer}, {O'Shea}, \&
  {Voit}}]{Kopenhafer23}
{Kopenhafer}, C., {O'Shea}, B.~W., \& {Voit}, G.~M. 2023, \apj, 951, 107,
  \dodoi{10.3847/1538-4357/accbb7}

\bibitem[{{Lagos} {et~al.}(2017){Lagos}, {Theuns}, {Stevens}, {Cortese},
  {Padilla}, {Davis}, {Contreras}, \& {Croton}}]{Lagos17}
{Lagos}, C. d.~P., {Theuns}, T., {Stevens}, A. R.~H., {et~al.} 2017, \mnras,
  464, 3850, \dodoi{10.1093/mnras/stw2610}

\bibitem[{{Lehner} {et~al.}(2021){Lehner}, {Kopenhafer}, {O'Meara}, {Howk},
  {Fumagalli}, {Prochaska}, {Acharyya}, {O'Shea}, {Peeples}, {Tumlinson}, \&
  {Hummels}}]{lehner21}
{Lehner}, N., {Kopenhafer}, C., {O'Meara}, J., {et~al.} 2021, arXiv e-prints,
  arXiv:2112.03304.
\newblock \doarXiv{2112.03304}

\bibitem[{{Li} {et~al.}(2022){Li}, {Obreschkow}, {Power}, \& {Lagos}}]{LiJ22}
{Li}, J., {Obreschkow}, D., {Power}, C., \& {Lagos}, C. d.~P. 2022, \mnras,
  515, 437, \dodoi{10.1093/mnras/stac1740}

\bibitem[{{Liu} {et~al.}(2024){Liu}, {Guo}, {Wang}, {Xu}, {Lu}, {Cui}, \&
  {Dav'e}}]{Liu24}
{Liu}, K., {Guo}, H., {Wang}, S., {et~al.} 2024, arXiv e-prints,
  arXiv:2409.09379, \dodoi{10.48550/arXiv.2409.09379}

\bibitem[{{Lochhaas} {et~al.}(2021){Lochhaas}, {Tumlinson}, {O'Shea},
  {Peeples}, {Smith}, {Werk}, {Augustin}, \& {Simons}}]{lochhaas21}
{Lochhaas}, C., {Tumlinson}, J., {O'Shea}, B.~W., {et~al.} 2021, \apj, 922,
  121, \dodoi{10.3847/1538-4357/ac2496}

\bibitem[{{Lochhaas} {et~al.}(2019){Lochhaas}, {Mathur}, {Frank}, {Som},
  {Krongold}, {Kulkarni}, {Weinberg}, {Nicastro}, \& {Gupta}}]{lochhaas19}
{Lochhaas}, C., {Mathur}, S., {Frank}, S., {et~al.} 2019, \mnras, 489, 78,
  \dodoi{10.1093/mnras/stz2108}

\bibitem[{{Lochhaas} {et~al.}(2023){Lochhaas}, {Tumlinson}, {Peeples},
  {O'Shea}, {Werk}, {Simons}, {Juno}, {Kopenhafer}, {Augustin}, {Wright},
  {Acharyya}, \& {Smith}}]{lochhaas23}
{Lochhaas}, C., {Tumlinson}, J., {Peeples}, M.~S., {et~al.} 2023, \apj, 948,
  43, \dodoi{10.3847/1538-4357/acbb06}

\bibitem[{{Lockman} {et~al.}(2023){Lockman}, {Benjamin}, {Pichette}, \&
  {Thibodeau}}]{Lockman23}
{Lockman}, F.~J., {Benjamin}, R.~A., {Pichette}, N., \& {Thibodeau}, C. 2023,
  \apj, 943, 55, \dodoi{10.3847/1538-4357/aca764}

\bibitem[{{Lopez} {et~al.}(2018){Lopez}, {Tejos}, {Ledoux}, {Barrientos},
  {Sharon}, {Rigby}, {Gladders}, {Bayliss}, \& {Pessa}}]{Lopez18}
{Lopez}, S., {Tejos}, N., {Ledoux}, C., {et~al.} 2018, \nat, 554, 493,
  \dodoi{10.1038/nature25436}

\bibitem[{{Lopez} {et~al.}(2020){Lopez}, {Tejos}, {Barrientos}, {Ledoux},
  {Sharon}, {Katsianis}, {Florian}, {Rivera-Thorsen}, {Bayliss}, {Dahle},
  {Fernandez-Figueroa}, {Gladders}, {Gronke}, {Hamel}, {Pessa}, \&
  {Rigby}}]{Lopez20}
{Lopez}, S., {Tejos}, N., {Barrientos}, L.~F., {et~al.} 2020, \mnras, 491,
  4442, \dodoi{10.1093/mnras/stz3183}

\bibitem[{{Ma} {et~al.}(2017){Ma}, {Hopkins}, {Wetzel}, {Kirby},
  {Angl{\'e}s-Alc{\'a}zar}, {Faucher-Gigu{\`e}re}, {Kere{\v{s}}}, \&
  {Quataert}}]{2017MNRAS.467.2430M}
{Ma}, X., {Hopkins}, P.~F., {Wetzel}, A.~R., {et~al.} 2017, \mnras, 467, 2430,
  \dodoi{10.1093/mnras/stx273}

\bibitem[{{Mandelker} {et~al.}(2017){Mandelker}, {Dekel}, {Ceverino}, {DeGraf},
  {Guo}, \& {Primack}}]{2017MNRAS.464..635M}
{Mandelker}, N., {Dekel}, A., {Ceverino}, D., {et~al.} 2017, \mnras, 464, 635,
  \dodoi{10.1093/mnras/stw2358}

\bibitem[{{Mandelker} {et~al.}(2014){Mandelker}, {Dekel}, {Ceverino}, {Tweed},
  {Moody}, \& {Primack}}]{2014MNRAS.443.3675M}
---. 2014, \mnras, 443, 3675, \dodoi{10.1093/mnras/stu1340}

\bibitem[{{Martin} {et~al.}(2019){Martin}, {Ho}, {Kacprzak}, \&
  {Churchill}}]{Martin19}
{Martin}, C.~L., {Ho}, S.~H., {Kacprzak}, G.~G., \& {Churchill}, C.~W. 2019,
  \apj, 878, 84, \dodoi{10.3847/1538-4357/ab18ac}

\bibitem[{{McCluskey} {et~al.}(2023){McCluskey}, {Wetzel}, {Loebman}, {Moreno},
  \& {Faucher-Giguere}}]{2023arXiv230314210M}
{McCluskey}, F., {Wetzel}, A., {Loebman}, S.~R., {Moreno}, J., \&
  {Faucher-Giguere}, C.-A. 2023, arXiv e-prints, arXiv:2303.14210,
  \dodoi{10.48550/arXiv.2303.14210}

\bibitem[{{McCluskey} {et~al.}(2024){McCluskey}, {Wetzel}, {Loebman}, {Moreno},
  {Faucher-Gigu{\`e}re}, \& {Hopkins}}]{McCluskey24}
{McCluskey}, F., {Wetzel}, A., {Loebman}, S.~R., {et~al.} 2024, \mnras, 527,
  6926, \dodoi{10.1093/mnras/stad3547}

\bibitem[{{Mo} {et~al.}(1998){Mo}, {Mao}, \& {White}}]{Mo98}
{Mo}, H.~J., {Mao}, S., \& {White}, S. D.~M. 1998, \mnras, 295, 319,
  \dodoi{10.1046/j.1365-8711.1998.01227.x}

\bibitem[{{Oppenheimer}(2018)}]{Oppenheimer18}
{Oppenheimer}, B.~D. 2018, \mnras, 480, 2963, \dodoi{10.1093/mnras/sty1918}

\bibitem[{{Peebles}(1969)}]{peebles69}
{Peebles}, P.~J.~E. 1969, \apj, 155, 393, \dodoi{10.1086/149876}

\bibitem[{{Peeples} {et~al.}(2019){Peeples}, {Corlies}, {Tumlinson}, {O'Shea},
  {Lehner}, {O'Meara}, {Howk}, {Earl}, {Smith}, {Wise}, \&
  {Hummels}}]{Peeples19}
{Peeples}, M.~S., {Corlies}, L., {Tumlinson}, J., {et~al.} 2019, \apj, 873,
  129, \dodoi{10.3847/1538-4357/ab0654}

\bibitem[{{Pillepich} {et~al.}(2019){Pillepich}, {Nelson}, {Springel},
  {Pakmor}, {Torrey}, {Weinberger}, {Vogelsberger}, {Marinacci}, {Genel}, {van
  der Wel}, \& {Hernquist}}]{2019MNRAS.490.3196P}
{Pillepich}, A., {Nelson}, D., {Springel}, V., {et~al.} 2019, \mnras, 490,
  3196, \dodoi{10.1093/mnras/stz2338}

\bibitem[{{Pontzen} \& {Tremmel}(2018)}]{pontzen2018}
{Pontzen}, A., \& {Tremmel}, M. 2018, \apjs, 237, 23,
  \dodoi{10.3847/1538-4365/aac832}

\bibitem[{{Ristea} {et~al.}(2022){Ristea}, {Cortese}, {Fraser-McKelvie},
  {Brough}, {Bryant}, {Catinella}, {Croom}, {Groves}, {Richards}, {van de
  Sande}, {Bland-Hawthorn}, {Owers}, \& {Lawrence}}]{Ristea22}
{Ristea}, A., {Cortese}, L., {Fraser-McKelvie}, A., {et~al.} 2022, \mnras, 517,
  2677, \dodoi{10.1093/mnras/stac2839}

\bibitem[{{Rodriguez} {et~al.}(2024){Rodriguez}, {Merch{\'a}n}, \&
  {Artale}}]{Rodriguez24}
{Rodriguez}, F., {Merch{\'a}n}, M., \& {Artale}, M.~C. 2024, arXiv e-prints,
  arXiv:2405.02398, \dodoi{10.48550/arXiv.2405.02398}

\bibitem[{{Rodriguez-Gomez} {et~al.}(2022){Rodriguez-Gomez}, {Genel}, {Fall},
  {Pillepich}, {Huertas-Company}, {Nelson}, {P{\'e}rez-Monta{\~n}o},
  {Marinacci}, {Pakmor}, {Springel}, {Vogelsberger}, \& {Hernquist}}]{RG22}
{Rodriguez-Gomez}, V., {Genel}, S., {Fall}, S.~M., {et~al.} 2022, \mnras, 512,
  5978, \dodoi{10.1093/mnras/stac806}

\bibitem[{{Romanowsky} \& {Fall}(2012)}]{RF12}
{Romanowsky}, A.~J., \& {Fall}, S.~M. 2012, \apjs, 203, 17,
  \dodoi{10.1088/0067-0049/203/2/17}

\bibitem[{{S{\'a}nchez} {et~al.}(2023){S{\'a}nchez}, {Galbany}, {Walcher},
  {Garc{\'\i}a-Benito}, \& {Barrera-Ballesteros}}]{sanchez23}
{S{\'a}nchez}, S.~F., {Galbany}, L., {Walcher}, C.~J., {Garc{\'\i}a-Benito},
  R., \& {Barrera-Ballesteros}, J.~K. 2023, \mnras, 526, 5555,
  \dodoi{10.1093/mnras/stad3119}

\bibitem[{{Schaye} {et~al.}(2015){Schaye}, {Crain}, {Bower}, {Furlong},
  {Schaller}, {Theuns}, {Dalla Vecchia}, {Frenk}, {McCarthy}, {Helly},
  {Jenkins}, {Rosas-Guevara}, {White}, {Baes}, {Booth}, {Camps}, {Navarro},
  {Qu}, {Rahmati}, {Sawala}, {Thomas}, \& {Trayford}}]{2015MNRAS.446..521S}
{Schaye}, J., {Crain}, R.~A., {Bower}, R.~G., {et~al.} 2015, \mnras, 446, 521,
  \dodoi{10.1093/mnras/stu2058}

\bibitem[{{Semenov} {et~al.}(2023{\natexlab{a}}){Semenov}, {Conroy}, {Chandra},
  {Hernquist}, \& {Nelson}}]{Semenov23a}
{Semenov}, V.~A., {Conroy}, C., {Chandra}, V., {Hernquist}, L., \& {Nelson}, D.
  2023{\natexlab{a}}, arXiv e-prints, arXiv:2306.09398,
  \dodoi{10.48550/arXiv.2306.09398}

\bibitem[{{Semenov} {et~al.}(2023{\natexlab{b}}){Semenov}, {Conroy}, {Chandra},
  {Hernquist}, \& {Nelson}}]{Semenov23b}
---. 2023{\natexlab{b}}, arXiv e-prints, arXiv:2306.13125,
  \dodoi{10.48550/arXiv.2306.13125}

\bibitem[{{Sharma} \& {Steinmetz}(2005)}]{sharma05}
{Sharma}, S., \& {Steinmetz}, M. 2005, \apj, 628, 21, \dodoi{10.1086/430660}

\bibitem[{{Sheng} {et~al.}(2023){Sheng}, {Zhu}, {Yu}, {Ma}, {Li}, {Wang}, \&
  {Kang}}]{Sheng23}
{Sheng}, M.-J., {Zhu}, L., {Yu}, H.-R., {et~al.} 2023, arXiv e-prints,
  arXiv:2311.07969, \dodoi{10.48550/arXiv.2311.07969}

\bibitem[{{Simons} {et~al.}(2017){Simons}, {Kassin}, {Weiner}, {Faber},
  {Trump}, {Heckman}, {Koo}, {Pacifici}, {Primack}, {Snyder}, \& {de la
  Vega}}]{simons17}
{Simons}, R.~C., {Kassin}, S.~A., {Weiner}, B.~J., {et~al.} 2017, \apj, 843,
  46, \dodoi{10.3847/1538-4357/aa740c}

\bibitem[{{Simons} {et~al.}(2020){Simons}, {Peeples}, {Tumlinson}, {O'Shea},
  {Smith}, {Corlies}, {Lochhaas}, {Zheng}, {Augustin}, {Prasad}, {Snyder}, \&
  {Tollerud}}]{simons20}
{Simons}, R.~C., {Peeples}, M.~S., {Tumlinson}, J., {et~al.} 2020, \apj, 905,
  167, \dodoi{10.3847/1538-4357/abc5b8}

\bibitem[{{Stern} {et~al.}(2024){Stern}, {Fielding}, {Hafen}, {Su}, {Naor},
  {Faucher-Gigu{\`e}re}, {Quataert}, \& {Bullock}}]{stern24}
{Stern}, J., {Fielding}, D., {Hafen}, Z., {et~al.} 2024, \mnras, 530, 1711,
  \dodoi{10.1093/mnras/stae824}

\bibitem[{{Stern} {et~al.}(2021){Stern}, {Faucher-Gigu{\`e}re}, {Fielding},
  {Quataert}, {Hafen}, {Gurvich}, {Ma}, {Byrne}, {El-Badry},
  {Angl{\'e}s-Alc{\'a}zar}, {Chan}, {Feldmann}, {Kere{\v{s}}}, {Wetzel},
  {Murray}, \& {Hopkins}}]{Stern21}
{Stern}, J., {Faucher-Gigu{\`e}re}, C.-A., {Fielding}, D., {et~al.} 2021, \apj,
  911, 88, \dodoi{10.3847/1538-4357/abd776}

\bibitem[{{Stewart} {et~al.}(2013{\natexlab{a}}){Stewart}, {Brooks}, {Bullock},
  {Maller}, {Diemand}, {Wadsley}, \& {Moustakas}}]{2013ApJ...769...74S}
{Stewart}, K.~R., {Brooks}, A.~M., {Bullock}, J.~S., {et~al.}
  2013{\natexlab{a}}, \apj, 769, 74, \dodoi{10.1088/0004-637X/769/1/74}

\bibitem[{{Stewart} {et~al.}(2013{\natexlab{b}}){Stewart}, {Brooks}, {Bullock},
  {Maller}, {Diemand}, {Wadsley}, \& {Moustakas}}]{stewart13}
---. 2013{\natexlab{b}}, \apj, 769, 74, \dodoi{10.1088/0004-637X/769/1/74}

\bibitem[{{Tiley} {et~al.}(2021){Tiley}, {Gillman}, {Cortese}, {Swinbank},
  {Dudzevi{\v{c}}i{\={u}}t{\.{e}}}, {Harrison}, {Smail}, {Obreschkow}, {Croom},
  {Sharples}, \& {Puglisi}}]{tiley21}
{Tiley}, A.~L., {Gillman}, S., {Cortese}, L., {et~al.} 2021, \mnras, 506, 323,
  \dodoi{10.1093/mnras/stab1692}

\bibitem[{{Trapp} {et~al.}(2024){Trapp}, {Kere{\v{s}}}, {Hopkins}, \&
  {Faucher-Gigu{\`e}re}}]{trapp24}
{Trapp}, C.~W., {Kere{\v{s}}}, D., {Hopkins}, P.~F., \& {Faucher-Gigu{\`e}re},
  C.-A. 2024, arXiv e-prints, arXiv:2405.01632,
  \dodoi{10.48550/arXiv.2405.01632}

\bibitem[{{Trapp} {et~al.}(2022){Trapp}, {Kere{\v{s}}}, {Chan}, {Escala},
  {Hummels}, {Hopkins}, {Faucher-Gigu{\`e}re}, {Murray}, {Quataert}, \&
  {Wetzel}}]{trapp22}
{Trapp}, C.~W., {Kere{\v{s}}}, D., {Chan}, T.~K., {et~al.} 2022, \mnras, 509,
  4149, \dodoi{10.1093/mnras/stab3251}

\bibitem[{{Tumlinson} {et~al.}(2017){Tumlinson}, {Peeples}, \& {Werk}}]{tpw17}
{Tumlinson}, J., {Peeples}, M.~S., \& {Werk}, J.~K. 2017, \araa, 55, 389,
  \dodoi{10.1146/annurev-astro-091916-055240}

\bibitem[{{Turk} {et~al.}(2011){Turk}, {Smith}, {Oishi}, {Skory}, {Skillman},
  {Abel}, \& {Norman}}]{ytpaper}
{Turk}, M.~J., {Smith}, B.~D., {Oishi}, J.~S., {et~al.} 2011, The Astrophysical
  Journal Supplement Series, 192, 9, \dodoi{10.1088/0067-0049/192/1/9}

\bibitem[{{{\"U}bler} {et~al.}(2014){{\"U}bler}, {Naab}, {Oser}, {Aumer},
  {Sales}, \& {White}}]{ubler14}
{{\"U}bler}, H., {Naab}, T., {Oser}, L., {et~al.} 2014, \mnras, 443, 2092,
  \dodoi{10.1093/mnras/stu1275}

\bibitem[{{van de Voort} {et~al.}(2019){van de Voort}, {Springel}, {Mandelker},
  {van den Bosch}, \& {Pakmor}}]{vandevoort19}
{van de Voort}, F., {Springel}, V., {Mandelker}, N., {van den Bosch}, F.~C., \&
  {Pakmor}, R. 2019, \mnras, 482, L85, \dodoi{10.1093/mnrasl/sly190}

\bibitem[{{Virtanen} {et~al.}(2020){Virtanen}, {Gommers}, {Oliphant},
  {Haberland}, {Reddy}, {Cournapeau}, {Burovski}, {Peterson}, {Weckesser},
  {Bright}, {van der Walt}, {Brett}, {Wilson}, {Millman}, {Mayorov}, {Nelson},
  {Jones}, {Kern}, {Larson}, {Carey}, {Polat}, {Feng}, {Moore}, {VanderPlas},
  {Laxalde}, {Perktold}, {Cimrman}, {Henriksen}, {Quintero}, {Harris},
  {Archibald}, {Ribeiro}, {Pedregosa}, {van Mulbregt}, \& {SciPy 1. 0
  Contributors}}]{scipy2020}
{Virtanen}, P., {Gommers}, R., {Oliphant}, T.~E., {et~al.} 2020, Nature
  Methods, 17, 261, \dodoi{10.1038/s41592-019-0686-2}

\bibitem[{Walt {et~al.}(2011)Walt, Colbert, \& Varoquaux}]{walt2011numpy}
Walt, S. v.~d., Colbert, S.~C., \& Varoquaux, G. 2011, Computing in Science \&
  Engineering, 13, 22

\bibitem[{{Welker} {et~al.}(2014){Welker}, {Devriendt}, {Dubois}, {Pichon}, \&
  {Peirani}}]{welker14}
{Welker}, C., {Devriendt}, J., {Dubois}, Y., {Pichon}, C., \& {Peirani}, S.
  2014, \mnras, 445, L46, \dodoi{10.1093/mnrasl/slu106}

\bibitem[{{Werk} {et~al.}(2014){Werk}, {Prochaska}, {Tumlinson}, {Peeples},
  {Tripp}, {Fox}, {Lehner}, {Thom}, {O'Meara}, {Ford}, {Bordoloi}, {Katz},
  {Tejos}, {Oppenheimer}, {Dav{\'e}}, \& {Weinberg}}]{werk14}
{Werk}, J.~K., {Prochaska}, J.~X., {Tumlinson}, J., {et~al.} 2014, \apj, 792,
  8, \dodoi{10.1088/0004-637X/792/1/8}

\bibitem[{{White}(1984)}]{white84}
{White}, S.~D.~M. 1984, \apj, 286, 38, \dodoi{10.1086/162573}

\bibitem[{{Wilde} {et~al.}(2021){Wilde}, {Werk}, {Burchett}, {Prochaska},
  {Tchernyshyov}, {Tripp}, {Tejos}, {Lehner}, {Bordoloi}, {O'Meara}, \&
  {Tumlinson}}]{wilde21}
{Wilde}, M.~C., {Werk}, J.~K., {Burchett}, J.~N., {et~al.} 2021, \apj, 912, 9,
  \dodoi{10.3847/1538-4357/abea14}

\bibitem[{{Wilde} {et~al.}(2023){Wilde}, {Tchernyshyov}, {Werk}, {Tripp},
  {Burchett}, {Prochaska}, {Tejos}, {Lehner}, {Bordoloi}, {O'Meara},
  {Tumlinson}, \& {Howk}}]{wilde23}
{Wilde}, M.~C., {Tchernyshyov}, K., {Werk}, J.~K., {et~al.} 2023, arXiv
  e-prints, arXiv:2301.02718, \dodoi{10.48550/arXiv.2301.02718}

\bibitem[{{Wisnioski} {et~al.}(2015){Wisnioski}, {F{\"o}rster Schreiber},
  {Wuyts}, {Wuyts}, {Bandara}, {Wilman}, {Genzel}, {Bender}, {Davies},
  {Fossati}, {Lang}, {Mendel}, {Beifiori}, {Brammer}, {Chan}, {Fabricius},
  {Fudamoto}, {Kulkarni}, {Kurk}, {Lutz}, {Nelson}, {Momcheva}, {Rosario},
  {Saglia}, {Seitz}, {Tacconi}, \& {van Dokkum}}]{wisnioski15}
{Wisnioski}, E., {F{\"o}rster Schreiber}, N.~M., {Wuyts}, S., {et~al.} 2015,
  \apj, 799, 209, \dodoi{10.1088/0004-637X/799/2/209}

\bibitem[{{Wisnioski} {et~al.}(2019){Wisnioski}, {F{\"o}rster Schreiber},
  {Fossati}, {Mendel}, {Wilman}, {Genzel}, {Bender}, {Wuyts}, {Davies},
  {{\"U}bler}, {Bandara}, {Beifiori}, {Belli}, {Brammer}, {Chan}, {Davies},
  {Fabricius}, {Galametz}, {Lang}, {Lutz}, {Nelson}, {Momcheva}, {Price},
  {Rosario}, {Saglia}, {Seitz}, {Shimizu}, {Tacconi}, {Tadaki}, {van Dokkum},
  \& {Wuyts}}]{wisnioski19}
{Wisnioski}, E., {F{\"o}rster Schreiber}, N.~M., {Fossati}, M., {et~al.} 2019,
  \apj, 886, 124, \dodoi{10.3847/1538-4357/ab4db8}

\bibitem[{{Wright} {et~al.}(2024){Wright}, {Tumlinson}, {Peeples}, {O'Shea},
  {Lochhaas}, {Corlies}, {Smith}, {Binh}, {Augustin}, \& {Simons}}]{wright24}
{Wright}, A.~C., {Tumlinson}, J., {Peeples}, M.~S., {et~al.} 2024, \apj, 970,
  70, \dodoi{10.3847/1538-4357/ad49a3}

\bibitem[{{Yu} {et~al.}(2023){Yu}, {Bullock}, {Gurvich}, {Hafen}, {Stern},
  {Boylan-Kolchin}, {Faucher-Gigu{\`e}re}, {Wetzel}, {Hopkins}, \&
  {Moreno}}]{Yu23}
{Yu}, S., {Bullock}, J.~S., {Gurvich}, A.~B., {et~al.} 2023, \mnras, 523, 6220,
  \dodoi{10.1093/mnras/stad1806}

\bibitem[{{Zabl} {et~al.}(2019){Zabl}, {Bouch{\'e}}, {Schroetter}, {Wendt},
  {Finley}, {Schaye}, {Conseil}, {Contini}, {Marino}, {Mitchell}, {Muzahid},
  {Pezzulli}, \& {Wisotzki}}]{Zabl19}
{Zabl}, J., {Bouch{\'e}}, N.~F., {Schroetter}, I., {et~al.} 2019, \mnras, 485,
  1961, \dodoi{10.1093/mnras/stz392}

\bibitem[{{Zheng} {et~al.}(2020){Zheng}, {Peeples}, {O'Shea}, {Simons},
  {Lochhaas}, {Corlies}, {Tumlinson}, {Smith}, \& {Augustin}}]{zheng20}
{Zheng}, Y., {Peeples}, M.~S., {O'Shea}, B.~W., {et~al.} 2020, \apj, 896, 143,
  \dodoi{10.3847/1538-4357/ab960a}

\end{thebibliography}

\end{document}